\documentclass[useAMS,usenatbib]{mn2e}

\usepackage[english]{babel}
\usepackage{amsmath}
\usepackage{txfonts}
\usepackage{graphicx}
\usepackage{url}
\usepackage{amssymb}

\newcommand{\pd}{\partial}
\newcommand{\ud}{\ensuremath{\mathrm{d}}}

\title[Non-Radial Instabilities Progenitor Asphericities and in Core-Collapse Supernovae]{Non-Radial Instabilities and Progenitor Asphericities in Core-Collapse Supernovae}
\author[B.~M\"uller and H.-Th.~Janka]{B.~M\"uller$^{1}$\thanks{E-mail:
bernhard.mueller@monash.edu} and H.-Th.~Janka$^{2}$\\
$^{1}$Monash Centre for Astrophysics, School of
   Mathematical Sciences, Building 28, Monash University, Victoria
   3800, Australia
\\
$^{2}$Max-Planck-Institut f\"ur Astrophysik,
   Karl-Schwarzschild-Str.~1, D-85748 Garching, Germany}

\begin{document}

\maketitle

\begin{abstract}
Since core-collapse supernova simulations still struggle to produce
robust neutrino-driven explosions in 3D, it has been proposed that
asphericities caused by convection in the progenitor might facilitate
shock revival by boosting the activity of non-radial hydrodynamic instabilities
in the post-shock region. We investigate this scenario in depth using
42 relativistic 2D simulations with multi-group neutrino
transport to examine the effects of velocity and density perturbations
in the progenitor for different perturbation geometries that obey 
fundamental physical constraints (like the anelastic condition).
As a framework for analysing our results, we
introduce semi-empirical scaling laws relating neutrino heating,
average turbulent velocities in the gain region, and the shock
deformation in the saturation limit of non-radial instabilities.
The squared turbulent Mach number, $\langle\mathrm{Ma}^2\rangle$,
reflects the violence of aspherical motions in the gain
layer, and explosive runaway occurs for $\langle\mathrm{Ma}^2\rangle
\gtrsim 0.3$, corresponding to a reduction of the critical
neutrino luminosity by $\mathord{\sim}25\%$ compared to 1D. In the light
of this theory, progenitor asphericities aid shock revival mainly by
creating anisotropic mass flux onto the shock: Differential infall
efficiently converts velocity perturbations in the progenitor into
density perturbations $\delta \rho/\rho$ at the shock of the order of
the initial convective Mach number $\mathrm{Ma}_\mathrm{prog}$. The anisotropic mass flux
and ram pressure deform the shock and thereby amplify post-shock
turbulence. Large-scale ($\ell=2$, $\ell=1$) modes prove most
conducive to shock revival, whereas small-scale perturbations require
unrealistically high convective Mach numbers. Initial density
perturbations in the progenitor are only of order
$\mathrm{Ma}_\mathrm{prog}^2$ and therefore play a subdominant role.
\end{abstract}




\section{Introduction}
\label{sec:intro}

The core-collapse supernova explosion mechanism has remained one of
the outstanding challenges in theoretical astrophysics for
decades. Among the various mechanisms that have been proposed over the
years to explain supernova explosions (see \citealt{janka_12} and
\citealt{burrows_13} for an up-to-date summary), the delayed
neutrino-driven mechanism currently remains the best explored and most
promising scenario, at least for core-collapse supernovae with
explosion energies not exceeding $\mathord{\sim} 10^{51} \ \mathrm{erg}$.
Throughout its long history, the idea of shock revival due to
neutrino-heating has seen many refinements; in its modern form it
relies on the joint action of neutrino heating and multi-dimensional
hydrodynamic instabilities in the post-shock region, such as
convection
\citep{herant_92,burrows_92,herant_94,burrows_95,janka_96,mueller_97}
and the standing accretion shock instability (SASI;
\citealp{blondin_03,marek_09} to revive the supernova shock. Over the recent
years, an impressive set of successful two-dimensional (2D)
multi-group neutrino hydrodynamics simulations
\citep{buras_06_b,marek_09,yakunin_10,suwa_10,mueller_12a,mueller_12b,janka_12b,bruenn_13,suwa_13}
has lent further credence to the idea of neutrino-driven supernova
explosions.

However, even in the light of these recent successes of
first-principle modelling in 2D, there are indications that our
understanding of the neutrino-driven mechanism is still incomplete. It
has yet to be demonstrated that the current 2D simulations are
compatible with observed supernova explosion energies, nickel masses
\citep{tanaka_09,smartt_09b,utrobin_11}, and neutron star masses
\citep{schwab_10,valentim_11,oezel_12,kiziltan_13}, although the
explosion energies obtained by \citet{bruenn_13} already fall roughly
within the expected range. More importantly, parametrised studies of
shock revival in core-collapse supernovae in 3D
\citep{hanke_12,couch_12b}, as well as fully-fledged 3D neutrino
hydrodynamics simulations \citep{hanke_13,takiwaki_14} indicate that
explosions may be harder to obtain in 3D (see, however,
\citealt{dolence_13} for a differing opinion). Among the available 3D
neutrino hydrodynamics simulations, those \citep{hanke_13,tamborra_14}
relying on what is currently the most rigorous approach
to neutrino transport and the most
comprehensive treatment of the neutrino microphysics still fail to
show explosions. While it would be exaggerated to state that supernova
theory has reached an impasse after progressing to 3D, this suggests
that some important element for robust neutrino-driven explosions may
yet be missing.

Motivated by these recent results on 3D effects in supernova cores,
\citet{couch_13} proposed (extending ideas of
\citealt{arnett_11}) that \emph{large-scale asymmetries in the
  progenitor} might support shock revival by instigating more violent
aspherical motions in the post-shock region as they collapse and fall
through the shock, thus adding a new twist to a problem at the
interface between late-stage stellar evolution and supernova theory
that has so far been studied primarily as a possible explanation for
pulsar kicks
\citep{burrows_96,goldreich_97,lai_00,fryer_04,murphy_04}.  The idea
of \citet{couch_13} certainly has a firm basis in late-stage stellar
evolution: Seed asphericities will inevitable be present in the
silicon and oxygen shell due to convective burning. A number of
multi-D simulations of the pre-supernova phase suggest that convection
in these shells is violent enough to produce significant density and
velocity perturbations
\citep{arnett_94,bazan_94,bazan_98,asida_00,meakin_07,meakin_07_b,arnett_11},
and those covering large 2D wedges
\citep{meakin_06,meakin_07_b,arnett_11} or the full solid angle in 3D
\citep{kuhlen_03} show that large-scale modes indeed dominate the
flow. In addition, the excitation of g-modes by shell burning
\citep{goldreich_97} may produce asymmetries even within the iron
core, although the analysis of \citet{murphy_04} suggests that the
growth time-scales for such unstable modes may be too long to produce
large deviations from spherical symmetry.

\citet{couch_13} studied the effect of large-scale seed perturbations
using 3D core-collapse simulations with a neutrino leakage scheme.  In
order to asses the impact of the progenitor asphericities, they
computed models with and without initial perturbations, and with two
different settings for a multiplicative factor regulating the neutrino
heating term (not the net neutrino heating) in their leakage
scheme. Among their four simulations, they found an explosion only for
the case with initial perturbations with slightly enhanced neutrino
heating. The similarity of their unperturbed model with enhanced
neutrino heating (by $2\%$) and their perturbed model with the
standard heating rate suggested a slight decrease of the critical
luminosity required for shock revival.

While the idea of \citet{couch_13} is interesting, their work leaves a
number of unanswered questions. Arguably, they find only a rather
modest effect, which is just sufficient to tilt the balance in favour
of an explosion in a marginal case. Moreover, \citet{couch_13}
restricted their attention to a single perturbation pattern. 
In a recent follow-up paper \citep{couch_14}, they
  simulated the same progenitor using ten different perturbation
  patterns, but a systematic exploration of the role of the
  perturbation geometry and amplitude is still lacking.

Furthermore, while taking some constraints on the convective
velocities and the typical spatial scales from multi-D simulations of
supernova progenitors into account (motivated
by \citealt{arnett_11}), \citet{couch_13,couch_14} use
purely transverse velocity perturbations which hardly resemble any
convective flow pattern and even violate the important physical
constraint of near-anelasticity for subsonic flow.  Due to their use
of a neutrino leakage scheme, the heating conditions also deviate
considerably from simulations using a more elaborate neutrino
treatment, at least during the later accretion phase after more than
$150 \ \mathrm{ms}$ after bounce.

In this study, we extend the work of \citet{couch_13,couch_14} with a more
systematic investigation of the role of progenitor asphericities from
convective burning in the neutrino-driven mechanism. As we still lack
multi-D progenitor models evolved up to the onset of collapse, we try
to incorporate more physical constraints in our setup of the initial
perturbation and also explore the effect of different
perturbation amplitudes and geometries in detail.  Unlike
\citet{couch_13,couch_14}, we use a newly developed multi-group transport
scheme based on a one-moment closure of the Boltzmann equation to ensure
reasonable quantitative agreement with the most advanced multi-D
neutrino hydrodynamics simulations. However, in order to explore the
parameter space in depth with over 40 simulations, we restrict
ourselves to axisymmetric 2D models. With this extensive
parameter study, we attempt to address a number of questions
concerning the role of progenitor asphericities in the
explosion mechanism:
\begin{enumerate}
\item Can we better quantify the impact of progenitor asphericities
  on the conditions for shock revival?
\item How relevant is the spatial scale of the convective seed
  perturbations?
\item What are the minimum perturbation amplitudes required
  for an appreciable effect on the heating conditions?
\item Can we better understand the physical mechanism whereby
  perturbations facilitate shock revival?
\end{enumerate}

In addition, we also pursue a second, subsidiary goal: In order to
fully grasp the role of non-radial
instabilities\footnote{In this paper, the term
  ``non-radial'' refers to all modes/instabilities that are not
  \emph{purely} radial, following the widespread usage of this term in
  the literature on hydrodynamic instabilities and stellar
  pulsations.} in the core-collapse supernova explosion mechanism, it
is imperative that we develop a more quantitative understanding of the
interplay of neutrino heating on the one hand and convection and/or
the SASI on the other hand (regardless of whether seed asphericities
in the progenitor are present or not). Several authors have already
proposed theories to explain the saturation properties of the SASI
\citep{guilet_10} and of convection \citep{murphy_11,murphy_12}, but,
with the exception of \citet{murphy_12}, no attempt has yet been made
to break down these theories to simple scaling laws for
volume-integrated quantities that can easily be extracted from multi-D
simulations (total kinetic energies in non-spherical motions,
volume-integrated neutrino heating rate, etc.). Furthermore, only
\citet{murphy_12} undertook the first steps to explain the feedback of
non-radial instabilities on the heating conditions quantitatively by
analysing the effect of turbulent stresses on the average shock
radius. We follow up on some of the ideas enunciated in the
aforementioned papers better by formulating semi-empirical scaling
laws relating neutrino heating, the violence of non-spherical
instabilities, and the deformation of the shock.  We also present some
ideas about quantifying the effect of the non-spherical instabilities
on the neutrino heating conditions and the critical luminosity for
shock revival. Although we cannot hope to fully anatomise the
interplay between neutrino heating and non-radial instabilities in
this paper, we believe that the ideas formulated here may turn out
helpful for the conceptual and quantitative understanding of the role
of non-spherical instabilities in the supernova core in the future.

Our paper is organised as follows: In Section~\ref{sec:convection},
we discuss a number of constraints on the multi-D structure of
(non-rotating) supernova progenitors to provide some background
information for the numerical setup of our simulations, which is
detailed in Section~\ref{sec:setup}.  We then provide a cursory and
descriptive overview of the effects of progenitor asphericities in our
simulations in Section~\ref{sec:overview}. The detailed analysis of
our results is split in two Sections: In Section~\ref{sec:baseline}, we
discuss the evolution of the unperturbed baseline model and establish
quantitative relations governing the interplay between neutrino
heating, non-radial motions in the post-shock region, and large-scale
deformations of the shock. In Section~\ref{sec:perturbations}, we then
present a quantitative analysis of how progenitor asphericities modify
the approach to an explosive runaway, describe the mechanism whereby
progenitor asphericities facilitate shock revival, and discuss the
dependence on the character and geometry of the initial
perturbations. In Section~\ref{sec:conclusions}, we summarise our
results, discuss uncertainties and outline central questions
for future research on the role of progenitor asphericities in
core-collapse supernovae. Our paper also contains two appendices,
the first of which (Appendix~\ref{sec:numerics}) provides a
detailed description of the fast multi-group transport (FMT)
scheme used in our simulations. In Appendix~\ref{sec:toy_model},
we present a simple toy model for the effect of non-spherical
instabilities on the explosion conditions.

\section{Seed Perturbations in the Progenitor}
\label{sec:convection}

While there have been a handful of multi-D simulations of \mbox{Si-,} \mbox{O-,} and
C-shell burning during the late pre-collapse evolution
\citep{arnett_94,bazan_94,bazan_98,asida_00,kuhlen_03,meakin_07,meakin_07_b,arnett_11},
we still lack multi-D progenitor models evolved all the way to
collapse.  For this reason, we are presently forced to impose seed
perturbations onto 1D stellar evolution models by hand. Nevertheless
multi-D simulations of the pre-collapse burning phases, mixing-length
theory, and general physical principles still allow an informed
judgement about the amplitude and the geometry of aspherical seed
perturbations in the progenitor. In the following, we review a few of
these principles in order to obtain some guidelines for constructing
multi-D initial models before describing the initial perturbations
used in our simulations.

\subsection{Properties of Convective Regions in the Progenitor -- Perturbation Amplitudes}
\label{sec:amplitudes}
Both mixing-length theory and multi-D simulations of convective
burning in massive stars furnish estimates for the typical velocity
and density perturbations in the silicon and oxygen shells.  Depending
on the dimensionality, the numerical methodology, and the inclusion or
non-inclusion of multiple burning shells, the magnitude of the
perturbations varies considerably: \citet{bazan_98} reported
relatively large typical Mach numbers of the order of $\mathrm{Ma}_\mathrm{prog}
\sim 0.1 \ldots 0.2$ and density fluctuations of up to $8\%$ in their
2D simulations of oxygen shell burning,  whereas \citet{kuhlen_03} found much
smaller typical Mach numbers $\mathrm{Ma}_\mathrm{prog} \sim 0.01$ and density
fluctuations $\delta \rho/\rho \sim (2 \ldots 3) \times 10^{-3}$ in
their pseudospectral 3D simulations relying on the anelastic
approximation. In a detailed comparison of their compressible
2D and 3D models with the results of \citet{kuhlen_03},
\citet{meakin_07_b} ascribe these differences to the choice of boundary
conditions, which, as they argue, prevented \citet{kuhlen_03} from
capturing large density perturbations at the convective boundaries
associated with convective overshoot. Furthermore,
\citet{meakin_07_b} found smaller typical Mach numbers in 3D compared
to 2D (by a factor of $2\ldots 3$). However, the 3D results of
\citet{meakin_07_b} were limited to a wedge of $30^\circ \times
30^\circ$ and did not include the interaction of multiple burning
shells. As demonstrated by \citet{arnett_11}, shell interactions could
again lead to more violent convection: Their 2D simulations show
considerably higher convective velocities (up to $\mathord{\sim} 2 \times 10^8
\ \mathrm{cm} \ \mathrm{s}^{-1}$) than 2D models without an active
silicon burning shell. Naturally, it remains to be seen whether this
finding is also borne out by full $4 \pi$ simulations in 3D.  Weighing
the limitations of the available multi-D simulations of the
pre-collapse phase, we feel that they still justify the assumption of
convective velocities $\gtrsim 10^8 \ \mathrm{cm} \ \mathrm{s}^{-1}$,
maximum convective Mach numbers $\gtrsim 0.1$ and density fluctuations
of a few percent for exploratory studies.

In spite of all its demerits, one should also consider 1D
mixing-length theory for obtaining a complementary estimate of the
convective velocities in the progenitor. None of the available multi-D
simulations of the pre-collapse phase has been evolved right to the
onset of collapse, although structural changes during the last minutes
may still affect the violence of convective motions in the shells
around the iron core; this is at least suggested by some of the
Kippenhahn diagrams in \citet{heger_00} (e.g.\ their Figs. 18 and
19).  Unlike the presently available multi-D simulations of supernova
progenitors, mixing-length theory allows us to estimate the turbulent
velocities at the onset of collapse taking into account the structural
changes during the last minutes of the pre-collapse phase.

In principle, convective velocities and density perturbations can be
estimated directly from the progenitor profile at the onset of
collapse: Up to a small factor of order unity, the typical turbulent
velocity $\delta v$ is given in terms of the mixing length
$l_\mathrm{mix}$, the local gravitational acceleration $g$, the
density $\rho$, the pressure $P$, and the sound speed $c_s$
as\footnote{Note that the form of $\delta v$ and $\delta \rho$ given
  here is equivalent to the formulation in terms of the temperature
  and composition gradients usually found in textbooks on stellar
  evolution \citep{kippenhahn,cox}, which can be obtained by applying
  simple thermodynamic identities to express the deviation
  $\rho^{-1} (\pd \rho/\pd r-c_s^{-2} \pd P/\pd r)$ from an adiabatic
  density stratification in terms of these variables.}
\begin{equation}
\label{eq:mlt_velocity}
\delta{v}
\approx
\sqrt{
g
\frac{l_\mathrm{mix}^2}{\rho}
\left(\frac{\pd \rho}{\pd r} - \frac{1}{c_s^2}\frac{\pd P}{\pd r}\right)
},
\end{equation}
and the typical density perturbation $\delta \rho/\rho$ is given by
\begin{equation}
\label{eq:mlt_density}
\frac{\delta{\rho}}{\rho}
\approx
\frac{l_\mathrm{mix}}{\rho}
\left(\frac{\pd \rho}{\pd r} - \frac{1}{c_s^2}\frac{\pd P}{\pd r}\right).
\end{equation}
The mixing length $l_\mathrm{mix}$ is typically assumed to be of the
order of the pressure scale-height $\ud r / \ud \ln P$, which implies
$l_\mathrm{mix} \approx P/(\rho g)$ in hydrostatic equilibrium. 

In practice, equations~(\ref{eq:mlt_velocity}) and
(\ref{eq:mlt_density}) are difficult to handle since the density
gradient is typically close to adiabatic (because convection is very
efficient) so that any inconsistency with the equation of state and
the finite-difference representation used in the stellar evolution
model introduces considerable numerical errors. Nevertheless, even
naive estimates of $\delta{v}$ using equation~(\ref{eq:mlt_velocity})
yield convective Mach number of the order of $10^{-2}\ldots 10^{-1}$
(see below for examples), and are not in gross disagreement with
multi-D simulations of the pre-collapse phase.

\subsection{Properties of Convective Regions in the Progenitor -- Flow Geometry}
\label{sec:geometry}
There are likewise a few indications about the flow geometry (in the
broadest sense) of convective motions in the progenitor. The subsonic
character of convection implies that the flow is only weakly
compressible (or, more precisely, almost anelastic), at least in the
interior of the convective zones. If the deviations of the density
field from the spherical background stratification are to remain small,
the velocity field $\mathbf{v}$ must fulfil the condition,
\begin{equation}
\frac{\pd \rho}{\pd t} \approx 0,
\end{equation}
or,
\begin{equation}
\nabla \cdot (\rho \mathbf{v}) \approx 0,
\end{equation}
i.e.\ $\rho \mathbf{v}$ should be a \emph{solenoidal} vector
field. However, the divergence-free condition may be violated
at convective boundaries due to convective overshoot.

Furthermore, simulations indicate that convection is dominated by
\emph{large-scale, low-$\ell$} modes. These correspond to the
fastest-growing modes in the linear regime, i.e.\ convective eddies
that extend over the entire width $\delta r$ of the convective zone
and over a distance $(2 \ldots 3) \delta r$ in the angular direction
(\citealp{foglizzo_06}, cf.\ also \citealp{chandrasekhar_61}).  The
dominant angular wavenumber $\ell$ is therefore given by
\begin{equation}
\ell \sim \frac{\pi}{4}\frac{r_i+r_o}{\delta r},
\end{equation}
in terms of $\delta r$ and the radii $r_i$ and $r_o$ of the inner and
outer boundaries of the convective zones. The very extended oxygen shell seen
in many progenitor models favours the lowest-$\ell$ modes. Consequently,
 \citet{arnett_11} observed a dominant $\ell=4$ mode,
which was the lowest possible mode allowed due to their imposition of
equatorial symmetry, and the simulations of \citet{kuhlen_03} even
showed the presence of an $\ell=2$ mode with two updrafts and
two downdrafts. However, the width of convective regions
in stellar evolution models shows considerable variation across
different progenitors, and low-$\ell$ modes may not be dominant 
in all cases.

\begin{figure*}
\includegraphics[width=0.48 \linewidth]{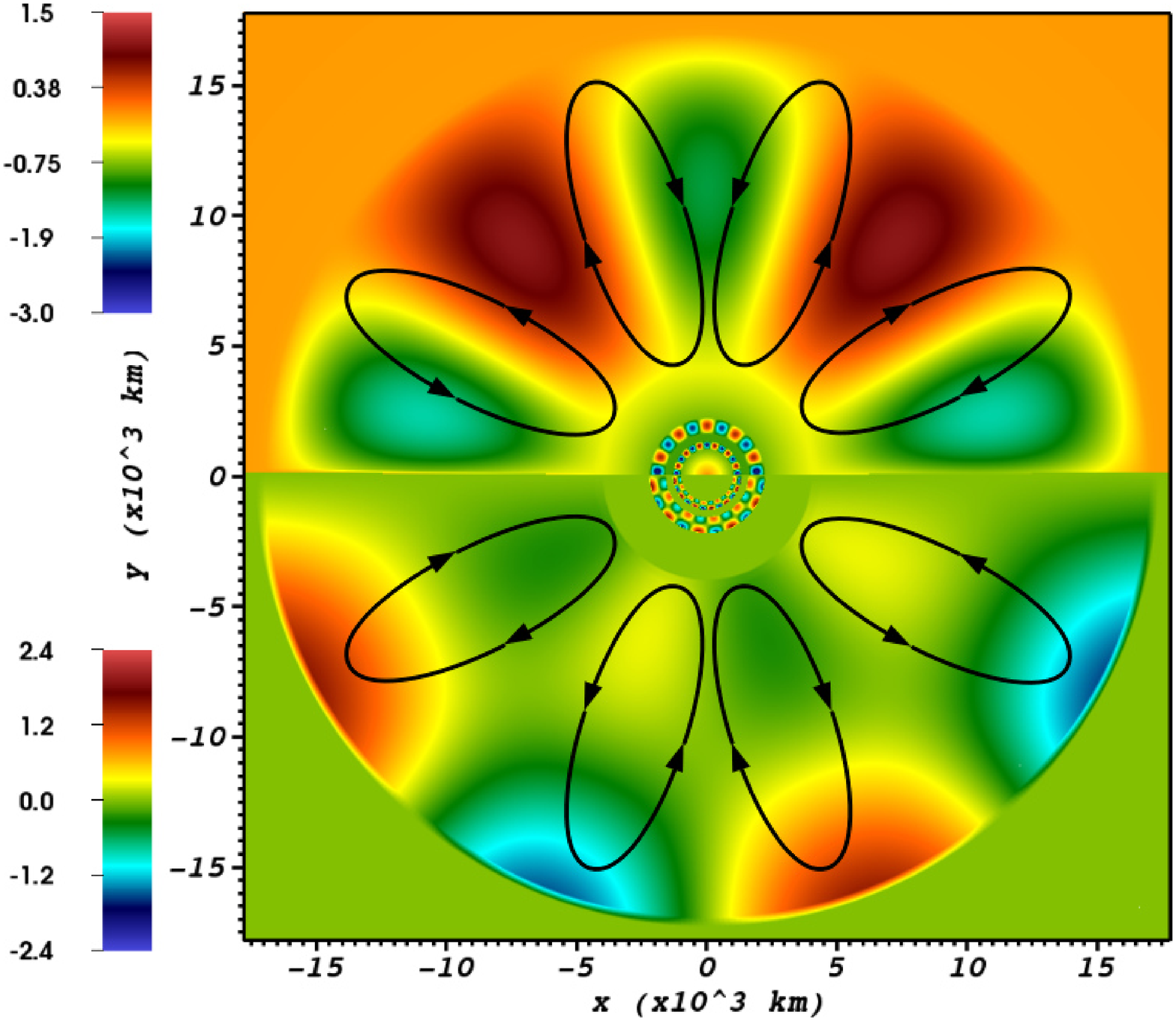}
\hspace{0.03 \linewidth}
\includegraphics[width=0.48 \linewidth]{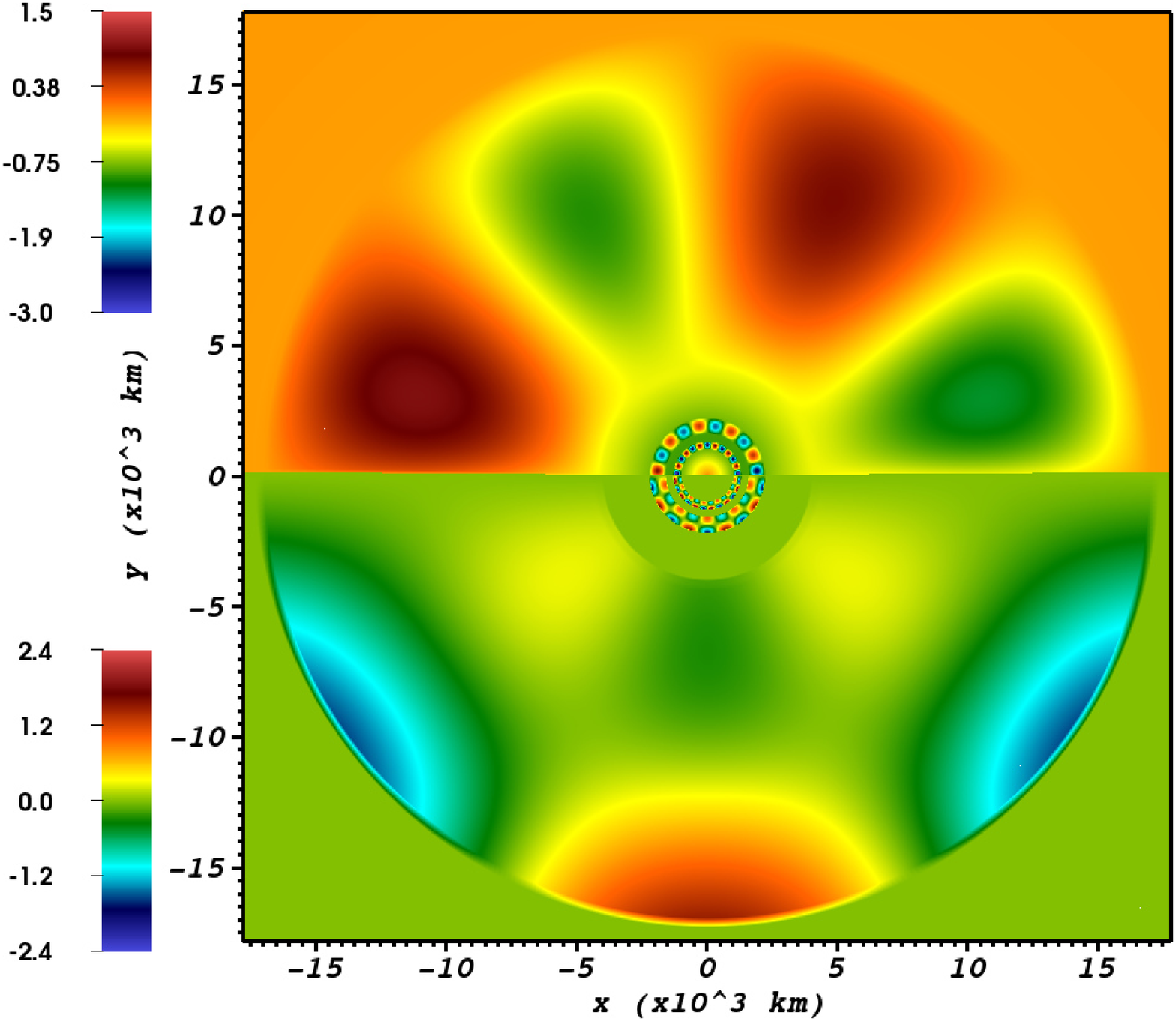}
\\
\includegraphics[width=0.48 \linewidth]{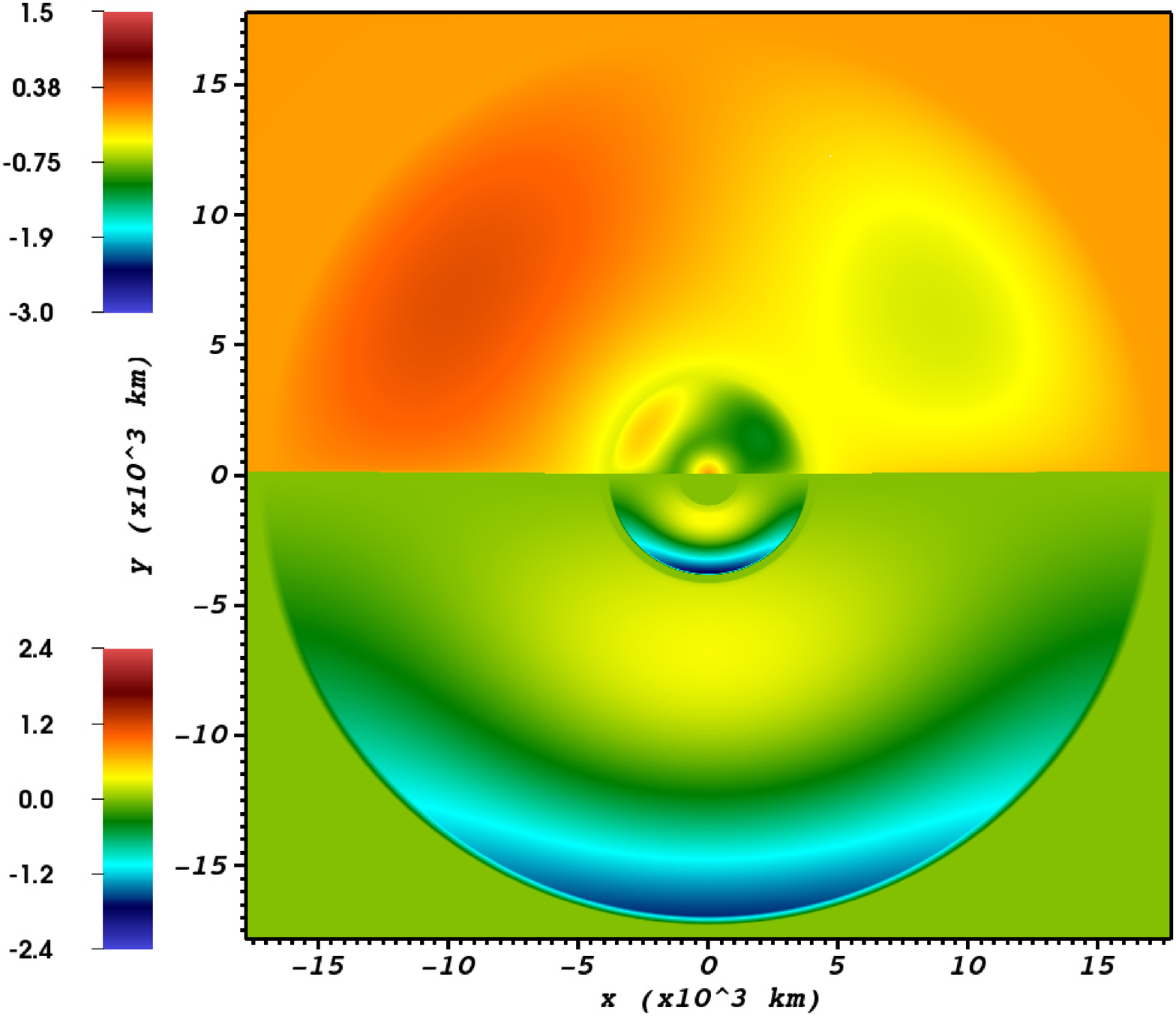}
\hspace{0.03 \linewidth}
\includegraphics[width=0.48 \linewidth]{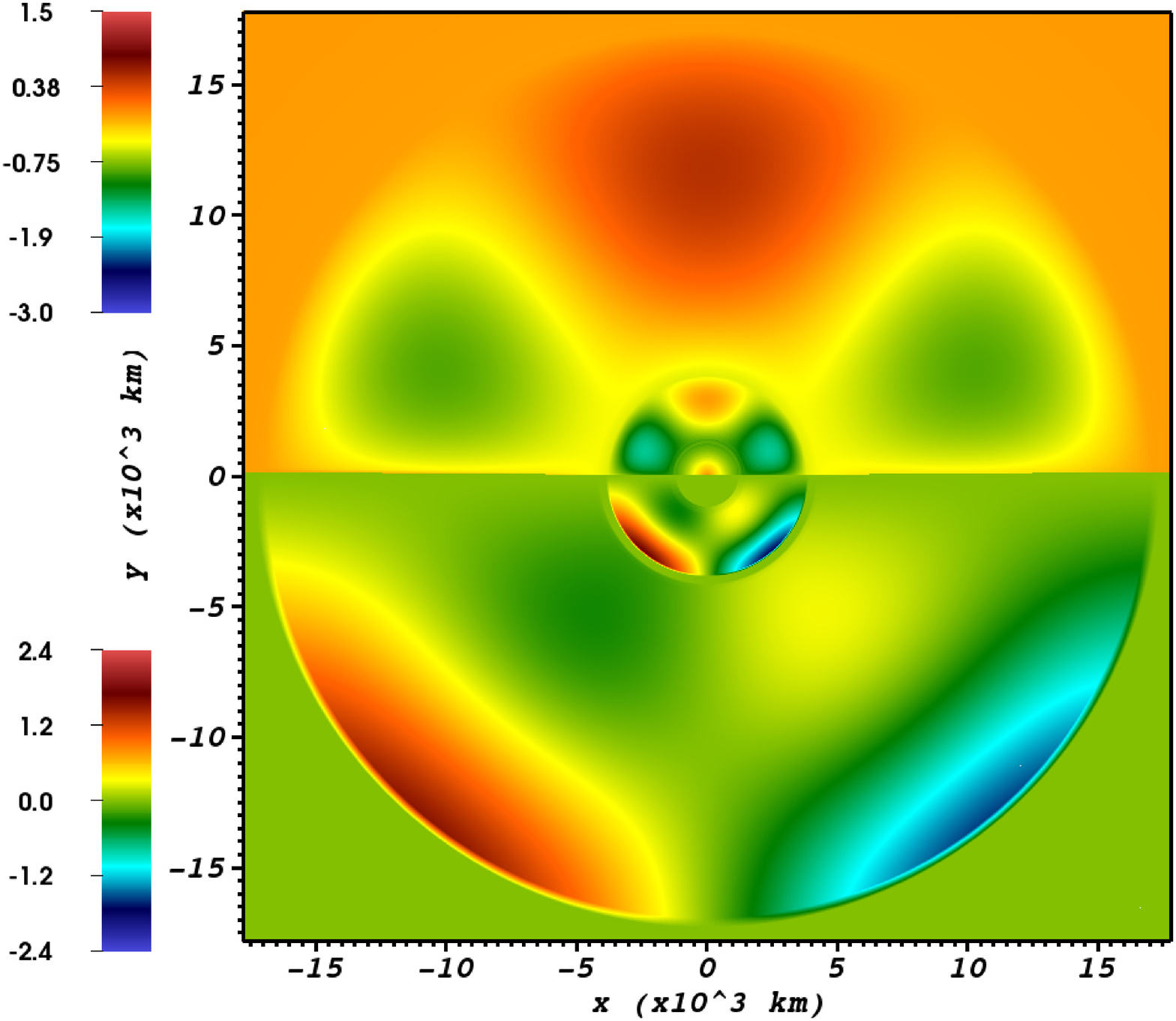}
\\
\includegraphics[width=0.48 \linewidth]{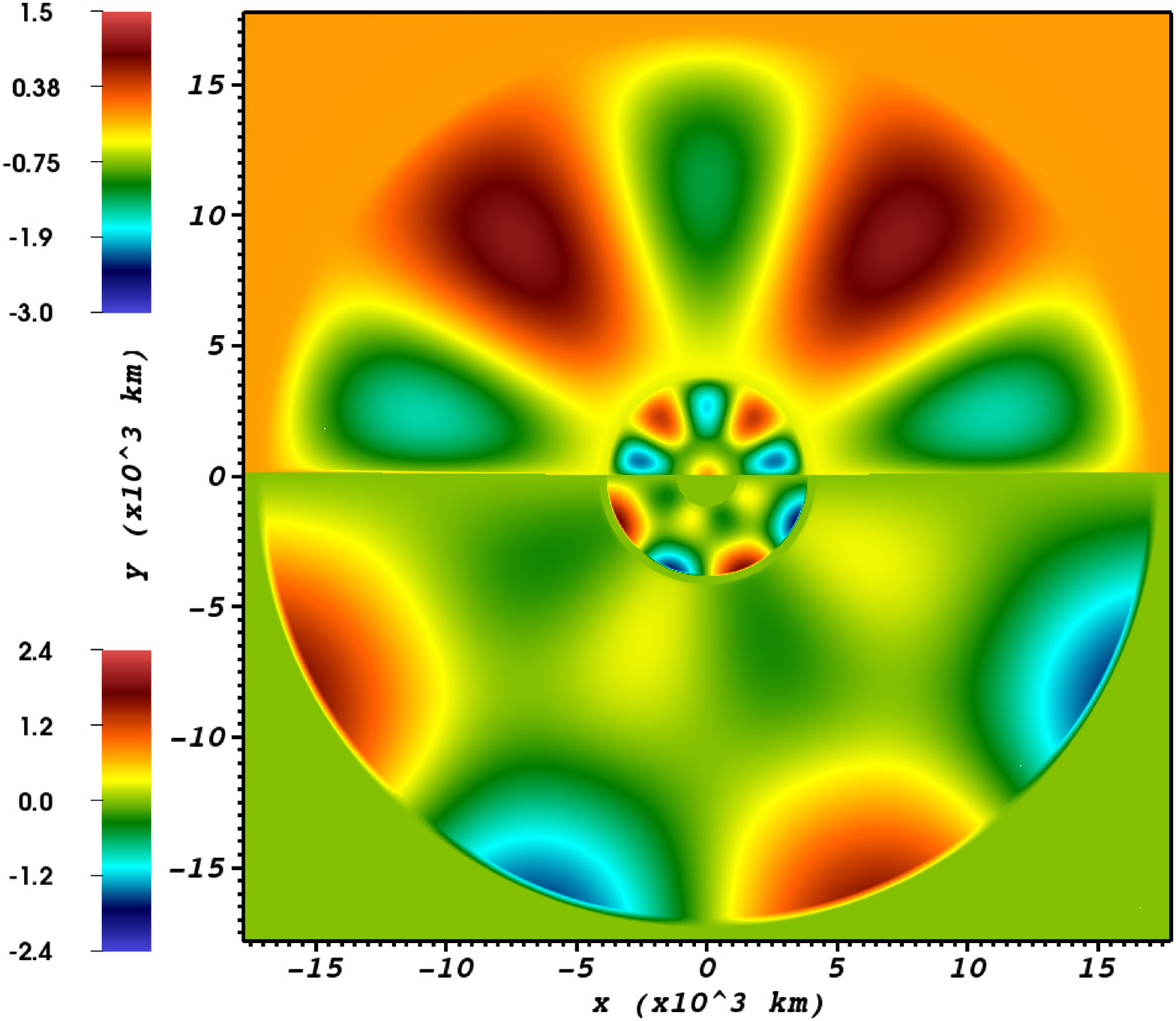}
\hspace{0.03 \linewidth}
\includegraphics[width=0.48 \linewidth]{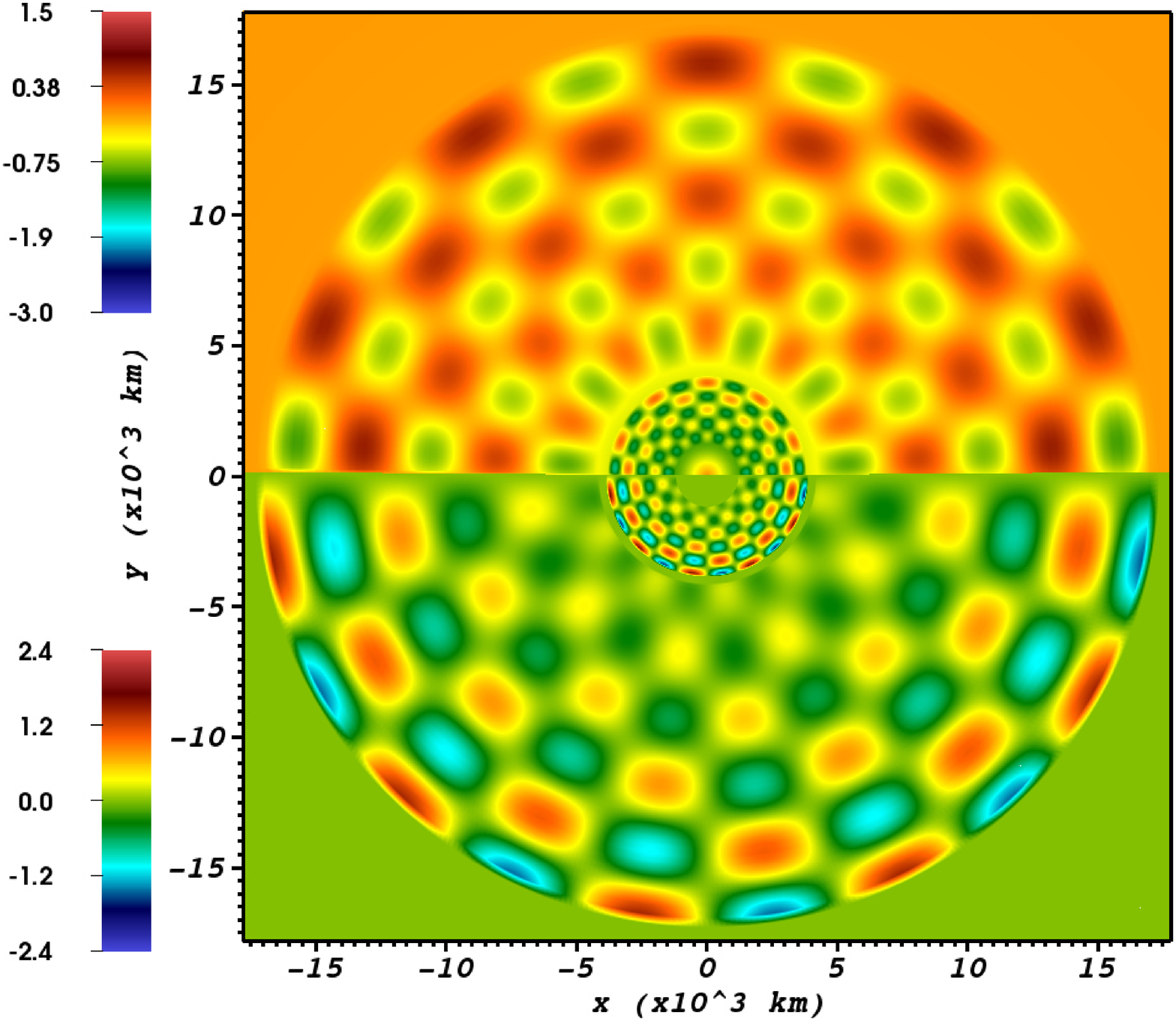}
\caption{Colour plots of perturbations patterns (velocity): The panels
  show the radial velocity (top half of panels) and lateral velocity
  (bottom half of panels) in units of $10^{8}\, \mathrm{cm} \,
  \mathrm{s}^{-1}$ for models pPSa1, pPAa1, pL1a1, pL2a1, pL4a1, and
  pL10a1 (top left to bottom right in zigzag order). The $x$-axis is the symmetry axis of the spherical polar grid.
For model pPSa1, we have included arrows to indicate
(albeit schematically) the direction of the flow in the
convective eddies.
\label{fig:perturbation_patterns1}
}
\end{figure*}

\begin{figure*}
\includegraphics[width=0.48 \linewidth]{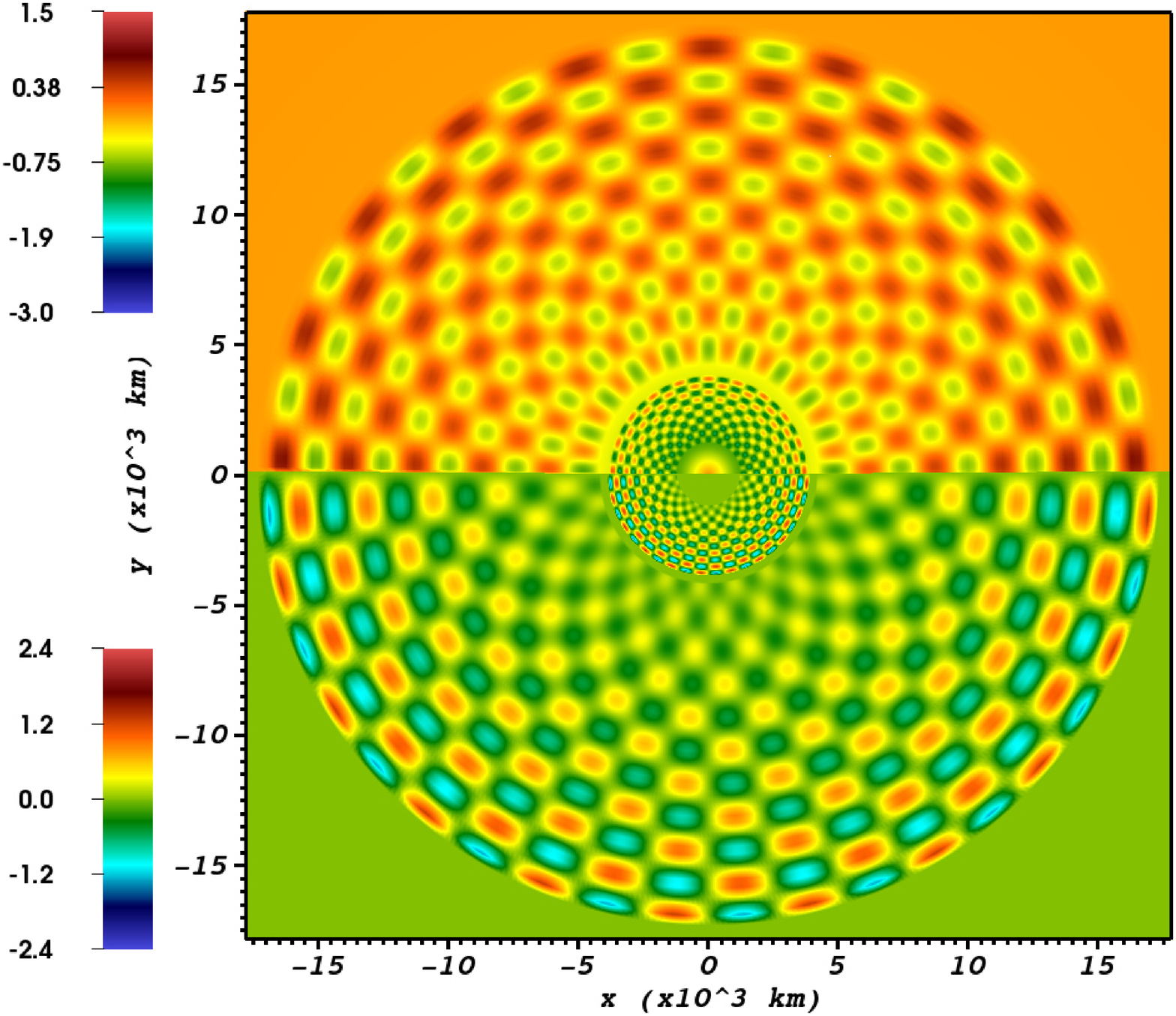}
\hspace{0.03 \linewidth}
\includegraphics[width=0.48 \linewidth]{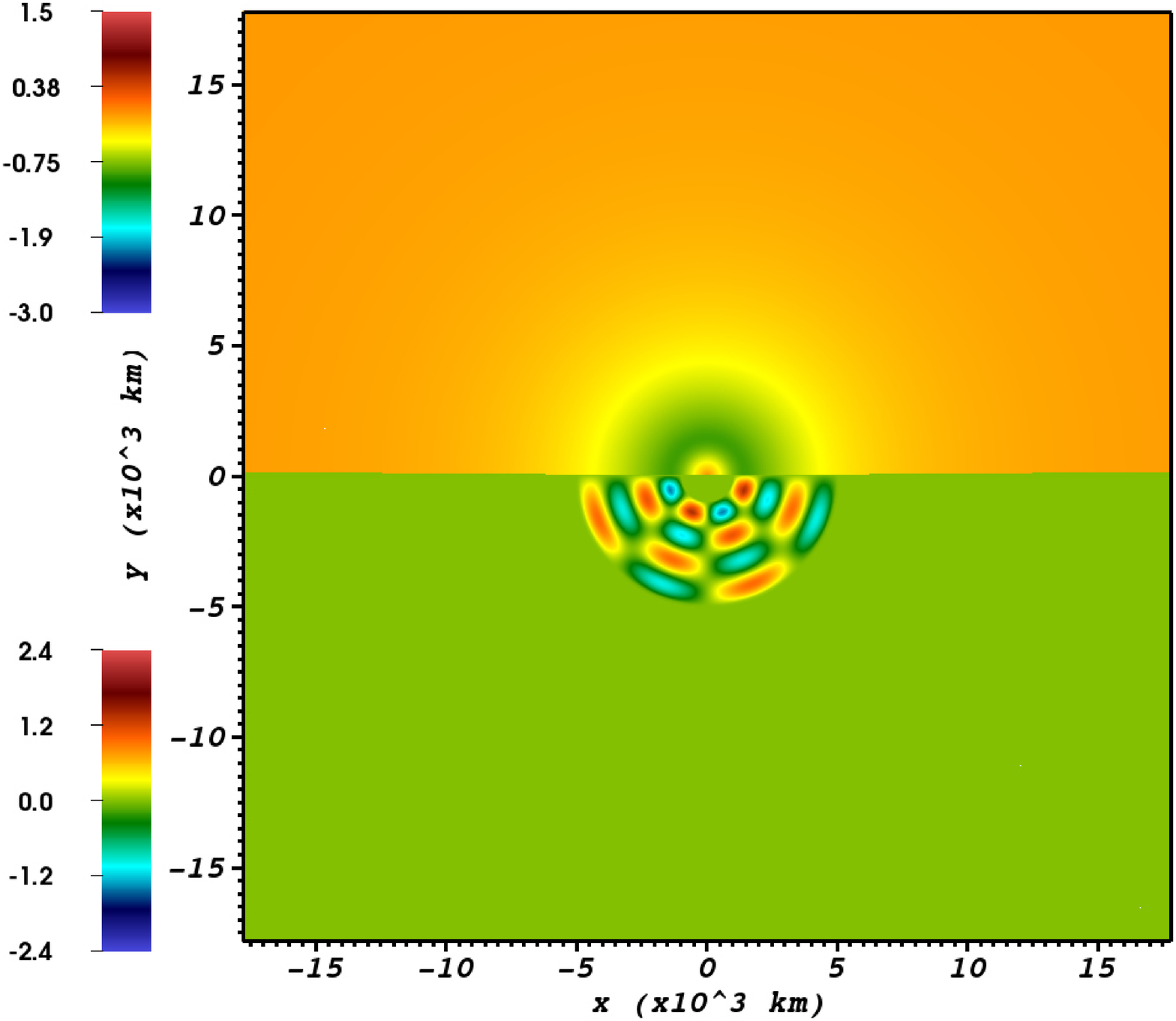}
\caption{Colour plots of perturbation patterns (ctd.):
The panels show the radial velocity (top half of panels)
and lateral velocity (bottom half of panels) in units of $10^{8}\, \mathrm{cm} \, \mathrm{s}^{-1}$ for
models pL20a1 and pCOa1. The $x$-axis
is the axis of the spherical polar grid.
\label{fig:perturbation_patterns2}
}
\end{figure*}

\begin{figure*}
\includegraphics[width=0.48 \linewidth]{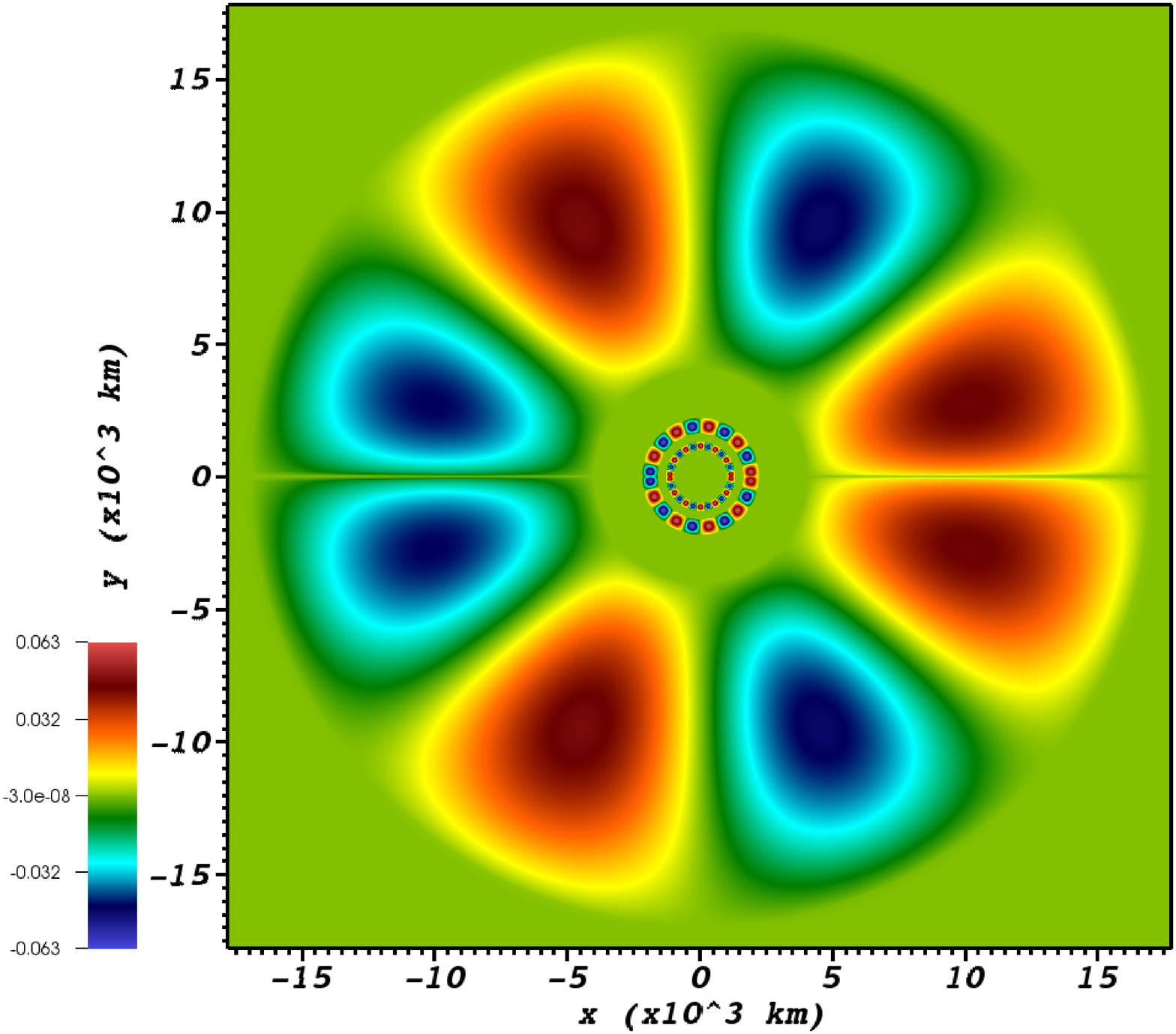}
\hspace{0.03 \linewidth}
\includegraphics[width=0.48 \linewidth]{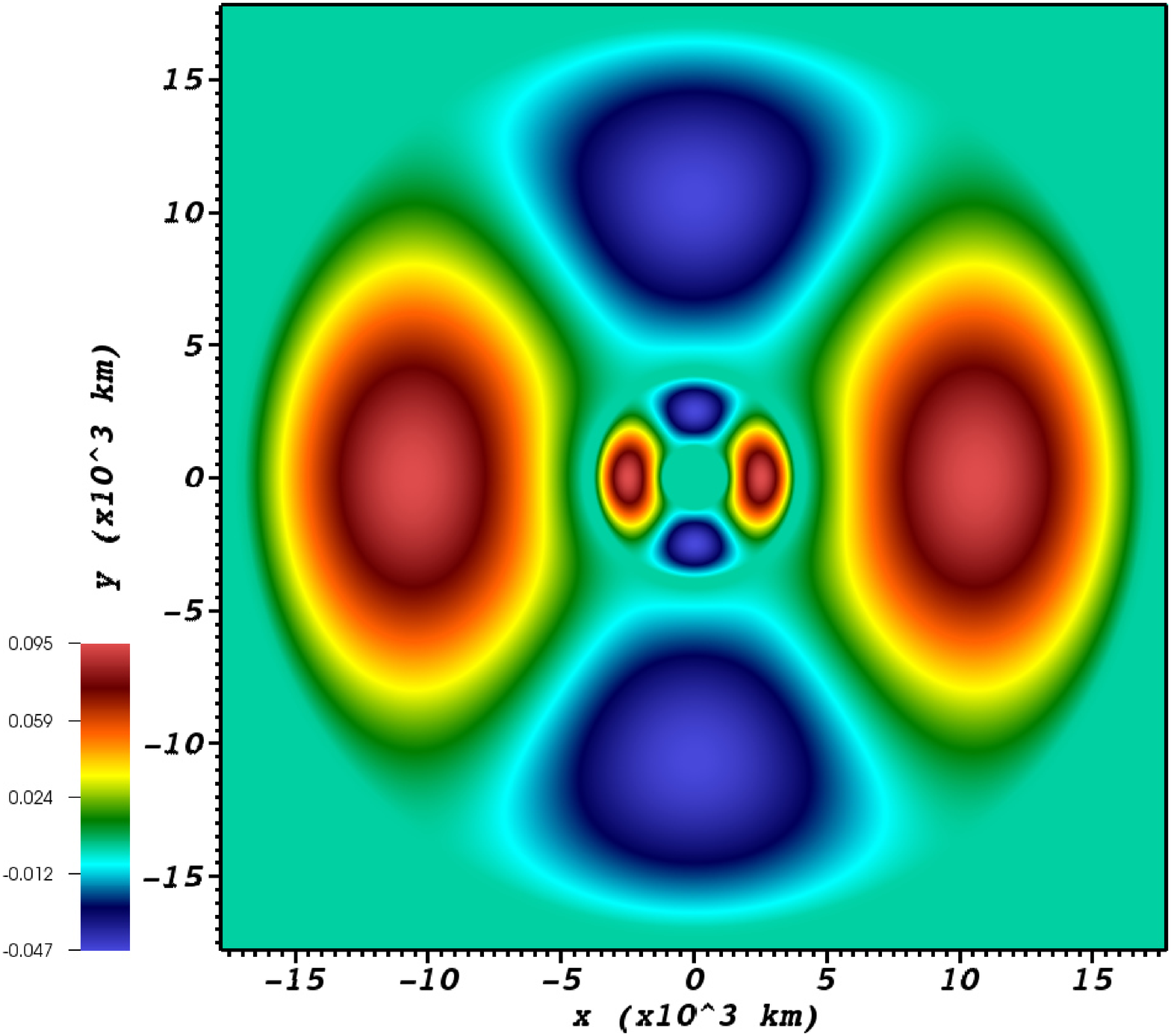}
\caption{Colour plots of perturbation patterns (density): The panels
  show the relative density perturbation $\delta \rho/\rho$ for models
  pPDa1 (left) and pDL2a1. The $x$-axis is the axis of the spherical
  polar grid.
\label{fig:perturbation_patterns3}
}
\end{figure*}

\begin{table*}
  \begin{center}
  \caption{Model Setup -- Solenoidal Perturbations (velocity)
    \label{tab:solenoidal_perturbations}}
  \begin{tabular}{cccccccccccccccc}
    \hline
    perturbation & $\ell_1$ & $n_1$ & $C_1$ &$r_\mathrm{min,1}$ & $r_\mathrm{max,1}$
                   & $\ell_2$ & $n_2$ & $C_2$ &$r_\mathrm{min,2}$ & $r_\mathrm{max,2}$
                   & $\ell_3$ & $n_3$ & $C_3$ &$r_\mathrm{min,3}$ & $r_\mathrm{max,3}$  \\
      pattern & & & $(\mathrm{g}\ \mathrm{cm}^{-1} \ \mathrm{s}^{-1})$ & $(\mathrm{km}) $ & $(\mathrm{km}) $ 
              & & & $(\mathrm{g}\ \mathrm{cm}^{-1} \ \mathrm{s}^{-1})$ & $(\mathrm{km}) $ & $(\mathrm{km}) $ 
              & & & $(\mathrm{g}\ \mathrm{cm}^{-1} \ \mathrm{s}^{-1})$ & $(\mathrm{km}) $ & $(\mathrm{km}) $ \\
      \hline
      pPSa1   & 12 & 1  & $3 \times 10^{23}$  & 1000 & 1300 
              & 10  & 1  & $10 \times 10^{23}$ & 1600 & 2200 
              & 4  & 1  & $80 \times 10^{23}$ & 4000 & 17000 \\
      pPAa1   & 12 & 1  & $3  \times 10^{23}$ & 1000 & 1300 
              & 9  & 1  & $10 \times 10^{23}$ & 1600 & 2200 
              & 3  & 1  & $80 \times 10^{23}$ & 4000 & 17000 \\
\noalign{\smallskip}                                 
      pL1a1     & 1  & 1  & $  2 \times 10^{23}$ & 1200 & 3800 
                & 1  & 1  & $ 80 \times 10^{23}$ & 4200 & 17000 \\
      pL2a1     & 2  & 1  & $  2 \times 10^{23}$ & 1200 & 3800 
                & 2  & 1  & $ 80 \times 10^{23}$ & 4200 & 17000 \\
      pL4a1     & 4  & 1  & $  2 \times 10^{23}$ & 1200 & 3800 
                & 4  & 1  & $ 80 \times 10^{23}$ & 4200 & 17000 \\
      pL10a1    & 10 & 5  & $0.4 \times 10^{23}$ & 1200 & 3800 
                & 10 & 5  & $1.6 \times 10^{23}$ & 4200 & 17000 \\
      pL20a1    & 20 & 10 & $0.2 \times 10^{23}$ & 1200 & 3800 
                & 20 & 10 & $8 \times 10^{23}$ & 4200 & 17000 \\
      \hline
  \end{tabular}

\medskip
See equation~(\ref{eq:solenoidal_velocity_field})
for the definition of the perturbed velocity field
in terms of the parameters
$\ell_i$, $n_i$,  $C_i$, $r_\mathrm{min,i}$, and $r_\mathrm{max,i}$.
Indices from 1 to 3 denote values of the respective
parameters for the two or three different ``convective'' regions
in the model. 
\end{center}
\end{table*}

\begin{table*}
  \begin{center}
    \caption{Solenoidal Perturbations -- Supplementary Information
      \label{tab:solenoidal_properties}}
    \begin{tabular}{cccc}
      \hline
      perturbation & $v_{r,\mathrm{max}}/v_{\theta,\mathrm{max}}$ & $\mathrm{Ma}_{r,\mathrm{max}}/\mathrm{Ma}_{\theta,\mathrm{max}}$ & $E_{\mathrm{kin},r,\mathrm{ini}}/E_{\mathrm{kin},\theta,\mathrm{ini}}$ \\
      pattern &  \\
      \hline
      pPSaY   & 0.94 & 0.72 & 1.55 \\
      pPAaY   & 0.94 & 0.57 & 1.12 \\
      \noalign{\smallskip}                                 
      pL1aY     & 0.24 & 0.22 & 0.18 \\
      pL2aY     & 0.45 & 0.40 & 0.55 \\
      pL4aY     & 0.83 & 0.74 & 1.80 \\
      pL10aY    & 0.53 & 0.54 & 0.46 \\
      pL20aY    & 0.56 & 0.59 & 0.50 \\
      \hline
    \end{tabular}

    \medskip
     $v_{r,\mathrm{max}}$ and $v_{\theta,\mathrm{max}}$ are
    the maximum absolute values of the $\theta$- and $r$-component of
    the perturbation velocity, $\mathrm{Ma}_{r,\mathrm{max}}$ and
    $\mathrm{Ma}_{\theta,\mathrm{max}}$ are the Mach numbers
    corresponding to this velocity.
    $E_{\mathrm{kin},r,\mathrm{ini}}$
    $E_{\mathrm{kin},\theta,\mathrm{ini}}$ are the kinetic energies
    contained in the radial and lateral components of the velocity
perturbations. Note that $v_{\theta,\mathrm{max}}$ and
    $\mathrm{Ma}_{\theta,\mathrm{max}}$ are also the maximum values of
    the \emph{total} velocity and the corresponding Mach number.
  \end{center}
\end{table*}

\begin{table*}
  \begin{center}
  \caption{Model Setup -- Density Perturbations
    \label{tab:density_perturbations}}
    \begin{tabular}{cccccccccccccccc}
      \hline
      perturbation & $\ell_1$ & $n_1$ & $D_1$ &$r_\mathrm{min,1}$ & $r_\mathrm{max,1}$
                   & $\ell_2$ & $n_2$ & $D_2$ &$r_\mathrm{min,2}$ & $r_\mathrm{max,2}$
                   & $\ell_3$ & $n_3$ & $D_3$ &$r_\mathrm{min,3}$ & $r_\mathrm{max,3}$  \\
      pattern & & & & $(\mathrm{km}) $& $(\mathrm{km}) $ 
              & & & & $(\mathrm{km}) $& $(\mathrm{km}) $ 
              & & & & $(\mathrm{km}) $& $(\mathrm{km}) $ \\
\hline
      pPDa1   & 12 & 1  & 0.15  & 1000 & 1300 
              & 9  & 1  & 0.2 & 1600 & 2200 
              & 3  & 1  & 0.2 & 4000 & 17000 \\

      pDL2a1    & 2 & 1  & 0.15 & 1200 & 3800 
                & 2 & 1  & 0.15 & 4200 & 17000 \\
      \hline
    \end{tabular}

\medskip
See equation~(\ref{eq:density_perturbation})
for the definition of the density perturbation in
terms of the parameters $\ell_i$, $n_i$, $D_i$, $r_\mathrm{min,i}$, and $r_\mathrm{max,i}$.
Indices from 1 to 3 denote values of the respective
parameters for the two or three different ``convective'' regions
in each model.
\end{center}
\end{table*}

\section{Model Setup and Numerical Methods}
\label{sec:setup}

\subsection{Initial Perturbations}
Based on the constraints and uncertainties enunciated in
Section~\ref{sec:convection}, we set up a suite of models to explore
the sensitivity of neutrino-driven shock revival to the progenitor
asphericities. As we presently lack multi-D progenitor models at the
onset of collapse, we impose artificial velocity and density
perturbations on the progenitor model s15-2007 of \citet{woosley_07}.

Even if we disregard the obvious lack of self-consistency of this
procedure, these models are bound to remain deficient in other
respects as well: For example, we presently do not attempt to
reproduce the turbulence spectrum of convection as \citet{chen_13} and
\citet{chatzopoulos_14} suggested,\footnote{The shape of the
  turbulence spectrum cannot be easily predicted anyway. There will be
  deviations from a Kolmogorov spectrum, particularly at large scales,
  where the turbulence is driven by the Rayleigh-Taylor instability
  and will be anisotropic.}  and we neither attempt to construct
\emph{consistent} perturbation patterns for velocity, density,
pressure, and composition (which are related in reality, because the
density, pressure and composition contrasts drive convection in the
first place). Despite these deficiencies, however, our models allow us
to explore the impact of progenitor asphericities more systematically
than the recent studies of \citet{couch_13,couch_14}.  At
  this stage, our goal must obviously be limited to studying
  sensitivities by exploring both the regime that realistic models
  could be expected to cover, as well as somewhat less plausible
  regions in parameter space to quantify how strong possible null
  results really are.

In total, we study ten different perturbation patterns for velocity
and density, and consider a number of different amplitudes for each
pattern. We refer to the individual models as pXaY, where X denotes
the perturbation pattern and Y denotes the amplitude relative to an
arbitrarily chosen reference amplitude. The initial configurations for
the reference models pXa1 are visualised in
Figs.~\ref{fig:perturbation_patterns1}
--\ref{fig:perturbation_patterns3}. All other models are obtained by
rescaling the perturbation amplitude by a factor of $Y$. In the
following, we describe the individual perturbation patterns in more
detail.

\subsubsection{Solenoidal Momentum Density Field
(Models pPSaY, pPAaY, pLZaY)}
For the majority of our simulations, we consider pure velocity
perturbations that obey the divergence-free condition $\nabla \cdot (\rho
\mathbf{v})=0$ (Fig.~\ref{fig:perturbation_patterns1}
and left panel of Fig.~\ref{fig:perturbation_patterns2}). These are generated by expressing the
(vectorial) velocity perturbation $\delta\mathbf{v}$ in terms
of the curl of a generalised stream function $\boldsymbol{\psi}$,
\begin{equation}
\label{eq:solenoidal_velocity_field}
\mathbf{\delta \mathbf{v}} =
\left\{
\begin{array}{ll}
\frac{C}{\rho}  \nabla \times \boldsymbol{\psi}, & r_\mathrm{min} \leq r \leq r_\mathrm{max} \\
0, & \ \mathrm{else}
\end{array}
\right.
\end{equation}
where $r_\mathrm{min}$ and $r_\mathrm{max}$ are the inner and
outer boundary of the convective layer. For a single convective layer,
we parameterise $\boldsymbol{\psi}$ as
\begin{equation}
\label{eq:psi_vel}
\boldsymbol{\psi} = \mathbf{e}_\varphi
\frac{\sqrt{\sin \theta}}{r} \sin \left(n \pi \frac{r-r_\mathrm{min}}{r_\mathrm{max}-r_\mathrm{min}}\right)
Y_{\ell,1} (\theta,0),
\end{equation}
where $n$ and $\ell$ denote the number of convective cells in the
radial and angular direction, respectively. 
We use the following definition for the spherical
harmonics $Y_{\ell,m}$ for non-negative $m$ in terms of the associated
Legendre polynomials $P_\ell^m$:
\begin{equation}
Y_{\ell,m}(\theta,\varphi)=\sqrt{\frac{2 \ell+1}{4\pi} \frac{(\ell-m)!}{(\ell+m)!}}
P_\ell^m(\cos \theta) e^{i m \varphi}.
\end{equation}

The factor $\sqrt{\sin
  \theta}$ and the choice of $Y_{\ell,1}$ for the angular
dependence of $\boldsymbol{\psi}$ instead of $Y_{\ell,0}$ is critical
to avoid singularities at $\theta=0$ and $\theta=\pi$, and also
guarantees $\delta v_\theta(0)=0$ and $\delta v_\theta(\pi)=0$, as well as $\delta v_r=0$ at
the boundaries of the convective layer (i.e.\ convective overshoot is
not included).  Furthermore, this definition ensures that velocity
perturbations are isotropic in the sense that the maximum velocities
in the convective eddies do not depend on latitude for $\ell
\rightarrow \infty$.  For a logarithmic density gradient in the
progenitor close to $\pd \ln \rho / \pd \ln r \approx -2$ , our
prescription also results in roughly constant maximum radial velocities
in the case of multiple convective eddies stacked onto each other
in a convective shell.

The formalism presented here provides a convenient and simple way to
generate solenoidal momentum perturbations with a preferred spatial
scale. Capturing the full spectrum of turbulent eddies (with a
prescribed power spectrum) is less straightforward. The approach of
\citet{chatzopoulos_14} (decomposition into divergence-free
eigenfunctions of the vector Helmholtz equation expressed in terms of
vector spherical harmonics) may be better adapted to handle this more
general case.

Several distinct convective shells may reach the shock during the
first $\mathord{\sim} 1 \ \mathrm{s}$ of the post-bounce evolution
depending on the progenitor structure, and we therefore combine
several perturbation patterns computed according to
equations~(\ref{eq:solenoidal_velocity_field}) and (\ref{eq:psi_vel})
in our models.  The different perturbation geometries are summarised
in Table~\ref{tab:solenoidal_perturbations}, and some
  supplementary information about the relative strength of radial and
  lateral motions is provided in
  Table~\ref{tab:solenoidal_properties}.

\begin{figure}
\includegraphics[width=\linewidth]{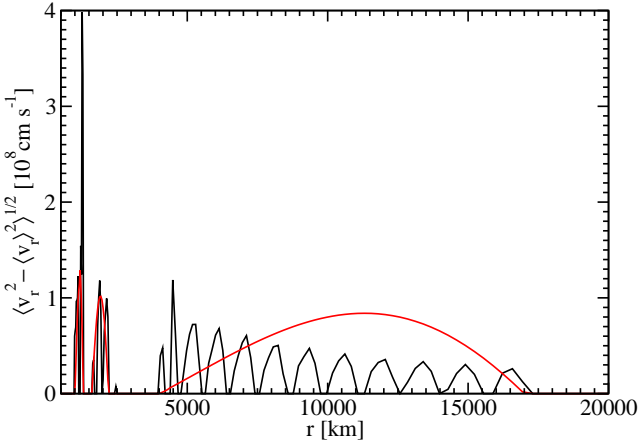}
\caption{ RMS deviation $\delta v = \sqrt{\langle v_r^2 -\langle
    v_r^2\rangle \rangle}$ of the radial velocity $v_r$ from its
  angular average for perturbation pattern pPSa1 (red line) compared
  to a mixing-length estimate (black) computed from the progenitor
  profile using equation~(\ref{eq:mlt_velocity}).
  \label{fig:progenitor_velocity_perturbations}
}
\end{figure}

The setup of models pPSa1
and pPAa1 is based on a mixing-length estimate of the unstable regions and the
typical convective velocities in the progenitor determined
using  equation~(\ref{eq:mlt_velocity}), which indicates the existence of two
relatively narrow convection zones and a more extended convective
layer driven by neon burning.  The RMS (root mean square) deviation of
the radial velocity $v_r$ from its spherical average is compared to
the (noisy) mixing-length estimate in
Fig.~\ref{fig:progenitor_velocity_perturbations}; only a rough
agreement of the average convective velocities can be reached.  The
angular wavenumber is chosen such that the angular and radial extent
of the convective eddies similar. Model pPSa1 differs from model pPAa1
in that the total momentum of the perturbed
configuration vanishes in the former case.

Despite potential numerical problems in discretising
equation~(\ref{eq:mlt_velocity}), the location and extent of the
unstable regions obtained from equation~(\ref{eq:mlt_velocity}) is
roughly compatible with the original stellar evolution calculation in
the \textsc{Kepler} code \citep{woosley_07}. The region between $1000
\ \mathrm{km}$ and $1300 \ \mathrm{km}$ is not Ledoux-unstable in the
original stellar evolution model (except for a single zone flagged as
convective), but there is a thermohaline convection layer between $900
\ \mathrm{km}$ and $1500 \ \mathrm{km}$.  Convective instability is
indicated for $1500 \ \mathrm{km}\leq r < 2300  \mathrm{km}$ and
$3900 \ \mathrm{km} \leq r \leq 18000 \ \mathrm{km}$, i.e.\ these
convective zone are only a little wider in \textsc{Kepler} than predicted
by our post-processing approach. Thus, even a naive, straightforward
discretisation of equation~(\ref{eq:mlt_velocity}) seems to yield
acceptable predictions for convective instability.

The pLZ models (where Z denotes the angular wavenumber) are geared towards a more systematic exploration of
possible flow geometries. For these models, we assume a wide
convective layer encompassing both the Si and Si/O shells and another
convective zone in the O/Ne/Mg shell. We vary the scale of the
convective eddies, covering angular wavenumbers from $\ell=1$ to
$\ell=20$, as well as the amplitude of the velocity perturbation
$\mathbf{\delta v}$.  To a lesser extent, these
  variations in flow geometry are motivated by the possibility that
  strong shell interactions \citep{arnett_11} could lead to a merger
  of different convection zones, which is one among many
  uncertainties in the structure of pre-supernova cores.  A more
  important reason, however, lies in the fact that the extent of the
  convective shells varies greatly between progenitor models (see
  again the Kippenhahn diagrams in \citealt{heger_00})

The pLZ perturbations are also set up such that the convective eddies
are of similar extent in the radial and lateral direction, but because
of the broader ``convection zones'' this cannot be perfectly
accomplished for $\ell=1,2$ and only only for a restricted range in
radius for
large $\ell > 4$ with our functional ansatz for the generalised stream
function. As a result, there is some variation in the relative
strength of radial and lateral motions (see
Table~\ref{tab:solenoidal_properties}), and some perturbation patterns
do not conform to the expectation \citep{arnett_09} of equipartition
between the kinetic energy in radial motions and the combined energy
in transverse motions in all other directions (which, however, need not
be universally valid).

\subsubsection{Purely Transverse Velocity Perturbations
(Models pCOaY)}
In addition, we also study a perturbation pattern similar to the one
used by \citet{couch_13} (right panel of
Fig.~\ref{fig:perturbation_patterns2}). Different from
\citet{couch_13}, the perturbation pattern pCO is axisymmetric; it is
essentially a meridional cut through their model n5m2 that we revolve
around the symmetry axis of the spherical polar grid. Furthermore, we perturb the initial model at the
onset of collapse whereas \citet{couch_13} imposed perturbations at
bounce. Only the lateral velocity component $v_\theta$ is perturbed.
For model pCOaX, $v_\theta$ is given by
\begin{equation}
v_\theta=
0.2 \times 
c_s \sin \left(4 \theta\right) \sin \left(4 \pi \frac{r-1000 \ \mathrm{km}}{4000 \ \mathrm{km}}\right), 
\end{equation}
in terms of the local sound speed $c_s$ for $ 1000 \ \mathrm{km} \leq r \leq 4000 \ \mathrm{km}$.

It must be emphasised that this perturbation pattern hardly resembles
a convective flow. The radial velocity field is spherically symmetric,
i.e.\ convective updrafts and downdrafts are absent. The velocity
field also violates the divergence-free condition. This results in a
strong (and probably unphysical) excitation of acoustic waves in the
perturbed (and supposedly ``convective'') region as the
model evolves.  We nevertheless include this perturbation pattern in
our model suite as the closest possible analogue to the models of
\citet{couch_13} in 2D.

\subsubsection{Density Perturbations (Models pPDaY and pDL2aY)}
In addition to velocity perturbations, we also explore a smaller
set of models with density perturbations (Fig.~\ref{fig:perturbation_patterns3}). We confine ourselves
to one series (pPD) where the convective regions are estimated
using equation~(\ref{eq:mlt_density}) for the mixing-length density
contrast and to a series (pDL2) with two large convective
zones and large-scale $\ell=2$ density perturbations.
Within each zone, the density contrast is calculated
as follows,
\begin{equation}
\label{eq:density_perturbation}
\frac{\delta \rho}{\rho}=
D 
\sin \left(n \pi \frac{r-r_\mathrm{min}}{r_\mathrm{max}-r_\mathrm{min}}\right)
Y_{\ell,0} (\theta,0),\quad \mathrm{for}\
r_\mathrm{min}\leq r \leq r_\mathrm{max}.
\end{equation}
The values for $D$, $r_\mathrm{min}$, $r_\mathrm{max}$, and $\ell$ for
the individual zones are shown in
Table~\ref{tab:density_perturbations}.  Note that we also include one
model (pPDLa2m) with a \emph{negative} value for the normalised
perturbation amplitude, i.e.\ the maxima and minima of the density
perturbation $\delta \rho$ are interchanged in this model
compared to model pPDLa2.

\begin{table*}
  \centering
  \begin{minipage}{0.85\textwidth}
  \caption{Simulation Results -- Velocity Perturbations
    \label{tab:perturbations}}
  \begin{tabular}{cccccccc}
    \hline 
    perturbation & $v_{\theta,\mathrm{max}}$ & $\mathrm{Ma}_{\theta,\mathrm{max}}$ & $E_{\mathrm{kin},\theta,\mathrm{ini}}$ & $\zeta$ & comment & explosion & 
    $\mathrm{Ma}_{\theta,\mathrm{expl}}$ \\
    pattern   &  $(10^8 \ \mathrm{cm} \ \mathrm{s}^{-1})$ & & $(10^{48} \ \mathrm{erg})$  & &  & time &  \\
    \hline
    p0    & 0 & 0 & 0 & 0 & unperturbed baseline model & --- & ---\\
    \noalign{\medskip}
    pPSa1 & $2.37$ & $0.59$ & $2.78$ & $0.9\%$ & Ledoux-unstable zones, symmetric & --- & ---\\
    pPSa2 & $4.74$ & $1.19$ & $11.2$ & $3.5\%$ & Ledoux-unstable zones, symmetric & $680 \ \mathrm{ms}$ & $0.07$ \\
    pPSa4 & $9.49$ & $2.37$ & $44.6$ & $14\%$ & Ledoux-unstable zones, symmetric & --- & ---\\
    \noalign{\smallskip}
    pPAa1 & $2.37$ & $0.58$ & $2.83$ & $0.9\%$ & Ledoux-unstable zones, asymmetric & --- & ---\\
    pPAa2 & $4.74$ & $1.16$ & $11.3$ & $3.5\%$ & Ledoux-unstable zones, asymmetric & $820 \ \mathrm{ms}$ & $0.08$ \\
    pPAa4 & $9.49$ & $2.32$ & $45.2$ & $14\%$ & Ledoux-unstable zones, asymmetric & $830 \ \mathrm{ms}$ & $0.16$\\
    \noalign{\medskip}
    pL1a0.25 & $0.45$ & $0.16$ & $0.19$ & $0.06\%$ & two large zones, $\ell=1$ & --- & ---  \\
    pL1a0.5 & $0.89$ & $0.32$ & $0.75$ & $0.23\%$ & two large zones, $\ell=1$ & $810 \ \mathrm{ms}$ & $0.19$\\
    pL1a1 & $1.78$ & $0.63$ & $3.00$ & $0.9\%$ & two large zones, $\ell=1$ & $530 \ \mathrm{ms}$  & $0.25$ \\
    pL1a2 & $3.56$ & $1.26$ & $12.0$ & $3.8\%$ & two large zones, $\ell=1$ & $480 \ \mathrm{ms}$  & $0.35$\\
    \noalign{\smallskip}
    pL2a0.0625 & $0.11$ & $0.04$ & $0.01$ & $0.003\%$ & two large zones, $\ell=2$ & --- & --- \\
    pL2a0.125 & $0.21$ & $0.08$ & $0.04$ & $0.01\%$ & two large zones, $\ell=2$ & $750 \ \mathrm{ms}$ & $0.05$\\
    pL2a0.25 & $0.43$ & $0.15$ & $0.16$ & $0.05\%$ & two large zones, $\ell=2$ & $740 \ \mathrm{ms}$ & $0.09$\\
    pL2a0.5 & $0.86$ & $0.30$ & $0.62$ & $0.19\%$ & two large zones, $\ell=2$ & $720 \ \mathrm{ms}$ & $0.18$ \\
    pL2a1 & $1.72$ & $0.61$ & $2.50$ & $0.79\%$ & two large zones, $\ell=2$ & $480 \ \mathrm{ms}$ & $0.20$ \\
    pL2a2 & $3.43$ & $1.22$ & $10.0$ & $3.1\%$ & two large zones, $\ell=2$ & $420 \ \mathrm{ms}$ & $0.25$\\
    pL2a4 & $6.87$ & $2.43$ & $40.0$ & $13\%$ & two large zones, $\ell=2$ & $300 \ \mathrm{ms}$ & $0.20$\\
    \noalign{\smallskip}
    pL4a0.5 & $0.85$ & $0.30$ & $0.57 $ & $0.18\%$ & two large zones, $\ell=4$ & --- & --- \\
    pL4a1 & $1.70$ & $0.60$ & $2.29 $ & $0.72\%$ & two large zones, $\ell=4$ & $700 \ \mathrm{ms}$ & $0.35$ \\
    pL4a2 & $3.40$ & $1.20$ & $9.2$ & $2.9\%$ & two large zones, $\ell=4$ & $480 \ \mathrm{ms}$ &  $0.35$\\
    pL4a4 & $6.81$ & $2.41$ & $36.6$ & $11\%$ & two large zones, $\ell=4$ & $290 \ \mathrm{ms}$ &  $0.20$\\
    \noalign{\smallskip}
    pL10a1 & $0.82$ & $0.29$ & $0.47$ & $0.15\%$ & two large zones, $\ell=10$ & --- & ---\\
    pL10a1 & $1.63$ & $0.58$ & $1.87$ & $0.58\%$ & two large zones, $\ell=10$ & $790 \ \mathrm{ms}$ & $0.35$\\
    pL10a2 & $3.27$ & $1.16$ & $7.49$ & $2.3\%$ & two large zones, $\ell=10$ & $800 \ \mathrm{ms}$ & $0.70$ \\
    pL10a4 & $6.54$ & $2.32$ & $30.0$ & $9.4\%$ & two large zones, $\ell=10$ & $510 \ \mathrm{ms}$ & $1.00$ \\
    \noalign{\smallskip}
    pL20a0.5 & $0.73$ & $0.26$ & $0.40$ & $0.13\%$ & two large zones, $\ell=20$ & --- & --- \\
    pL20a1 & $1.45$ & $0.52$ & $1.61$ & $0.5\%$ & two large zones, $\ell=20$ & $710 \ \mathrm{ms}$ & $0.30$\\
    pL20a2 & $2.90$ & $1.04$ & $6.42$ & $2.0\%$ & two large zones, $\ell=20$ & $740 \ \mathrm{ms}$ & $0.60$\\
    pL20a4 & $5.81$ & $2.07$ & $25.7$ & $8.0\%$ & two large zones, $\ell=20$ & $560 \ \mathrm{ms}$ & $0.80$\\
    \noalign{\medskip}
    pCOa0.25 & $0.35$ & $0.05$ & $0.13$ & $0.02\%$ & transverse velocity perturbations & $800 \ \mathrm{ms}$ & $0.05$\\
    pCOa0.5 & $0.69$ & $0.10$ & $0.53$ & $0.08\%$ & transverse velocity perturbations & $700 \ \mathrm{ms}$ & $0.10$\\
    pCOa1 & $1.39$ & $0.20$ & $2.14$ & $0.34\%$ & transverse velocity perturbations & $650 \ \mathrm{ms}$ & $0.20$\\
    pCOa2 & $2.77$ & $0.40$ & $8.55$ & $1.36\%$ & transverse velocity perturbations & $270 \ \mathrm{ms}$ & $0.40$\\
    pCOa4 & $5.55$ & $0.80$ & $34.3$ & $5.4\%$ & transverse velocity perturbations & $240 \ \mathrm{ms}$ & $0.80$ \\
    \hline 
  \end{tabular}

\medskip
$v_{\theta,\mathrm{max}}$ is the maximum absolute value of the
$\theta$-component of the perturbation velocity,
$\mathrm{Ma}_{\theta,\mathrm{max}}$ is the Mach number corresponding
to this velocity, $E_{\mathrm{kin},\theta,\mathrm{ini}}$ is the
lateral kinetic energy contained in the velocity perturbations,
$\zeta$ is the ratio of this energy to the binding energy of the
perturbed mass shells, and $\mathrm{Ma}_{\theta,\mathrm{expl}}$ is the
maximum lateral Mach number in the initial model inside the mass shell
that reaches the shock at the onset of the explosion.
    \end{minipage}
\end{table*}

\begin{table*}
  \centering
  \begin{minipage}{0.55\textwidth}
  \caption{Simulation Results -- Density Perturbations
    \label{tab:perturbations_den}}
  \begin{tabular}{ccccc}
    \hline
    perturbation &  & & explosion \\
     pattern& $(\delta \rho/\rho)_{\mathrm{max}}$ & comment & time \\
    \hline
    pPDa1 & $0.07$ & Ledoux-unstable zones & --- \\
    pPDa2 & $0.13$ & Ledoux-unstable zones & $880 \ \mathrm{ms}$ \\
    pPDa4 & $0.25$ & Ledoux-unstable zones & $880 \ \mathrm{ms}$ \\
    \noalign{\smallskip}
    pDL2a1 & $0.09$ & two large zones, $\ell = 2$ & $780 \ \mathrm{ms}$ \\
    pDL2a2 & $0.18$ & two large zones, $\ell = 2$ & $710 \ \mathrm{ms}$ \\
      pDL2a4 & $0.36$ & two large zones, $\ell = 2$ & $630 \ \mathrm{ms}$ \\
      pDL2a2m & $0.18$ & two large zones, $\ell = 2$ & --- \\
      \hline
  \end{tabular}

\medskip
  $(\delta \rho/\rho)_{\mathrm{max}}$ is the maximum relative
deviation of the density from its spherical average in the initial
model.
  \end{minipage}
\end{table*}

\subsection{Numerical Methods}
We evolve the perturbed progenitor models as well as an unperturbed
baseline model (p0) with very small inherent numerical seed
perturbations from the onset of collapse to at least $800
\ \mathrm{ms}$ after bounce using the relativistic hydrodynamics code
\textsc{CoCoNuT} \citep{dimmelmeier_02_a}. During the collapse, we
apply the deleptonisation scheme of \citet{liebendoerfer_05_b}. At
bounce, we switch to a newly developed fast multi-group neutrino
transport (FMT) scheme based on an approximate solution of the
neutrino energy equation.  The required closure relation is provided
by solving the Boltzmann equation in a two-stream approximation (which
yields an accurate flux factor at high optical depths). At larger
distances from the neutrinosphere, we match to a solution for the flux
factor derived from an analytic variable Eddington factor closure.  We
take all the relevant charged-current reactions as well as
isoenergetic neutral-current neutrino interactions with
  nucleons and nuclei into account (see Appendix~\ref{sec:numerics}
  for details). Furthermore, we include an effective one-particle
rate for nucleon-nucleon bremsstrahlung and approximately account for
the energy exchange of $\mu$ and $\tau$ neutrinos with the medium due
to nucleon recoil in neutral-current scattering.\footnote{The lack of
  an efficient method to account for energy-exchanging scattering
  reactions (i.e.\ neutrino-electron scattering) of \emph{electron
    neutrinos} during collapse is the primary reason for resorting to
  the deleptonisation scheme of \citet{liebendoerfer_05_b} up to
  bounce.  The multi-group scheme presented here would lead to weaker
  deleptonization during collapse and hence to a more massive
  homologous core at bounce compared to more sophisticated neutrino
  transport schemes
    (cf.\ \citealt{bruenn_85,bruenn_86}). This would alter the early
  post-bounce dynamics quite noticeably.} A comparison with results
from the \textsc{Vertex-Prometheus} code and its relativistic offshoot
\textsc{Vertex-CoCoNuT} shows that our new scheme allows us to achieve
reasonable qualitative and quantitative agreement with more
sophisticated methods for multi-group neutrino transport at a fraction
of the computational cost. For a detailed description of the neutrino
transport treatment in our simulations, we refer the reader to
Appendix~\ref{sec:numerics}.

 However, future users of our method should bear in mind
  that these savings come at a cost, and one must check the
  approximations inherent in the FMT scheme on a case-by-case basis:
  We already mentioned the limitations of the FMT scheme during the
  collapse phase (where neutrino-electron scattering cannot be
  neglected). Moreover, many of the complexities of neutrino-nucleon
  interactions at high densities are presently ignored, such as nucleon
  correlations \citep{burrows_98,burrows_99,reddy_99}, the effect of
  nucleon interaction potentials \citep{martinez_12,roberts_12c}, weak
  magnetism \citep{horowitz_97}, and the quenching of the axial-vector
  coupling at high densities \citep{carter_02}. This would, among
  other things, delay the cooling of the proto-neutron star
  considerably \citep{huedepohl_10}, and could affect the
  nucleosynthesis conditions after the onset of the explosion, which
  are very sensitive to the \emph{difference} of the electron neutrino
  and anti-neutrino luminosities and mean energies.  As with any
  transport scheme (flux-limited diffusion, the IDSA approximation of
  \citealt{liebendoerfer_09}, or even two-moment closure schemes) not
  based on a rigorous solution of the Boltzmann equation or a
  Boltzmann closure, the flux factor at intermediate optical depths
  $\lesssim 1$ (which is crucial for the neutrino heating and
cooling) requires
  careful checking.

We use a computational grid with $N_r \times N_\theta=550 \times 256$
zones to cover the innermost $10^5 \ \mathrm{km}$ of the progenitor.
In the innermost $10  \mathrm{km}$, the grid
spacing is fairly uniform and then transitions smoothly to a 
roughly logarithmic grid spacing
with $\Delta r/r \approx 1.5\%$ between $10\ \mathrm{km}$
and $400 \ \mathrm{km}$. Outside $400 \ \mathrm{km}$, 
$\Delta r/r$ gradually rises to  $2.2\%$ at the outer boundary.
An equidistant grid is used for the $\theta$-coordinate.
 
For
the high-density regime, we employ the equation of state of
\citet{lattimer_91} with a bulk incompressibility modulus of nuclear
matter of $K=220 \ \mathrm{MeV}$ (LS220).
 At densities lower than $5 \times 10^8 \ \mathrm{g} \ \mathrm{cm}^{-3}$
(prior to bounce) or $10^{11} \ \mathrm{g} \ \mathrm{cm}^{-3}$
(after bounce), we include the ideal gas contributions
of photons, electrons/positrons of arbitrary degeneracy, and
of 17 different nuclear species (protons, neutrons,
$\alpha$-particles, and 14 intermediate and heavy nuclei).
Nuclear burning is taken into account nuclear burning according to
the ``flashing'' treatment of Appendix~B.2 in \citet{rampp_02}. Above
a temperature of $T=0.5 \ \mathrm{MeV}$, we switch to nuclear statistical
equilibrium.

\section{Overview of Simulation Results}
\label{sec:overview}
Our simulations show that sufficiently strong asphericities in the
progenitor can indeed tip the scales in favour of an explosion for a
rather pessimistic model. While the shock is not revived in the
baseline model p0 at least until $1.4 \ \mathrm{s}$ after bounce, an
explosion develops in many of the simulations with initial
perturbations. The final fate of the different models is summarised in
the penultimate of Tables~\ref{tab:perturbations} (velocity
perturbations) and \ref{tab:perturbations_den} (density
perturbations).

These tables give the time of explosion for each model (if
applicable), which we define as the time when the critical ratio
between the advection time-scale and the heating time-scale (defined
as in \citealt{mueller_12a}) reaches unity.  For the models with
velocity perturbations, we also list the maximum lateral velocity
$v_{\theta,\mathrm{max}}$, the maximum Mach number
$\mathrm{Ma}_{\theta,\mathrm{max}}$ corresponding to this velocity,
and the total kinetic energy $E_{\mathrm{kin},\theta,\mathrm{ini}}$
contained in lateral motions in the initial model. The ratio $\zeta$
of this energy to the total binding energy (i.e.\ gravitational energy
+ internal energy)\footnote{ The kinetic energy contained in radial
  motions in the unperturbed model is not included here, as it is a
  minor contribution in these shells at the pre-collapse stage.} of
the perturbed mass shells is also provided.  Since the lateral Mach
number varies considerably with radius in the models with solenoidal
perturbations, we also give the maximum lateral Mach number
$\mathrm{Ma}_{\theta,\mathrm{expl}}$ in the initial model
\emph{inside} the mass shell that reaches the shock at the time of the
explosion. This number provides a better measure for the violence of
convective motions in the progenitor that is required to achieve shock
revival. For models with density perturbations, we provide the maximum
of $\delta \rho/\rho$ in the initial model.  When
  comparing to multi-D stellar evolution models or mixing-length
  estimates, the reader should always carefully consider whether
  average or maximum quantities are involved.

The numbers in these tables suggest some systematic trends, but also
show the dependence on the perturbation pattern and amplitude to be
non-trivial in some cases.

Perturbations restricted to regions where we diagnose convective
instability in the progenitor models based on the Ledoux criterion
appear to be relatively inefficient at boosting the heating
conditions.  For the model series pPAaX and pPSaX with velocity
perturbations, we only obtain explosions for
$E_{\mathrm{kin},\theta,\mathrm{ini}} > 1.1 \times 10^{49} \mathrm{erg}$, and
the strongly perturbed model pPSa4 even fails to explode. Shock revival
does not occur earlier than $680 \ \mathrm{ms}$ after bounce
(model pPSa2).

Explosions occur more readily in models with perturbations in the
entire silicon and oxygen shells, or with the perturbation pattern
pCOaX inspired by the setup of \citet{couch_13}.  Low-$\ell$ perturbations with
$\ell=2$ and $\ell=1$ emerge as most efficient in inducing an
explosion, which can occur around $500 \ \mathrm{ms}$ with reasonably
subsonic velocity ($\mathrm{Ma}_{\theta,\mathrm{expl}} \leq 0.25$ for
these cases). Perturbations with higher $\ell$ tend to be less
effective for a given maximum Mach number, but there are exceptions as
models pL20a1 and pL20a2 explode earlier than models pL10a1 and pL10a2.
Larger perturbations generally result in earlier explosions,
but there are also some non-monotonicities (pL10a2 explodes
later than pL10a1 and pL20a2 explodes later than pL20a1).

The non-solenoidal perturbation pattern pCOaX mimicking the setup of
\citet{couch_13} in 2D tends to give earlier explosions 
  for a given value of $\mathrm{Ma}_{\theta,\mathrm{max}}$ or
  $E_{\mathrm{kin},\theta,\mathrm{ini}}$ than the corresponding
solenoidal perturbation pattern with $\ell=4$. The earliest explosions
are found for this perturbation pattern with large initial amplitudes
($E_{\mathrm{kin},\theta,\mathrm{ini}}$ up to $3.4 \times 10^{49}
\ \mathrm{erg}$). 

Density perturbations in the progenitor likewise lead to shock revival
in some cases, but rather high perturbation amplitudes are required to
achieve explosions as early as for velocity perturbations with low
$\ell$. We find an appreciable effect only if the density contrast is
of the order of $\delta \rho/\rho \approx 0.1$, which would
require a convective Mach number of $\mathrm{Ma}_\mathrm{prog} \gtrsim 0.3$
in the O-burning shell.

\begin{figure}
\includegraphics[width=\linewidth]{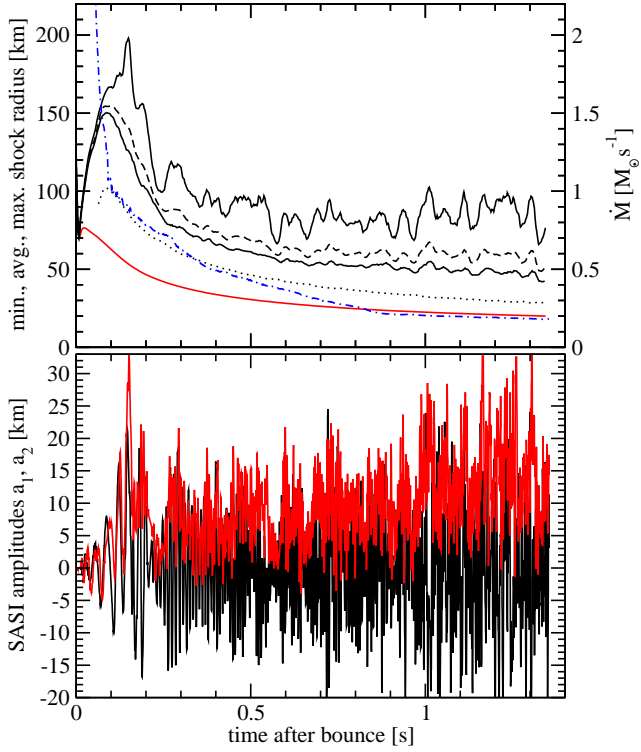}
\caption{
\label{fig:shock_p0}
\emph{Top:} Time evolution of the maximum (black solid curve), average
(black, dashed), and minimum (black, solid) shock radius for the baseline model p0. The proto-neutron star radius (red), defined by a fiducial density
of $10^{11} \ \mathrm{g} \ \mathrm{cm}^{-3}$, the gain radius (black, dotted),
 and the mass accretion rate $\dot{M}$ (measured at
a radius of $400 \ \mathrm{km}$, blue, dash-dotted,
scale on the right vertical axis) are also shown. 
\emph{Bottom:} Coefficients $a_1$ (black) and $a_2$ (red) for the
decomposition of the shock surface into Legendre polynomials.  }
\end{figure}

\begin{figure}
\includegraphics[width=\linewidth]{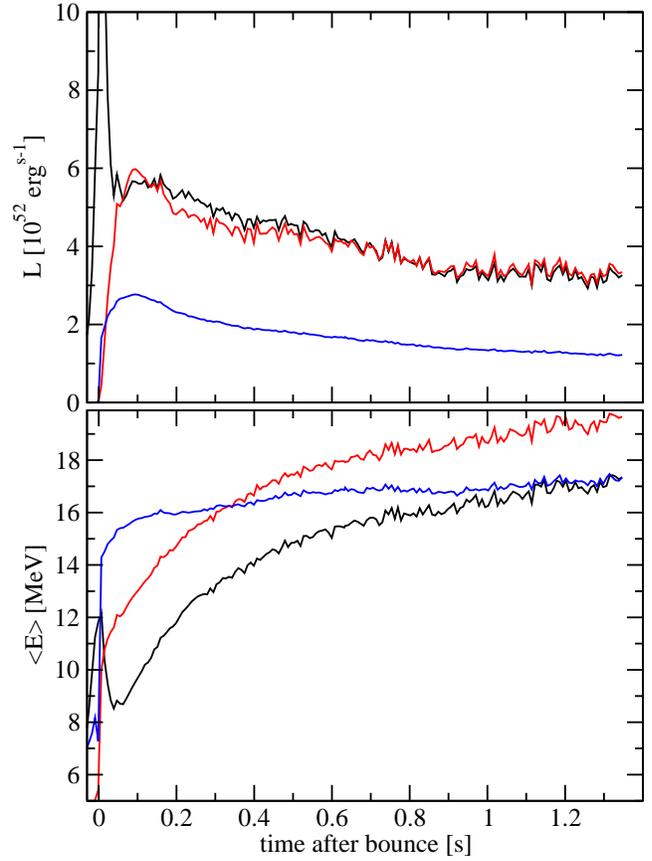}
\caption{
\label{fig:neutrino_p0}
Total neutrino luminosities (top panel) and angle-averaged mean energies (bottom panel) 
for the baseline model p0. Black, red, and blue curves are used for
$\nu_e$, $\bar{\nu}_e$, and $\nu_{\mu/\tau}$, respectively. All quantities
are measured at a radius of $400 \ \mathrm{km}$.
}
\end{figure}

\begin{figure}
\includegraphics[width=\linewidth]{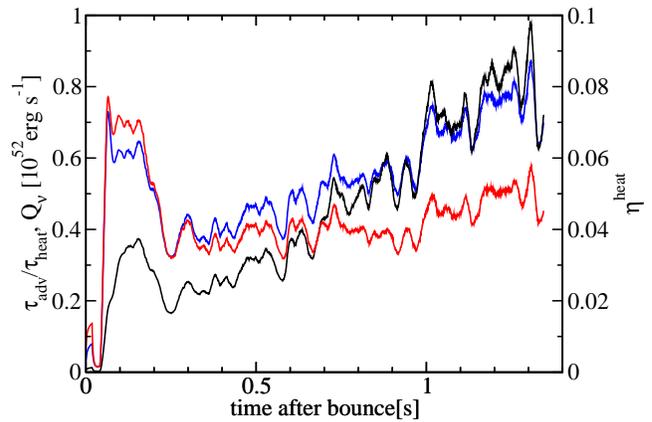}
\caption{Time-scale criterion $\tau_\mathrm{adv}/\tau_\mathrm{heat}$ (black),
volume-integrated neutrino heating rate in the gain region (red), and
heating efficiency $\eta$ (blue) for the baseline model p0.
\label{fig:heating_p0}
}
\end{figure}

\begin{figure}
\includegraphics[width=\linewidth]{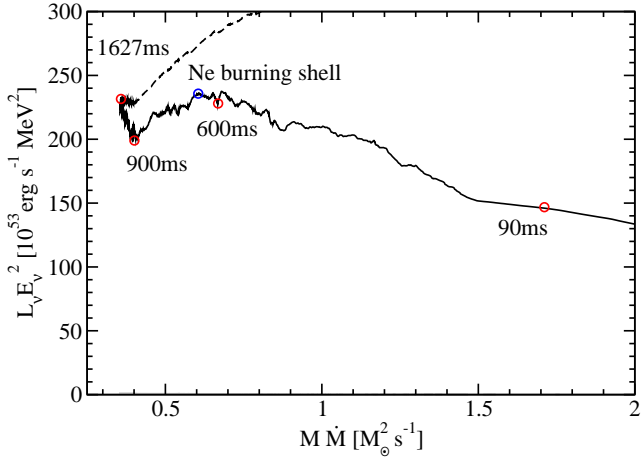}
\caption{Approach of the baseline model to the critical curve
in the $(M \dot{M},L_\nu E_\nu^2)$ plane (see Section~\ref{sec:critical_luminosity}).
Here, $L_\nu$ is the total electron flavor luminosity, and $E_\nu$ is
an appropriate average of electron neutrino and anti-neutrino mean energies
(equation~\ref{eq:avg_energy}). The black curve shows the
evolution of the model (with red circles indicating its
state at four selected post-bounce times), and a fiducial critical
curve (dashed) is anchored at the final location of the
model in the $(M \dot{M},L_\nu E_\nu^2)$-plane where
the time-scale ratio $\tau_\mathrm{adv}/\tau_\mathrm{heat}$
approaches unity. The blue circle roughly indicates the time
when the neon burning shell reaches the shock.
\label{fig:critical_curve_p0}
}
\end{figure}

\begin{figure}
  \includegraphics[width=\linewidth]{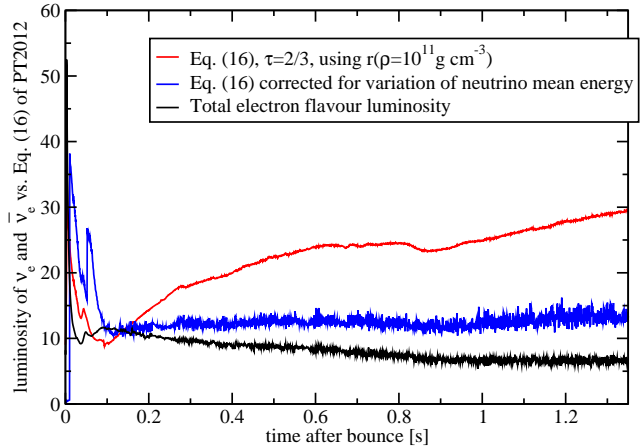}
  \caption{Comparison of the evolution of the total
    electron flavour luminosity (black) with the prediction of
    equation~(16) of \citet[PT2012 for short]{pejcha_12} for the
    critical luminosity (red). Note that this estimate for the critical
    luminosity \emph{increases} with time instead of decreasing, thus
    erroneously suggesting a deterioration in heating conditions for
    model p0.  A modified version of equation~(16) of PT2012 that
    includes a correction term $\propto E_{\bar{\nu}_e}^{-2}$
    (O.~Pejcha, private communication) is also shown in blue; it still
does not give the expected decrease.
    \label{fig:pejcha_thompson}
  }
\end{figure}

\section{Heating Conditions
and Multi-Dimensional Instabilities in the Baseline Model}
\label{sec:baseline}

In order to provide the required background for interpreting the
evolution of the perturbed models, we first analyse the baseline
model p0 in some detail. In particular, we introduce the concept
of the critical neutrino luminosity \citep{burrows_93} in a form best
suited for our further analysis
(Section~\ref{sec:critical_luminosity}). We also define quantities
needed later for comparing perturbed and unperturbed models, and
collect several useful scaling relations connecting the heating
conditions and the activity of non-radial hydrodynamic instabilities
(Section~\ref{sec:saturation}).

\subsection{Shock Evolution}
Figs.~\ref{fig:shock_p0}--\ref{fig:heating_p0} provide
a compact overview  of the evolution of the baseline model
p0. The shock trajectories, the evolution of non-radial
shock oscillations, the contraction of the proto-neutron star,
the neutrino emission, and the flow morphology are all in
good qualitative and quantitative agreement with
recent 2D simulations based on more ambitious multi-group
neutrino transport methods.

The top panel of Fig.~\ref{fig:shock_p0} shows the maximum, minimum,
and average shock radii ($r_\mathrm{sh,min}$, $r_\mathrm{sh,min}$ and
$r_\mathrm{sh}$) along with the gain radius and proto-neutron star
radius. The average shock radius reaches a maximum value of $\approx
150 \ \mathrm{km}$ at a time of $100 \ \mathrm{ms}$ after bounce and
steadily recedes thereafter, following the contraction of the
proto-neutron star. From around $100 \ \mathrm{ms}$, multi-dimensional
instabilities lead to a sizeable deformation of the shock. This is also
reflected by the Legendre coefficients $a_1$ and $a_2$ for the dipole
and quadrupole deformation of the shock (bottom panel of
Fig.~\ref{fig:shock_p0}). We define $a_\ell$ in terms of the
$\ell$-th Legendre polynomial $P_\ell$ and the angle-dependent shock
radius $r_\mathrm{sh}(\theta)$ as
\begin{equation}
\label{eq:legendre_expansion}
a_\ell = \frac{2 \ell + 1}{2} \int\limits_0^\pi
r_\mathrm{sh}(\theta) P_\ell(\cos \theta) \, \ud  \cos \theta.
\end{equation}
We also use $a_0$ as the average shock radius $r_\mathrm{sh}$ 
in Fig.~\ref{fig:shock_p0}. The dipole and quadrupole amplitudes
are typically in the range of $10 \ldots 20 \ \mathrm{km}$
throughout the simulation. It is noteworthy that
$a_2$ is almost invariably positive, i.e.\ the shock
always exhibits a prolate deformation. 

\subsection{Neutrino Emission}
The neutrino luminosities and mean energies
(Fig.~\ref{fig:neutrino_p0}) likewise show a familiar picture.
There is a steady and very gradual decline of the luminosity of all
flavors from $100 \ \mathrm{ms}$ after bounce onward. The mean
energies rise steadily with a crossing of $\bar{\nu}_e$ and
$\nu_{\mu/\tau}$ energies around $0.3 \ \mathrm{ms}$ that is well known
for more massive progenitors \citep{mueller_14,marek_08}. At very late
times (around $1.3 \ \mathrm{s}$),  there is even another crossing of
electron neutrino and heavy flavor neutrino mean energies
due to the ``accretion effect'' identified by \citet{mueller_14}
(i.e.\ a temperature inversion develops
in the accretion layer around the neutrinosphere so that the
effective temperature of the heavy flavor neutrinos originating
from deeper layers drops below that of the electron flavour
neutrinos and anti-neutrinos).

\subsection{Secular Evolution of the Heating Conditions and
Approach to the Critical Luminosity}
\label{sec:critical_luminosity}
Fig.~\ref{fig:heating_p0} shows the critical ratio of the advection
and heating time-scales $\tau_\mathrm{adv}$ and $\tau_\mathrm{heat}$
(defined as in \citealt{mueller_12a}), the volume-integrated neutrino
heating rate $\dot{Q}_\nu$ in the gain region, and the heating
efficiency $\eta_\mathrm{heat}$ (defined as the ratio between
$\dot{Q}_\nu$ and the sum of the electron flavor luminosities
$L_{\nu_e}+L_{\bar{\nu}_e}$). The critical time-scale ratio
$\tau_\mathrm{adv}/\tau_\mathrm{heat}$ remains below unity throughout
the entire simulation. During the first second it never exceeds $0.6$,
which indicates that the model is indeed relatively far from an
explosive runaway. However, $\tau_\mathrm{adv}/\tau_\mathrm{heat}$
comes close to the critical threshold towards the very end of the
simulation. We likewise observe a slow increase of
$\eta_\mathrm{heat}$ after some $250 \ \mathrm{ms}$. The
volume-integrated heating rate remains high ($\approx 4 \times 10^{51}
\ \mathrm{erg} \ \mathrm{s}^{-1}$) at late times, and also exhibits
a slight secular increase.

As shown by \citet{janka_12}, the time-scale criterion
$\tau_\mathrm{adv} / \tau_\mathrm{heat} \gtrsim 1$ can be
re-formulated as a condition for the ``critical luminosity''
\citep{burrows_93,murphy_08b,pejcha_12} required for a successful explosion for
a given mass accretion rate $\dot{M}$. This perspective will be useful
for understanding why the baseline model p0 as well as other general
relativistic and pseudo-relativistic models with ray-by-ray variable
Eddington factor transport \citep{marek_09,mueller_12a} show a secular
increase of the time-scale ratio $\tau_\mathrm{adv} /
\tau_\mathrm{heat}$ at late times.

For estimating the critical luminosity, we note that
$\tau_\mathrm{adv}$ scales roughly as \citep{janka_12},
\begin{equation}
\label{eq:tadv}
\tau_\mathrm{adv} \propto \frac{r_\mathrm{sh}^{3/2}}{\sqrt{M}},
\end{equation} 
where $r_\mathrm{sh}$ can in turn be expressed in terms of the total
electron flavor luminosity $L_\nu=L_{\nu_e}+L_{\bar{\nu}_e}$, the mean
energy $E_\nu$ of electron neutrinos and anti-neutrinos, the gain
radius $r_\mathrm{gain}$, the mass accretion rate $\dot{M}$, and the proto-neutron
star mass $M$ as (see \citealt{janka_12} and Appendix~\ref{sec:toy_model})
\begin{equation}
r_\mathrm{sh} \propto 
\frac{(L_\nu  E_\nu^2)^{4/9} r_\mathrm{gain}^{16/9}}
{\dot{M}^{2/3} M^{1/3}}.
\label{eq:shock_radius}
\end{equation}
Here $E_\nu$ is defined as a weighted average of
electron neutrino and anti-neutrino mean energies:
\begin{equation}
\label{eq:avg_energy}
E_\nu^2=
\frac{L_{\nu_e} E_{\nu_e}^2+L_{\bar{\nu}_e} E_{\bar{\nu}_e}^2}{L_{\nu_e}+L_{\bar{\nu}_e}}.
\end{equation}

The heating time-scale can be expressed in
terms of the mass in the gain region $M_\mathrm{gain}$,
the average mass-specific binding energy $|e_\mathrm{gain}|$,
the neutrino luminosity $L_\nu$, and the heating
efficiency $\eta_\mathrm{heat}$ as
\begin{equation}
\tau_\mathrm{heat}
\approx 
\frac{M_\mathrm{gain} |e_\mathrm{gain}|}{L_\nu \eta_\mathrm{heat}}.
\end{equation}
Since
\begin{equation}
\label{eq:heating_efficiency}
\eta_\mathrm{heat}
\propto
\frac{M_\mathrm{gain}
E_\nu^2}
{r_\mathrm{gain}^2}
\end{equation}
holds to very good approximation \citep{janka_12}, this implies
\begin{equation}
\label{eq:theat}
\tau_\mathrm{heat}
\propto
\frac{|e_\mathrm{gain}| r_\mathrm{gain}^2}{L_\nu E_\nu^2}.
\end{equation}
While \citet{janka_12} posits that $|e_\mathrm{gain}|$ scales with the
gravitational potential $G M/r_\mathrm{gain}$ at the gain radius,
 simply using either the gravitational potential energy or
  the internal energy to determine the heating time-scale may
  introduce uncertainties on the level of a few tens of percent in the
  critical threshold \citep{murphy_08b,pejcha_12}.  However, the
  time-scale criterion becomes a very accurate tracer for the runaway
  threshold if the actual binding energy, i.e.\ the difference of the
  internal+kinetic energy and the potential energy, is
  used \citep{fernandez_12}. Estimating this difference analytically is not
  straightforward, but simulations show that the assumption of a
  time-independent binding energy per baryon actually works rather
  well.  This leads to
\begin{equation}
\tau_\mathrm{heat} \propto 
\frac{r_\mathrm{gain}^2}{L_\nu  E_\nu^2}.
\end{equation}
Using these approximate scaling relations,
the time-scale criterion $\tau_\mathrm{adv}/\tau_\mathrm{heat} \sim 1$ then 
translates into a critical condition for $L_\nu E_\nu^2$
as a function of $\dot{M}$, $M$, and $r_\mathrm{gain}$ (which is mostly
determined by $M$ through the mass-radius relation for
hot neutron stars),
\begin{equation}
\label{eq:critical_luminosity}
(L_\nu  E_\nu^2)_\mathrm{crit}
\propto
\left(\dot{M} M\right)^{3/5} r_\mathrm{gain}^{-2/5}.
\end{equation}
Note that unlike \citet{janka_12} we do not eliminate the gain
radius $r_\mathrm{gain}$ and the neutrino mean energy $E_\nu$ from this
relation.

Fig.~\ref{fig:critical_curve_p0} shows the evolution of $L_\nu
E_\nu^2$ versus $\dot{M} M$ along with a critical
curve given by equation~(\ref{eq:critical_luminosity}) anchored at a
point towards the end of the simulation where
$\tau_\mathrm{adv}/\tau_\mathrm{heat}$ approaches unity.  For about
$800 \ \mathrm{ms}$, the model continuously approaches the critical
curve. At this junction, the accretion rate drops slightly as the Neon
burning shell reaches the shock and then transitions into a more
shallow decline ($\dot{M} \propto t^{-1/3}$ as opposed to $\dot{M}
\propto t^{-1}$) at earlier times, reflecting the changing density
gradient in the progenitor. For a while, the model then moves parallel
to the critical curve, and again starts to approach it around $1
\ \mathrm{s}$.

Equation~(\ref{eq:critical_luminosity}) nicely shows the
underlying reason for the secular approach to the critical curve,
i.e.\ the increase of the ratio
\begin{equation}
\frac{L_\nu  E_\nu^2}{(L_\nu  E_\nu^2)_\mathrm{crit}}
\propto \frac{L_\nu  E_\nu^2 r_\mathrm{gain}^{2/5}}{(\dot M M)^{3/5}},
\end{equation}
and illustrates that it is potentially dangerous to reduce the
critical condition to a power law for $L_\mathrm{crit}$ in terms of
$\dot{M}$ and $M$, or even $\dot{M}$ alone
as in the classical form $L_\mathrm{crit}=L_\mathrm{crit} (\dot{M})$. If we consider the individual
quantities ($L_\nu$, $E_\nu$, $M$, $\dot M$, $r_\mathrm{gain}$) that
enter into the critical condition, we find that the ratio
$L_\nu/\dot{M}^{3/5}$ becoming more favorable at late times as
both the electron flavor luminosity $L_\nu$ and the accretion rate
$\dot{M}$ decrease. However, the contraction of the proto-neutron
  star and its growing mass also enter into the critical
  curve, and, somewhat astonishingly, the ratio $L_\nu
  r_\mathrm{gain}^{2/5}/(\dot{M} M)^{3/5}$ is almost constant in model
  p0 from $\mathord{\sim} 100 \ \mathrm{ms}$ after bounce. The approach
  towards the critical curve therefore hinges solely on the secular
  increase of the neutrino mean energy ($E_\nu \propto M$, see
  \citealt{mueller_14}). While the interplay of the different terms
  may be somewhat sensitive to the detailed treatment of the neutrino
  transport, the neutrino opacities, and the equation of state, this
  suggests that at least for massive progenitors without an early and
  abrupt drop of the mass accretion rate, this highlights the
  paramount importance of the mean energies for an explosive runaway
  due to neutrino heating.

As an aside, we note that the phenomenological scaling
  relations describing the 2D simulations result in somewhat different
  power-law exponents for the scaling of the critical luminosity than
  in the power-law fits of \citet{pejcha_12}, who also include the
  dependence on neutron star mass and radius (in their case, more,
  precisely the neutrinosphere radius) in equation~(16) of their
  paper: They obtain
\begin{equation}
L_{\nu,\mathrm{core}}^\mathrm{crit} =
8.18 \times 10^{52} \ \mathrm{erg} \ \mathrm{s}^{-1} \times
\tau_\nu^{-0.206} \left(\frac{M}{M_\odot}\right )^{1.84} 
\left(\frac{\dot M}{M_\odot \ \mathrm{s}^{-1}}\right)^{0.723}
\left(\frac{r_\nu}{10 \ \mathrm{km}}\right)^{-1.61}
\end{equation}
for the critical \emph{core} luminosity of electron neutrinos and
antineutrinos in terms of $M$, $\dot{M}$, and the optical depth
$\tau_\nu$ and radius $r_\nu$ of the neutrinosphere. The dependence on
$M$ and $r_\nu$ is noticeably steeper than in
equation~(\ref{eq:critical_luminosity}). Using $\tau_\nu=2/3$ and the
neutron star radius defined by a density of $10^{11} \ \mathrm{g}
\ \mathrm{cm}^{-3}$ as a proxy for $r_\nu$, we find that equation~(16)
of \citet{pejcha_12} actually predicts an \emph{increase} of
$L_{\nu,\mathrm{core}}^\mathrm{crit}$ with time
(Figure~\ref{fig:pejcha_thompson}). Even including a correction factor
$(15.5 \ \mathrm{MeV}/E_{\bar{\nu}_e})^{2}$, with $E_{\bar{\nu}_e}$
being the mean energy of electron antineutrinos (O.~Pejcha, private
communication), merely moderates this increase so that
$L_{\nu,\mathrm{core}}^\mathrm{crit}$ stays fairly constant, but
contrary to the other indicators discussed before does not reflect the
secular improvement in the heating conditions.  The fact that
\citet{pejcha_12} derived their criterion for the core luminosity
instead of the luminosity in the gain region does not help matters
much, since the core luminosity is, by definition, smaller than the
total luminosity and also decreases with time, thus moving away from
their estimate for $L_{\nu,\mathrm{core}}^\mathrm{crit}$. While this
shows the limitations of the fit formula of \citet{pejcha_12}, it does
not completely invalidate their underlying physical model, which may still
capture much (though not all) of the relevant physics. It does,
however, illustrate the need to bolster empirical scaling laws by
simulation data and physical arguments in the regime where they are to
be applied.

\begin{figure}
\includegraphics[width=\linewidth]{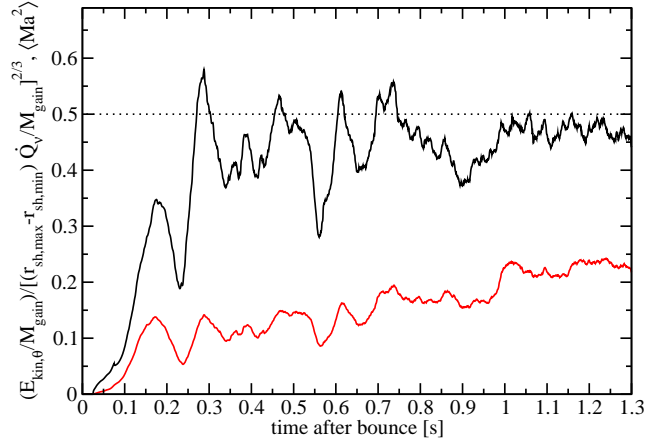}
\caption{Running average over $50 \ \mathrm{ms}$ of the ratio
$(E_{\mathrm{kin},\theta}/M_\mathrm{gain})/[(r_\mathrm{sh,min}-r_\mathrm{gain})\dot{Q}_\nu/M_\mathrm{gain}]^{2/3}$
(black) 
for the baseline model p0, illustrating the validity
of equation~(\ref{eq:kinetic_energy_vs_heating}) for
the kinetic energy in the gain region as a function of the
mass-specific heating rate in a time-averaged sense.
The plot also shows the average squared Mach number $\langle \mathrm{Ma}^2\rangle$ 
in the gain region (red).
\label{fig:kinetic_energy_vs_heating}
}
\end{figure}

\begin{figure}
\includegraphics[width=\linewidth]{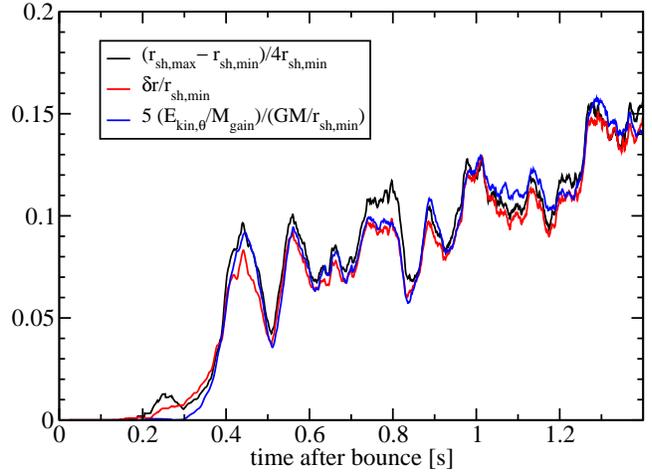}
\caption{ $(r_\mathrm{sh,max}-r_\mathrm{sh,min})/4$ (black), $\delta
  r$ (red), and the prediction for $\delta r$ according to
  equation~(\ref{eq:dr_sasi}) (blue) (all normalised by
  $r_\mathrm{sh,min}^{-1}$), smoothed over $50 \ \mathrm{ms}$,
for the baseline model p0.
\label{fig:shock_deformation_analytic}
}
\end{figure}

\subsection{Saturation of Non-Radial
Instabilities in the Baseline Model} 
\label{sec:saturation}

In the absence of large progenitor asphericities, the heating
conditions will eventually decide about the saturation of SASI and/or
convection in the non-linear regime.  This is more obvious in the case
of convection, where neutrino heating is itself the driving agent of
the instability. The SASI, on the other hand, is an instability of the
accretion flow that is not powered by neutrino heating, but its
\emph{saturation} by parasitic Rayleigh instabilities
\citep{guilet_10} is nonetheless regulated by the strength of neutrino
heating. Moreover, the growth conditions of the SASI are also
indirectly determined by neutrino heating and cooling as they depend
on the contraction of the proto-neutron star and the shock trajectory.

\subsubsection{Kinetic Energy Contained in Non-Radial
Instabilities}

One can formulate a simple model to describe the dependence of the
strength of non-radial instabilities on the heating conditions with
reasonable accuracy.  Several authors
\citep{thompson_00,murphy_11,murphy_12} have already addressed the
interplay of neutrino heating and turbulent motions in the gain
region, some of them \citep{murphy_11} with the very ambitious goal of
developing a full theory of turbulence in neutrino-driven supernovae.
Our approach relies on the observation of \citet{fernandez_14} that
the time-averaged stratification in the gain region adjusts itself to
achieve marginal stability to convection in the non-linear phase for a
wide range of initial conditions.

In order to maintain marginal stability to convection, the
volume-integrated energy input rate $\dot{Q}_\nu$ by neutrino heating 
in the gain layer must
be compensated by an outward ``turbulent luminosity''\footnote{In this
  context, we classify any deviation from a spherically symmetric flow
  pattern as ``turbulent'' for the sake of simplicity.}
$L_\mathrm{turb}$ of similar magnitude
(cf.\ \citealt{murphy_12}). Irrespective of whether the SASI or
convection dominate the dynamics of the post-shock flow, the net
turbulent energy flux is the result of hot matter moving outward and
cold matter flowing inward with respect to the spherically averaged
background flow. $L_\mathrm{turb}$ depends on the typical turbulent
velocity $\delta v$ (through the typical crossing time
$\tau_\mathrm{cross}\approx (r_\mathrm{sh,min}-r_\mathrm{gain})/\delta
v$ of a high-entropy ``bubble'' in a convection- or SASI-dominated
flow) and the enthalpy contrast $\delta h$:
\begin{equation}
L_\mathrm{turb} \propto \frac{M_\mathrm{gain} \, \delta h}{\tau_\mathrm{cross}}
\propto 
\frac{M_\mathrm{gain} \, \delta v \, \delta h}{r_\mathrm{sh,min}-r_\mathrm{gain}},
\end{equation}
where the enthalpy contrast should scale as 
$\delta h \propto \delta P/\rho\propto \delta v^2$ (see equation~(31.4) in
\citealt{landau_fluid}). The condition $L_\mathrm{turb} \sim \dot{Q}_\nu$
then leads to a scaling relation for the typical
turbulent velocity
\begin{equation}
\delta v
\propto
\left[\frac{(r_\mathrm{sh,min}-r_\mathrm{gain}) \dot{Q}_\nu}{M_\mathrm{gain}}\right]^{1/3}.
\end{equation}
If we take the RMS average $\sqrt{\langle v_\theta^2 \rangle}$ of the
$\theta$-component of the velocity as a proxy for $\delta v$
(which is a convenient measure because of rough equipartition
between the radial and lateral components
of the kinetic energy in 2D as found by \citealt{murphy_12}), we find
that the volume-averaged kinetic energy contained in lateral motions
should scale as:
\begin{equation}
\label{eq:kinetic_energy_vs_heating_prop}
\frac{E_{\mathrm{kin},\theta}}{M_\mathrm{gain}}
\propto
\left[\frac{(r_\mathrm{sh,min}-r_\mathrm{gain}) \dot{Q}_\nu}{M_\mathrm{gain}}\right]^{2/3}.
\end{equation}
The simulation data for model p0 suggests
\begin{equation}
\label{eq:kinetic_energy_vs_heating}
\frac{E_{\mathrm{kin},\theta}}{M_\mathrm{gain}}
\approx 
0.5 \times
\left[\frac{(r_\mathrm{sh,min}-r_\mathrm{gain}) \dot{Q}_\nu}{M_\mathrm{gain}}\right]^{2/3},
\end{equation}
as a good approximation for the time-averaged specific kinetic energy in the
gain region (Fig.~\ref{fig:kinetic_energy_vs_heating}) once
the SASI and/or convection reach their saturation level, although
there are considerable short-term excursions away from this value.

As an alternative to our argument based on the assumption of a
self-adjustment to marginal stability to convection, one can derive
the scaling law (\ref{eq:kinetic_energy_vs_heating}) based on the
picture of a Carnot engine operating in the supernova core
  with some crucial modifications of the original idea of
  \citet{herant_94}.  When popularised by \citet{herant_94} in the
  1990s, the heat engine in the supernova core was initially conceived
  of as powering the explosion itself through a continuous build-up of
  kinetic energy in the gain region. While this build-up is not observed
in modern multi-D simulations like ours, this does not render
the thermodynamic analogy with a heat engine invalid; instead a more careful
analysis shows why and how the original idea of \citet{herant_94}
and their conclusions about the explosion mechanism must be corrected.

In the heat engine picture,
  the overturn motions are viewed as an approximate Carnot cycle
  involving the (roughly) adiabatic transport of cold material from
  the shock to the gain radius, where it is heated, then expands as it
  is transported out to the shock and cools by mixing with cold
  material as the convective/SASI plume dissolves. The mechanical
  power $\dot{e}_\mathrm{kin}$ per unit mass pumped into the
  instabilities is then given in terms of the heating rate per unit
  mass $\dot{Q}_\nu/M_\mathrm{gain}$ as
\begin{equation}
\dot{e}_\mathrm{kin}=\frac{T_\mathrm{gain}-T_\mathrm{sh}}{T_\mathrm{gain}} \frac{\dot{Q}_\nu}{M_\mathrm{gain}},
\end{equation}
where the temperatures at the shock, $T_\mathrm{sh}$, and at the
gain radius, $T_\mathrm{gain}$, determine the Carnot efficiency.
Since we roughly have $T\propto r^{-1}$ in the gain region, we have
\begin{equation}
\dot{e}_\mathrm{kin}=\frac{r_\mathrm{sh}-r_\mathrm{gain}}{r_\mathrm{sh}} \frac{\dot{Q}_\nu}{M_\mathrm{gain}}.
\end{equation}
\citet{herant_94} proposed this continuous generation
of kinetic energy in the heat engine as a means to power the explosion,
and this is where their model breaks down: It overlooks the possibility 
that the Carnot engine reaches a (quasi-)stationary state where the input of mechanical energy
is balanced by turbulent dissipation. Dissipation happens at a rate
$\varepsilon$
of
\begin{equation}
\varepsilon \sim \frac{\delta v^3}{r_\mathrm{sh}},
\end{equation}
where $\delta v$ and $r_\mathrm{sh}$ enter as the typical velocity
scale and length scale of the turbulent flow (see \textsection~31 in
\citealt{landau_fluid}).\footnote{One can interpret this well-known
  result from turbulence theory as the rate of viscous dissipation for
  a (kinematic) eddy viscosity given by $\delta v\ r_\mathrm{sh}$ and
  a typical shear rate of $\delta v/r_\mathrm{sh}$. } The
  continuous loss of kinetic energy by downward advection to the
  cooling region will lead to a term of the same form (kinetic energy
  density $v^2/2$ divided by the advection time-scale
  $\tau_\mathrm{adv} \mathord{\sim} r_\mathrm{sh}/v$), and can be thought of as
  modifying the proportionality constant in the dissipation
  law.\footnote{The continuous exchange of the working
    substance is another issue that \citet{herant_94} ignored. Again,
    this complication does not undermine the thermodynamic analogy;
    many familiar heat engines use such an open cycle.}

 The condition $\dot{e}_\mathrm{kin} \mathord{\sim} \varepsilon $
then immediately leads to equation~(\ref{eq:kinetic_energy_vs_heating_prop}).
 Instead of the continuous build-up of kinetic energy
assumed by \citet{herant_94}, we therefore end up with a saturation
of the kinetic energy instead of a runaway. The Carnot engine
therefore cannot power continuous shock expansion, but it will
nevertheless have an effect on the \emph{threshold} for runaway
shock heating because it creates Reynolds stresses that alter
the post-shock stratification and the shock position as we shall
see in Section~\ref{sec:stresses}.

Sometimes we will use a simplified scaling law
based on the assumption that $r_\mathrm{sh,min}-r_\mathrm{gain} \propto r_\mathrm{sh}$,
\begin{equation}
\label{eq:kinetic_energy_vs_heating_simple}
\frac{E_{\mathrm{kin},\theta}}{M_\mathrm{gain}}
\propto
\left(\frac{r_\mathrm{sh} \dot{Q}_\nu}{M_\mathrm{gain}}\right)^{2/3}.
\end{equation}
This simpler form corresponds to the scaling law obtained by
\citet{thompson_00} (equation 29 in his paper) using a slightly
different derivation, which neglected turbulent dissipation and the
finite efficiency of the heat engine and instead assumed that all the
heating is converted into mechanical energy over one overturn
time-scale.

\subsubsection{Amplitude of Shock Oscillations}

The lateral kinetic energy, which can be taken as a convenient measure
for the \emph{overall} strength of lateral and radial turbulent
motions due because of rough equipartition between both components of
the turbulent kinetic energy in 2D \citep{murphy_12}, in turn
determines the amplitude of shock oscillations. We find a very tight
relation between the lateral kinetic energy per unit mass
$E_{\mathrm{kin},\theta}/M_\mathrm{gain}$, the gravitational potential
at the shock radius $G M / r_\mathrm{sh,min}$, and the typical
deviation $\delta r$ of the shock radius from its spherical average
$r_\mathrm{sh}=a_0$:
\begin{equation}
\delta r
= \sqrt{\int \left(r_\mathrm{sh}(\theta,\varphi)-a_0\right)^2 \, \ud \Omega}.
\end{equation}

$\delta r$ can be expressed in terms of the Legendre
coefficients $a_\ell$ as
\begin{equation}
\delta r
=  \sqrt{\int \left(r_\mathrm{sh}(\theta,\varphi)-a_0\right)^2 \, \ud \Omega}
=\sqrt{\frac{a_1^2}{3}+\frac{a_2^2}{5}+\frac{a_3^2}{7}+\ldots}.
\end{equation}
Fig.~\ref{fig:shock_deformation_analytic} shows that for
the baseline model
\begin{equation}
\label{eq:dr_sasi}
\delta r \approx
\frac{5 E_\mathrm{kin,\theta} r_\mathrm{sh,min}^2}{G M M_\mathrm{gain} }
\end{equation}
and
\begin{equation}
r_\mathrm{sh,max}-r_\mathrm{sh,min}
\approx 4 \delta r
\approx 
\frac{20 E_\mathrm{kin,\theta} r_\mathrm{sh,min}^2}{G M M_\mathrm{gain} }
\end{equation}
hold to good approximation. Note
that we have $r_\mathrm{sh,max}-r_\mathrm{sh,min} \approx 4 \delta r$ 
because the average deviation of the shock radius from its
angular average is roughly $(r_\mathrm{sh,max}-r_\mathrm{sh})/2$,
and in turn we have $r_\mathrm{sh,max}-r_\mathrm{sh} \approx
(r_\mathrm{sh,max}-r_\mathrm{sh,min})/2$.

One can construct a crude physical picture to motivate this scaling
relation. If a large bubble produced by SASI or convection is
to push out the shock by $\delta r$ in one hemisphere, it must
exert work against the pressure of the shocked material flowing
around it (which is roughly given by the post-shock pressure
$P_\mathrm{sh}$).
Expansion will cease once the entire kinetic energy of the bubble
has been consumed by $P \, \ud V$  work. If half of the
kinetic energy in lateral motions in the gain region is contained
in the hemispheric bubble, then we find
\begin{equation}
2 \pi r_\mathrm{sh}^2 P_\mathrm{sh} \delta r
=
\frac{1}{2}E_\mathrm{kin,\theta}.
\end{equation}
$P_\mathrm{sh}$ can be related to the pre-shock ram pressure
$\rho_\mathrm{pre} v_\mathrm{pre}^2$ using the jump conditions
for a stationary shock,
\begin{equation}
P_\mathrm{sh}=\frac{\beta-1}{\beta}  \rho_\mathrm{pre} v_\mathrm{pre}^2,
\end{equation}
where the pre-shock velocity $v_\mathrm{pre}$ is a large
fraction of the free-fall velocity ($v_\mathrm{pre}\approx \sqrt{2 G M/r_\mathrm{sh}}$),
$\rho_\mathrm{pre}=\dot{M}/(4\pi r^2 v_\mathrm{pre})$ is the pre-shock
density, and $\beta$ is the ratio of the post-and pre-shock
densities. Expressing $P_\mathrm{sh}$ in term
of the post-shock density $\rho_\mathrm{sh}$, we find
\begin{equation}
P_\mathrm{sh}\approx \frac{\beta-1}{\beta^2} \frac{\rho_\mathrm{sh} GM}{r_\mathrm{sh}},
\end{equation}
and hence obtain
\begin{equation}
\frac{
2 \pi r_\mathrm{sh}^2  \delta r (\beta-1) \rho_\mathrm{sh}
G M}{\beta^2 r_\mathrm{sh}}
=
\frac{1}{2}E_\mathrm{kin,\theta}.
\end{equation}
Since $\rho_\mathrm{sh} \propto M_\mathrm{gain}/(4/3 \pi r_\mathrm{sh}^3)$, this implies
\begin{equation}
\frac{
3 \delta r (\beta-1) M_\mathrm{gain}
G M}{2 \beta r_\mathrm{sh}^2 }
\propto
\frac{1}{2}E_\mathrm{kin,\theta},
\end{equation}
and hence a scaling law for $\delta r$ of the form of
equation~(\ref{eq:dr_sasi}),
\begin{equation}
\label{eq:dr_sasi_2}
\delta r \propto
\frac{E_\mathrm{kin,\theta} r_\mathrm{sh,min}^2}{G M M_\mathrm{gain} }.
\end{equation}
Alternatively, the scaling law may be written
in terms of the typical (squared) turbulent Mach number
of lateral motions in the post-shock region, $\langle \mathrm{Ma}^2\rangle$,
which we define as
\begin{equation}
\langle \mathrm{Ma}^2\rangle=\frac{\langle v_\theta^2 \rangle}{c_\mathrm{s,post}^2}=
\frac{2E_\mathrm{kin,\theta}/M_\mathrm{gain}}{c_\mathrm{s,post}^2}.
\end{equation}
Here, the post-shock sound speed $c_\mathrm{s,post}$
is given in terms of by $P_\mathrm{sh}$, $\rho_\mathrm{sh}$, 
and the adiabatic index $\Gamma$ as:
\begin{equation}
\label{eq:c_s_shock}
c_\mathrm{s,post}^2=\frac{\Gamma P_\mathrm{sh}}{\rho_\mathrm{sh}}
=\Gamma \frac{\beta-1}{\beta^2} v_\mathrm{pre}^2
=\Gamma \frac{\beta-1}{\beta^2} \frac{2 G M}{r_\mathrm{sh,min}}
\approx \frac{G M}{3r_\mathrm{sh,min}},
\end{equation}
if we use $\beta \approx 7$ and $\Gamma=4/3$. 
One can easily verify that equation~(\ref{eq:dr_sasi}) can thus
be written as
\begin{equation}
\label{eq:dr_sasi_mach}
\frac{\delta r}{r_\mathrm{sh,min}} \approx
\frac{5}{6}\langle \mathrm{Ma}^2\rangle.
\end{equation}

\begin{figure*}
\includegraphics[width=\linewidth]{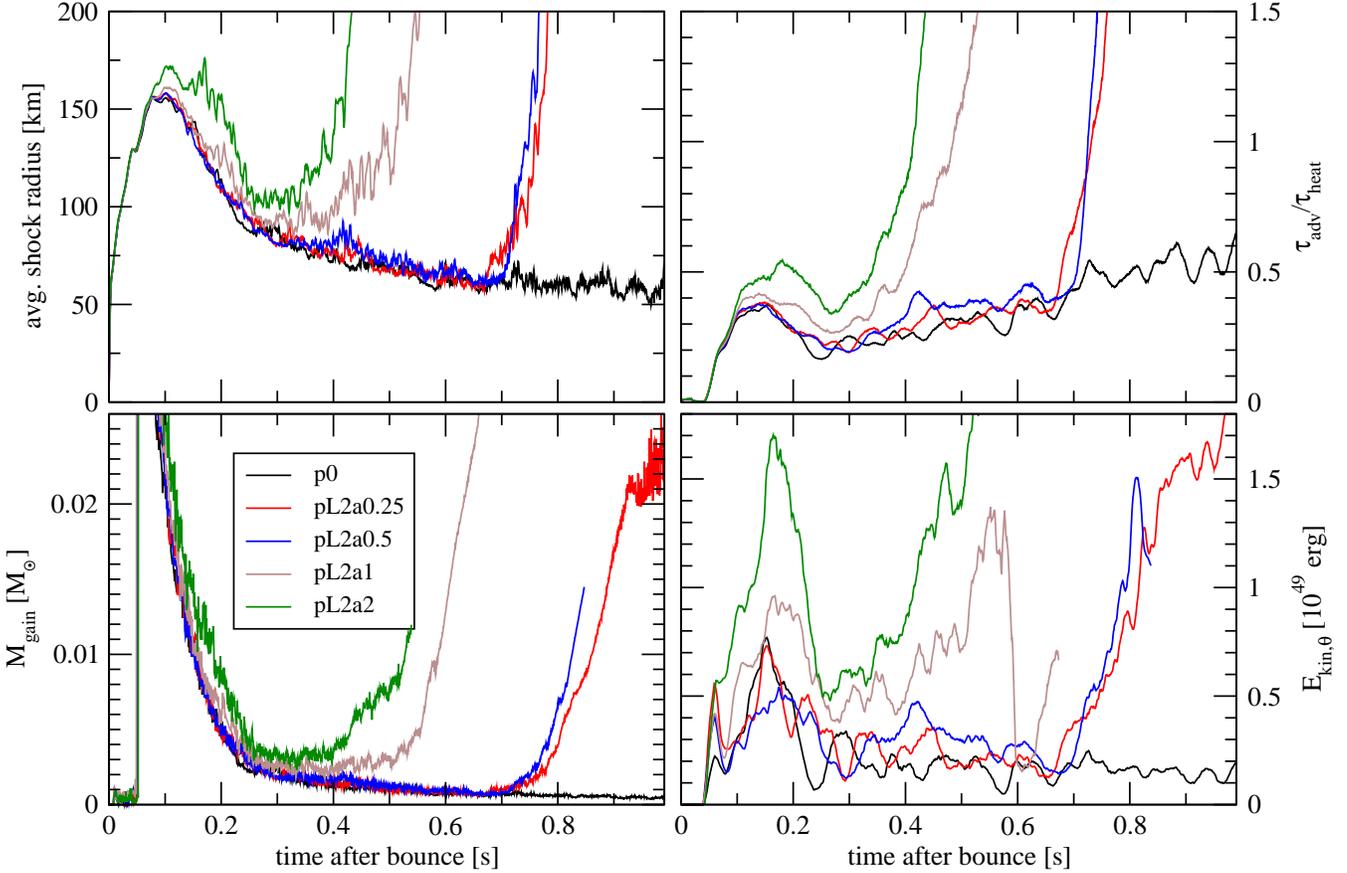}
\caption{
\label{fig:heating_pL2}
Comparison of the average shock radius (top left), the criticality
parameter $\tau_\mathrm{adv}/\tau_\mathrm{heat}$ (top right), the mass
in the gain region $M_\mathrm{gain}$ (bottom left), and the kinetic energy contained
in lateral fluid motions $E_{\mathrm{kin},\theta}$ (bottom right) for the baseline
model p0 and selected models from the pL2aX series.}
\end{figure*}

\subsection{The Impact of Non-Radial Instabilities on the
Heating Conditions}
\label{sec:stresses}
 Presently, we still lack a quantitative theory for
describing how multi-dimensional instabilities facilitate runaway
shock expansion. While both parameterized and first-principle
simulations have established that multi-dimensional effects are
beneficial for the heating conditions in the supernova core, different
concepts may be and have been used to interpret these findings, 
many of which date back already to the first generation of
multi-dimensional supernova models in the 1990s (see, e.g., \citealp{herant_94,burrows_95}). On the
one hand, it can be argued that turbulent stresses
\citep{murphy_12,mueller_12b,couch_14}, heating by secondary shocks
during the non-linear SASI phase \citep{mueller_12b} and convective
energy transport alter the (spherically averaged) post-shock flow,
enhance the mass in the gain region, and thereby bring the supernova
core closer to the critical condition
$\tau_\mathrm{adv}/\tau_\mathrm{heat}$ for runaway heating.  On the
other hand, one might view the possibility of asymmetric shock
expansion driven by large neutrino-heated bubbles
\citep{thompson_00,burrows_12,dolence_13,couch_12b,fernandez_14} as
the crucial factor for facilitating explosions in multi-D.

These different viewpoints are by no means incongruent with each
other, and it is difficult to decide which picture is most appropriate
as a \emph{causal} explanation on the basis of simulation
data. Nevertheless, it is desirable to have a rough diagnostic
quantity for the global influence of multi-dimensional effects in the
supernova core. We argue that the typical Mach number of non-radial
velocities in the gain region is an appropriate measure for assessing
how efficiently multi-dimensional instabilities assist neutrino
heating in the development of a runaway instability.

The pivotal role of the ``turbulent'' Mach number $\mathrm{Ma}$ in the
post-shock region can be illustrated by considering two very crude
models for an explosive runaway aided by multi-dimensional
instabilities.  If we incorporate the effect of turbulent stresses as
an additional isotropic pressure contribution $P_\mathrm{turb} \approx
\langle \delta v^2\rangle \rho \approx 4/3 \langle \mathrm{Ma}^2\rangle P$ throughout
the gain region into the derivation of the relation for the critical
luminosity (\ref{eq:critical_luminosity}), we obtain
\begin{equation}
\label{eq:critical_luminosity_stresses}
L_\nu E_\nu^2 \propto (\dot{M} M)^{3/5}
r_\mathrm{gain}^{-2/5} \left(1+\frac{4\langle \mathrm{Ma}^2\rangle}{3}\right)^{-3/5}
\end{equation}
because the higher post-shock pressure results in a
larger shock radius compared to equation~(\ref{eq:shock_radius}),
\begin{equation}
\label{eq:shock_radius_stress}
r_\mathrm{sh} \propto 
\frac{(L_\nu  E_\nu^2)^{4/9} r_\mathrm{gain}^{16/9}
\left(1+\frac{4\langle \mathrm{Ma}^2\rangle}{3}\right)^{2/3}}
{\dot{M}^{2/3} M^{1/3}}
\end{equation}
and a longer advection time-scale (see Appendix~\ref{sec:toy_model}). A large turbulent Mach number
therefore leads to a reduction of the critical luminosity in this
simple model. 
Note that
$\langle \mathrm{Ma}^2\rangle$ is a function of
the heating conditions here and hence depends $L_\nu$, $E_\nu$,
$r_\mathrm{gain}$, $\dot M$, $M$, and also $r_\mathrm{sh}$ itself.
Surprisingly, this simple model can roughly predict the
observed reduction of the critical luminosity by $\mathord{\sim} 20 \%$ in
multi-D compared to 1D, as we show in Appendix~\ref{sec:toy_model}.

On the other hand, if we view the emergence of a large, expanding
high-entropy bubble as the critical element in multi-dimensional
explosions \citep{thompson_00,burrows_12,dolence_13,couch_12b,fernandez_14},
we can formulate an alternative runaway condition.  If the buoyant
acceleration of such a bubble is to overcome the drag of the infalling
material flowing around it, the density contrast $\delta \rho /\rho$
to the surrounding matter must exceed a critical limit
\citep{fernandez_14},
\begin{equation}
\left(\frac{\delta \rho}{\rho} \right) \sim
\frac{C_D v_\mathrm{post}^2}{2 g l}.
\end{equation}
Here, $C_D$ is the drag coefficient, 
$v_\mathrm{post}$ is the post-shock velocity, $g=GM/r_\mathrm{sh}^2$
is the gravitational acceleration at the shock, and
$l$ is the ratio of the volume of the bubble to its
cross-section. Since $\delta \rho / \rho \sim \langle \mathrm{Ma}^2 \rangle$
(see \textsection~10 in \citealt{landau_fluid}), this
is again a critical condition involving the turbulent
Mach number:
\begin{equation}
\langle \mathrm{Ma}^2 \rangle 
\sim\frac{C_D v_\mathrm{post}^2}{2g l}.
\end{equation}
If we assume that the volume-to-surface ratio $l$ is
given by the shock radius (e.g. $l \sim r_\mathrm{sh}/2$)  when
runaway expansion sets in, this condition can be cast into
an even simpler form,
\begin{equation}
\label{eq:critical_mach_number}
\langle \mathrm{Ma}^2 \rangle 
\sim
\frac{C_D v_\mathrm{post}^2 r_\mathrm{sh}}{2 G M}
\sim C_D \beta^{-2},
\end{equation}
where $\beta$ is the ratio of the post- and pre-shock densities.  As
the turbulent velocities are related to the heating conditions, this
also implies a critical condition for the neutrino luminosity and mean
energy. Using
equations~(\ref{eq:shock_radius},\ref{eq:heating_efficiency},
\ref{eq:kinetic_energy_vs_heating_simple},\ref{eq:c_s_shock}) to
express the lateral kinetic energy $E_\mathrm{kin,\theta}$ in terms of
the neutrino heating $\dot{Q}_\nu$, and
$\dot{Q}_\nu=\eta_\mathrm{heat} L_\nu$ (with
  $\eta_\mathrm{heat}$ taken from equation~\ref{eq:heating_efficiency})
  in terms of $L_\nu$, $E_\nu$, and $r_\mathrm{gain}$, we obtain
\begin{equation}
\label{eq:bubble}
\langle \mathrm{Ma}^2\rangle
=
\frac{2E_\mathrm{kin,\theta}/M_\mathrm{gain}}{c_\mathrm{s,post}^2}
\propto
\frac{\left(r_\mathrm{sh} \dot{Q}_\nu/M_\mathrm{gain}\right)^{2/3}}{G M/r_\mathrm{sh}}
\propto
\left(
\frac{r_\mathrm{sh} L_\nu E_\nu^2}{r_\mathrm{gain}^2}
\right)^{2/3}
\left(\frac{r_\mathrm{sh}}{GM}\right)
\end{equation}
for the turbulent Mach number in terms of $L_\nu$, $E_\nu$, $r_\mathrm{gain}$,
$\dot{M}$ and $M$. 
Equation~(\ref{eq:shock_radius}) allows us to eliminate
the shock radius in favour of $L_\nu$, $E_\nu$, $r_\mathrm{gain}$,
$M$, and $\dot{M}$:
\begin{equation}
\label{eq:bubble2}
\langle \mathrm{Ma}^2\rangle
\propto
\frac{(L_\nu E_\nu^2)^{2/3}r_\mathrm{sh}^{5/3}}{r_\mathrm{gain}^{4/3} M}
\propto
\frac{(L_\nu E_\nu^2)^{38/27} r_\mathrm{gain}^{44/27}} {\dot{M}^{10/9}M^{14/9} }
\end{equation}
The critical condition $\langle \mathrm{Ma}^2\rangle \sim C_D \beta^{-2}$ will
thus be reached for a critical luminosity given by
\begin{equation}
\label{eq:critical_luminosity_bubbles}
L_\nu E_\nu^2 \propto C_D^{27/38} \beta^{-27/19}
\dot{M}^{15/19} M^{21/19} r_\mathrm{gain}^{-22/19}
\propto
 C_D^{0.71} \beta^{-1.42}
\dot{M}^{0.79} M^{1.11} r_\mathrm{gain}^{-1.16}.
\end{equation}
Hence, the critical condition for runaway bubble expansion implies that
we end up with a relation for the critical luminosity with a relatively
similar dependence on $\dot{M}$ as in
equation~(\ref{eq:critical_luminosity}), and a somewhat steeper dependence
on $M$ and $r_\mathrm{gain}$. Given the limited range of variation of
$M$ and $r_\mathrm{gain}$ and the approximations inherent in the derivation,
it seems unlikely that the shape of the critical curve alone (as inferred
from simulations) can distinguish between the two pictures of the
explosive runaway at shock revival.

The practical use of equation~(\ref{eq:critical_mach_number})
for a quantitative estimate of a ``critical Mach number'' is
limited, however. The geometry of the bubble is a major
uncertainty; arguably any value of $l$ in the
range $l=r_\mathrm{sh}/3\ldots r_\mathrm{sh}$ is defensible.
Likewise, the drag coefficient $C_D$ should be considered as highly uncertain:
Simply applying the subsonic drag coefficient $C_D\sim 0.5$ of a sphere
at intermediate Reynolds numbers may not be adequate, since the
bubble has to expand against a \emph{supersonic} flow from which
it is separated by a detached bow shock and a subsonic region of
colder, shocked material, which has to be taken into
account when computing an effective drag force. Nevertheless,
one should still expect the turbulent Mach number in the post-shock
region to be a critical factor in the balance between
the buoyancy and drag forces even though these complications modify
the picture quantitatively.

While admittedly based on two rather crude models for the effect of
multi-D instabilities,
equations~(\ref{eq:critical_luminosity_stresses}) and
(\ref{eq:critical_luminosity_bubbles}) are very suggestive: In both
cases, the reduction of the critical luminosity compared to the 1D
case is given by a simple scaling factor (roughly reflecting the
findings of \citealt{murphy_08b}, \citealt{hanke_12} and
\citealt{couch_12a,couch_12b}). The average squared Mach number
$\langle \mathrm{Ma}^2\rangle $ of aspherical motions in the gain
layer either enters directly as a critical parameter in
equation~(\ref{eq:critical_mach_number}), or regulates the reduction
of the critical luminosity by multi-D effects in
equation~(\ref{eq:critical_luminosity_stresses}).
In Section~\ref{sec:saturation}, we also found
that it is the crucial parameter that regulates the shock deformation. There is thus ample
motivation for considering $\langle \mathrm{Ma}^2 \rangle$ as an important measure for
the role of multi-D effects in shock revival. 
 Other factors, like the
effective drag coefficient for high-entropy bubbles (which depends
both on their form and on numerical viscosity) and their
surface-to-volume ratio may be of similar relevance for the runaway
threshold in multi-D, both are less readily quantifiable.

In order to define a useful volume-integrated measure for the
violence of non-radial instabilities, we compare the RMS
average of the lateral velocity $v_\theta$ to the post-shock sound speed, which is approximately
(see equation~(\ref{eq:c_s_shock}) for a derivation):
\begin{equation}
\label{eq:c_s_shock2}
c_\mathrm{s,post}^2
\approx
\frac{G M}{3r_\mathrm{sh,min}}.
\end{equation}
As in Section~\ref{sec:saturation}, we define the average
squared Mach number $\langle \mathrm{Ma}^2\rangle$ in the gain region using
this value for $c_\mathrm{s,post}^2$ as an average over the entire gain region:
\begin{equation}
\langle \mathrm{Ma}^2 \rangle
=
\frac{2 E_\mathrm{kin,\theta}}{M_\mathrm{gain} c_\mathrm{s,post}^2}
=
\frac{6 r_\mathrm{sh,min} E_\mathrm{kin,\theta}}{G M M_\mathrm{gain}}.
\end{equation}
$\langle \mathrm{Ma}^2\rangle$ is shown in Fig.~\ref{fig:kinetic_energy_vs_heating}
for the baseline model. There is a steady increase as the heating
conditions slowly improve with time.

\begin{figure}
\includegraphics[width=\linewidth]{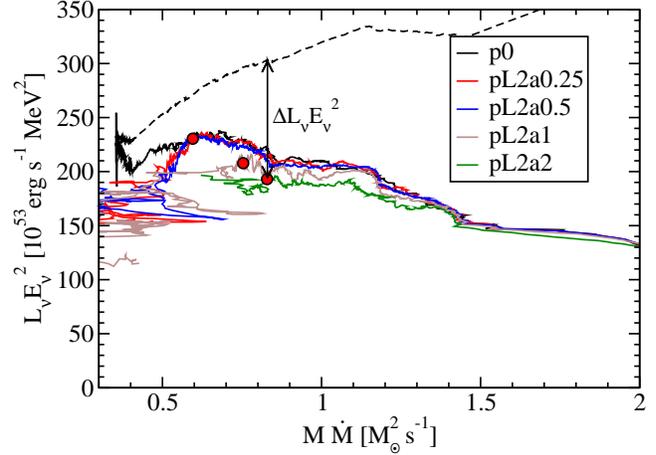}
\caption{
\label{fig:critical_curve_p5}
Modification of the critical luminosity due to seed aspheriticites for
the pL2aX series.  The approach of the baseline model p0 (solid black
curve) to a fiducial critical curve (dashed) in the $(M \dot{M},L_\nu
E_\nu^2)$ plane is shown as in Fig.~\ref{fig:critical_curve_p0},
i.e.\ the critical curve is anchored at the final location $(M
\dot{M},L_\nu E_\nu^2)$ of model p0.  The onset of the explosions
     (defined as the time when
      $\tau_\mathrm{adv}/\tau_\mathrm{heat}$ reaches unity) for
    models pL2a0.25 to pL2a2 (colored lines) is marked by a red circle
    on each trajectory. Note that the red circles for
      pL2a0.25 and pL2a0.5 lie on top of each other.
As indicated for model pL2a2, the threshold for the explosion
is reduced considerably by $\Delta L_\nu E_\nu^2$ for the perturbed models.  }
\end{figure}

\begin{figure}
\includegraphics[width=\linewidth]{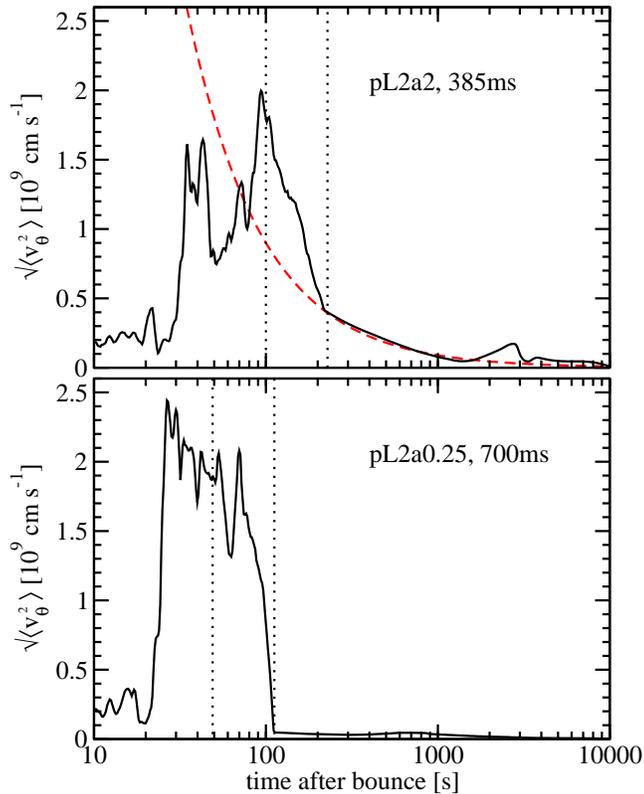}
\caption{
\label{fig:v_theta_rms}
Lateral velocity dispersion $\sqrt{\langle v_\theta^2\rangle}$ 
(equation \ref{eq:velocity_dispersion})
for
model pL2a2 at a post-bounce time of $385 \ \mathrm{ms}$ (top panel)
and for model pLa0.25 at a time of $700 \ \mathrm{ms}$.  The minimum
and maximum shock radii are indicated by dotted lines in both
panels. The red dashed curve in the top panel is given by the
functional expression $9 \times 10^{8} \times (r/100 \mathrm{km})^{-1}$ and
illustrates that $\sqrt{\langle v_\theta^2\rangle}$ roughly grows like
$r^{-1}$ during the infall of shells with similar initial
$\sqrt{\langle v_\theta^2\rangle}$ as predicted by linear perturbation
theory. Deviations from this behavior are mainly due to the fact
that $\sqrt{\langle v_\theta^2\rangle}$ is not exactly constant for the
initial perturbation pattern, and non-linear effects also play a role.}
\end{figure}

\begin{figure*}
\includegraphics[width=0.48 \linewidth]{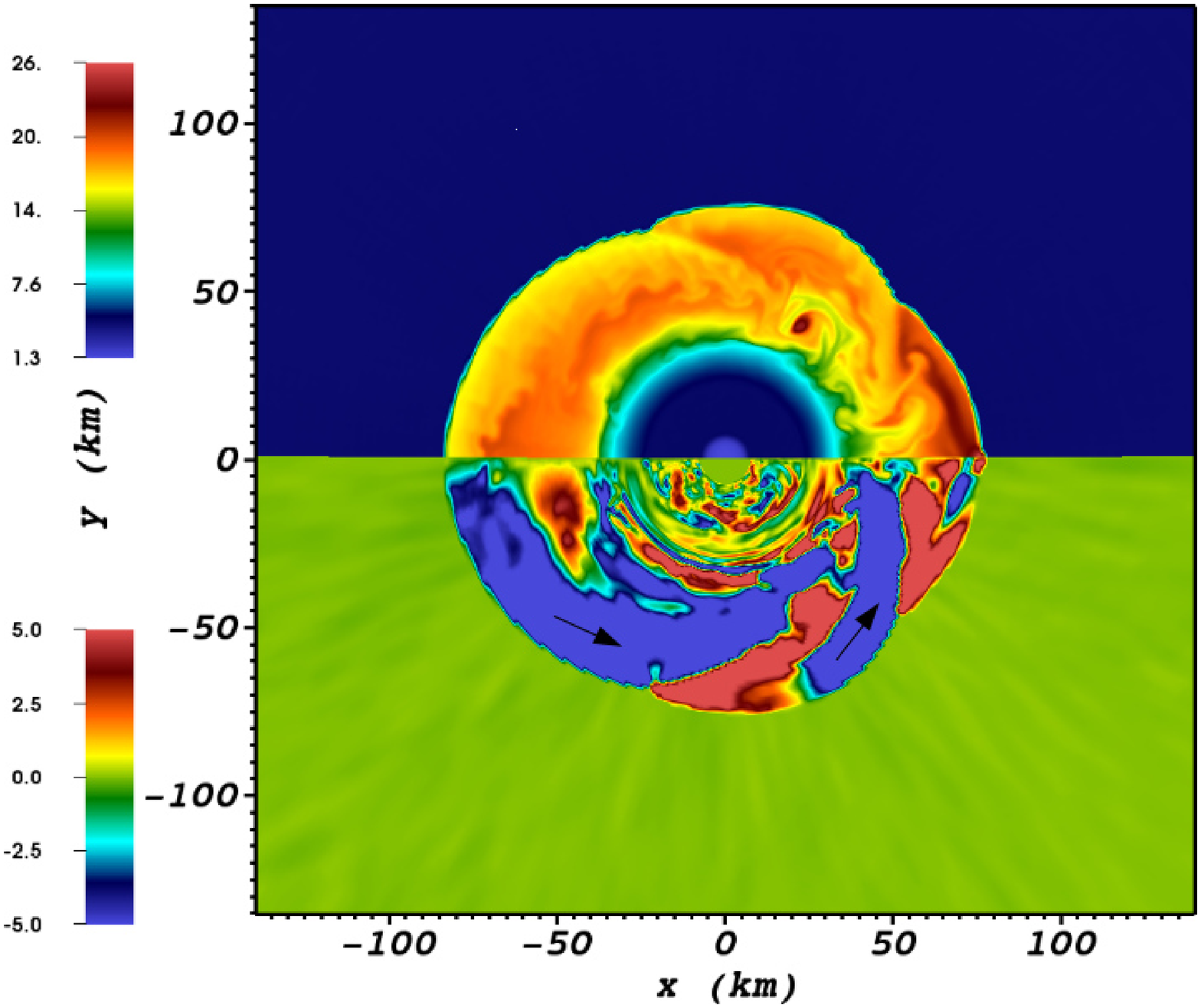}
\hspace{0.03 \linewidth}
\includegraphics[width=0.48 \linewidth]{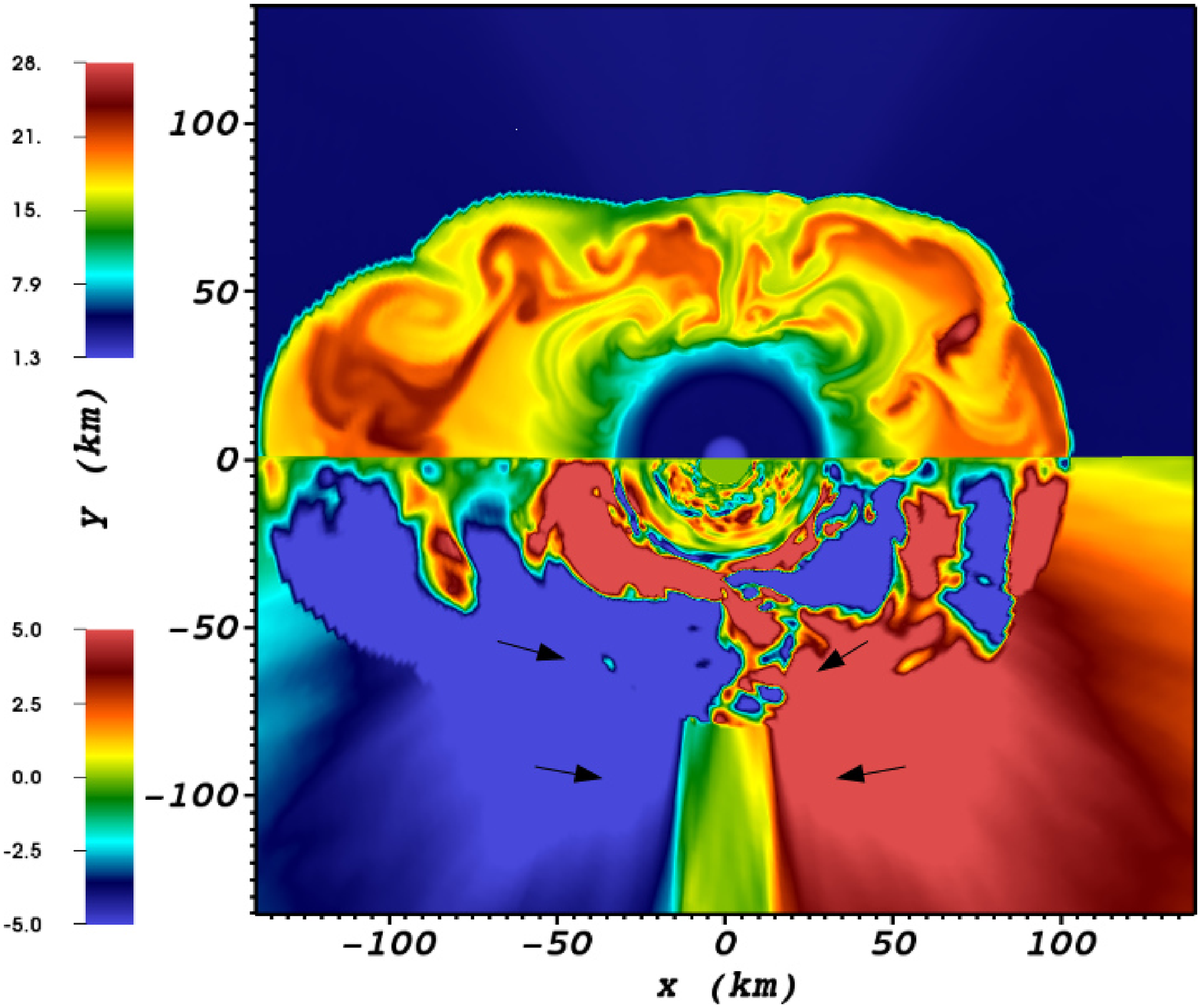}\\
\includegraphics[width=0.48 \linewidth]{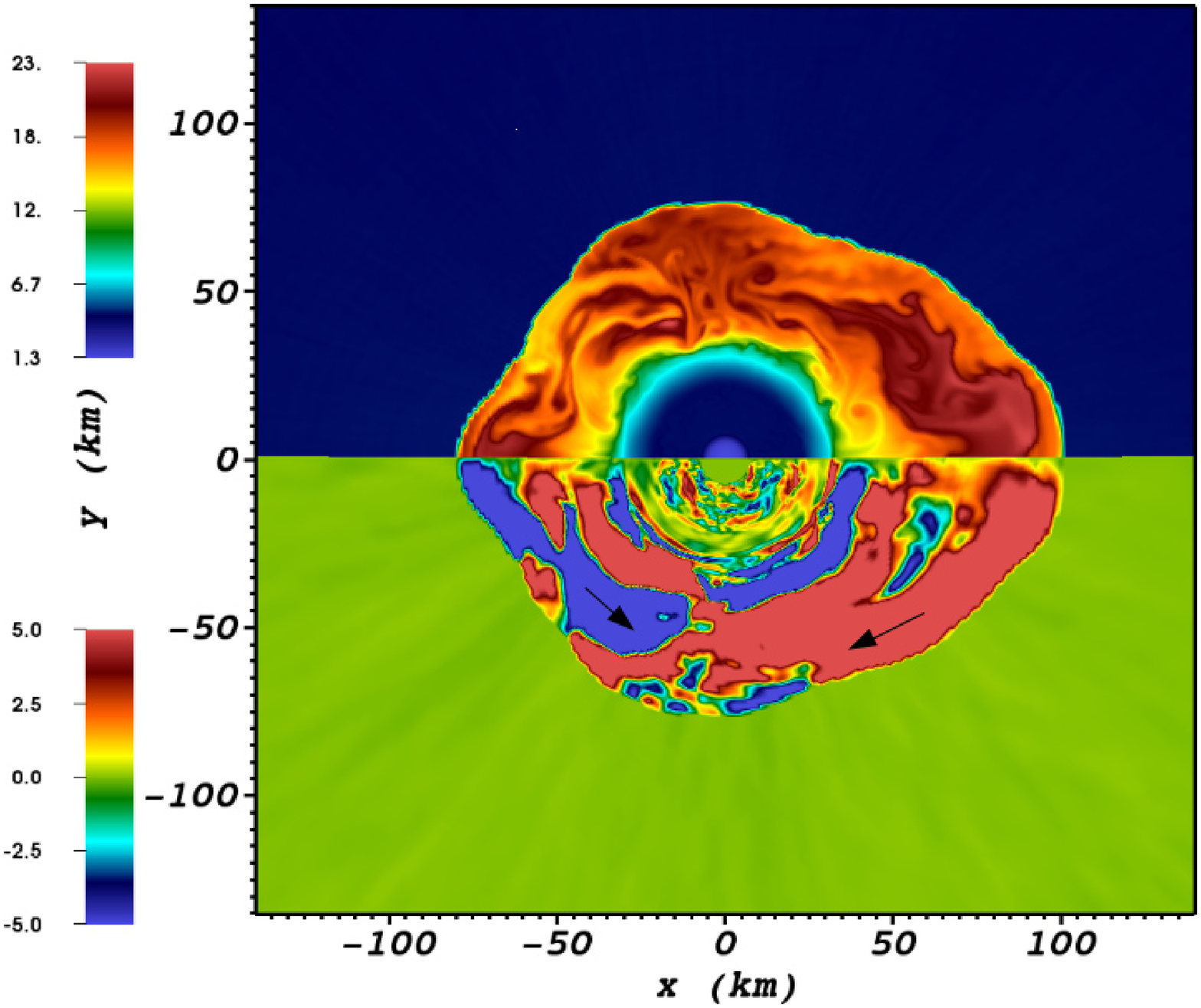}
\hspace{0.03 \linewidth}
\includegraphics[width=0.48 \linewidth]{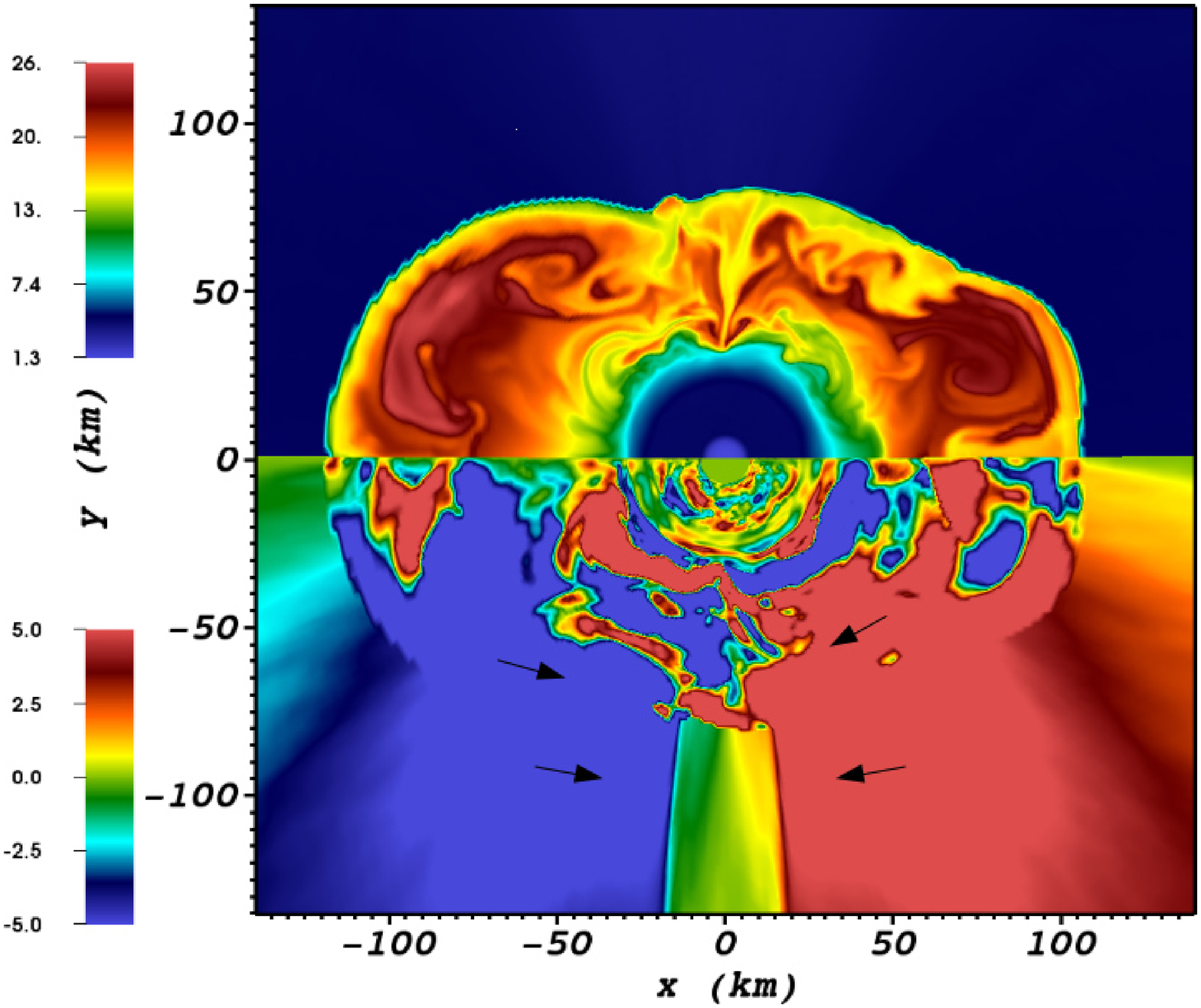}\\
\includegraphics[width=0.48 \linewidth]{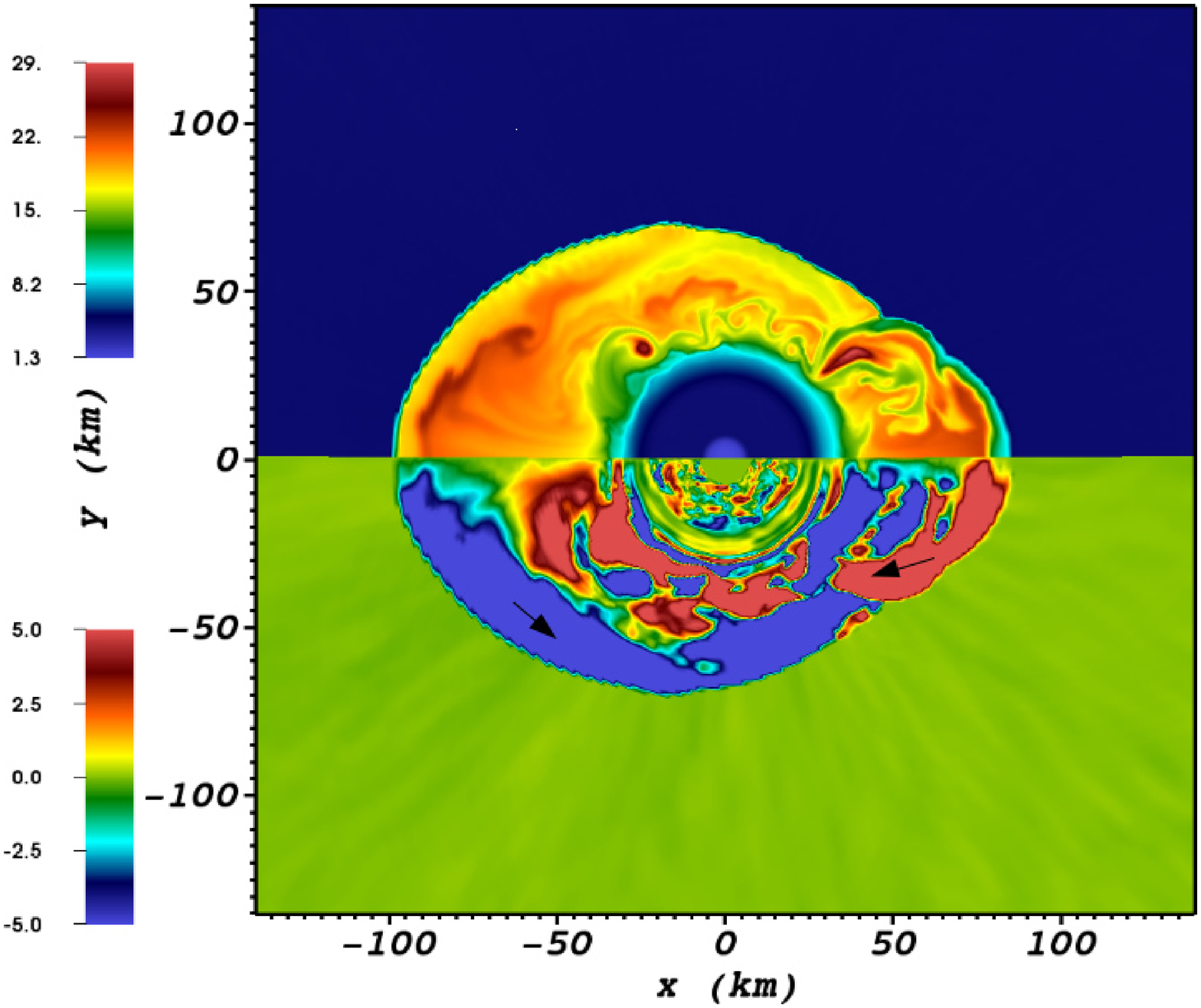}
\hspace{0.03 \linewidth}
\includegraphics[width=0.48 \linewidth]{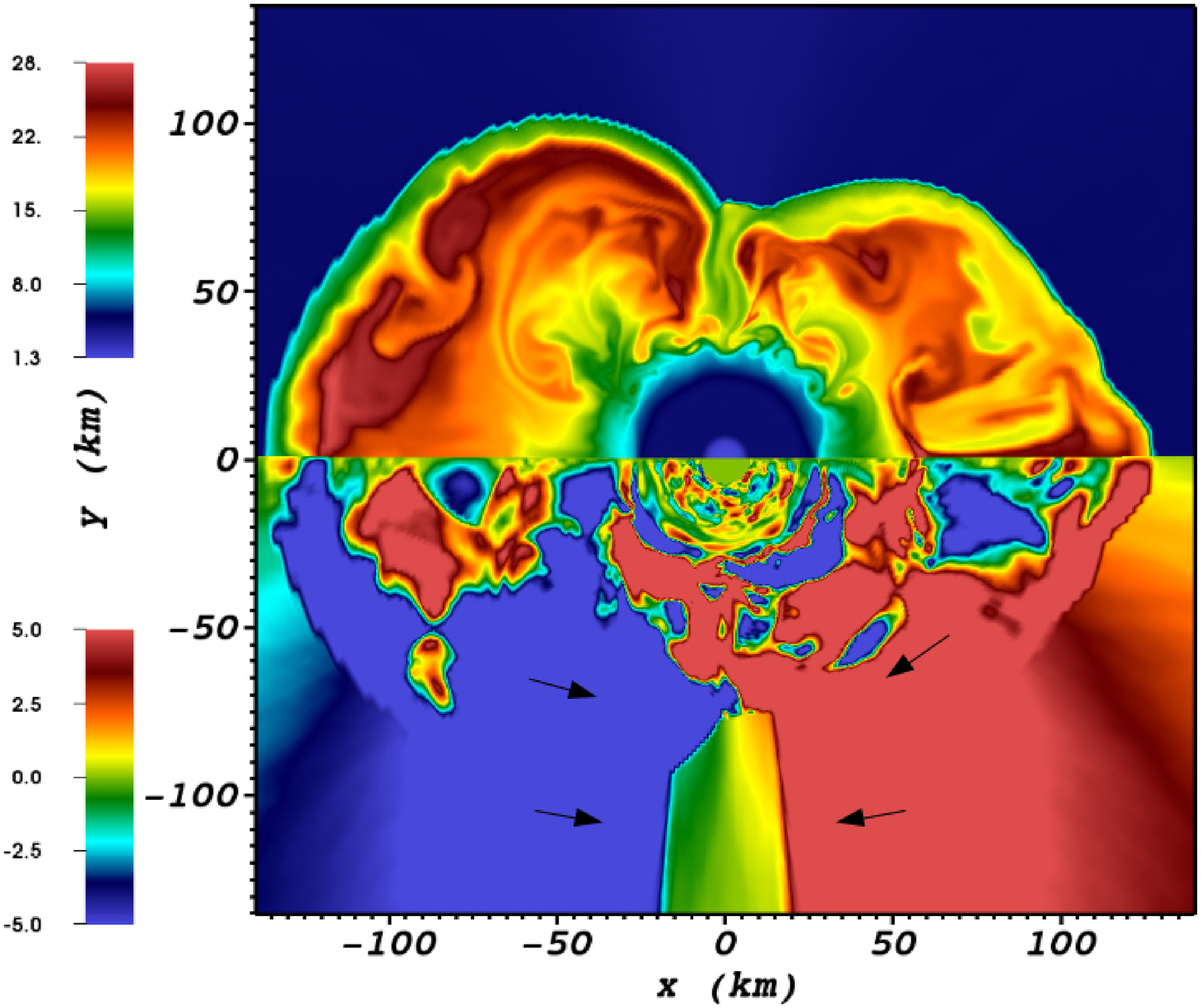}\\
\caption{Snapshots of the entropy $s$ in units of $k_b /
  \mathrm{nucleon}$ (top half of panels) and the lateral velocity
  $v_\theta$ in units of $10^{8}\, \mathrm{cm} \, \mathrm{s}^{-1}$
  (bottom half of panels) for model p0 (left column) and model pL2a1
  (right column) at post-bounce times of $363 \ \mathrm{ms}$, $383
  \ \mathrm{ms}$, and $403 \ \mathrm{ms}$ (top row to bottom
  row). At this stage, mass shells initially
located at a radius of $\mathord{\sim} 3000 \ \mathrm{km}$ reach the shock.
Note that we apply a cut-off for lateral velocities exceeding
  the minimum and maximum value of the color scale.
Arrows indicate the direction of the lateral
flow in the post-shock and pre-shock region (if applicable).
\label{fig:sasi_p0_vs_p5a1}
}
\end{figure*}

\section{The Effect of Seed Asphericities on the Explosion Conditions}
\label{sec:perturbations}

\subsection{Comparison of Heating Conditions for an Exemplary
Perturbation Pattern}

How do seed asphericities in the progenitor affect the
quasi-stationary evolution of the heating conditions and the
hydrodynamic instabilities described in Section~\ref{sec:baseline} in
such a manner as to shift the onset of the explosion by several
hundreds of milliseconds in some cases? A systematic comparison of the
heating conditions for model series pL2aX sheds some light on this
question. The quadrupolar perturbation pattern is best suited
for a such a comparison because the effects of the seed asphericities
are already noticeable for moderate seed amplitudes.

Fig.~\ref{fig:heating_pL2} shows the average shock radius, the
criticality parameter $\tau_\mathrm{adv}/\tau_\mathrm{heat}$, the mass
in the gain region, and the kinetic energy contained in lateral
fluid motions for selected perturbation amplitudes as well as for the
baseline model p0. It is evident that all these diagnostics show a
definite hierarchy, reflecting a systematic improvement of the heating
conditions and an increasing violence of aspherical motions in the
gain region with increasing perturbation amplitude.  It is noteworthy
that models with different perturbation amplitudes tend to diverge
already well before the onset of the explosion.

The increase of the pre-explosion value of the criticality parameter
already indicates that the critical luminosity required for the
initiation of an explosion is reduced dramatically in the more
strongly perturbed models. Since the critical curve is essentially
defined by $\tau_\mathrm{adv}/\tau_\mathrm{heat}=1$
(cf.\ Section~\ref{sec:critical_luminosity}), and
$\tau_\mathrm{adv}/\tau_\mathrm{heat} \propto (L_\nu E_\nu^2)^{5/3}$,
an increase of $\tau_\mathrm{adv}/\tau_\mathrm{heat}$ for a given
neutrino luminosity must be mirrored by a proportional decrease in the
critical luminosity. Models pL2a1 and pL2a2 reach
$\tau_\mathrm{adv}/\tau_\mathrm{heat} =1$ at a time when
$\tau_\mathrm{adv}/\tau_\mathrm{heat} \approx 0.3 \ldots 0.4$ in the
baseline model, i.e.\ the baseline model should have reached about
$40\% \ldots 60 \%$ of the critical luminosity at that junction. Based
on the pre-explosion value of $\tau_\mathrm{adv}/\tau_\mathrm{heat}$
in models pL2a1 and pL2a2, one would therefore estimate a decrease of
the critical luminosity of the order of a few tens of percent. This
rough estimate can be further corroborated by plotting the evolution
of the models in the $(M \dot{M},L_\nu E_\nu^2)$ plane
(Fig.~\ref{fig:critical_curve_p5}): The perturbed models clearly do
not hit the fiducial critical curve constructed for the baseline model
p0; instead they break off their approach to the unmodified critical
curve around the onset of the explosion and then fall \emph{below} the
$(M \dot{M},L_\nu E_\nu^2)$ trajectory of the baseline model. The
reduction of the critical luminosity can thus be estimated, it
appears to be lower by $15\%$ (pL2a0.25, pL2a0.5) to $40\%$ pL2a2
compared to the baseline model. Naturally, the exact value of the reduction
of the critical luminosity is difficult to determine, as it hinges
on the precise scaling relation between the time-scale ratio and
the neutrino luminosities and mean energies.

A closer inspection of Fig.~\ref{fig:critical_curve_p5} also reveals a
subtle higher-order effect: The trajectories of the more strongly
perturbed models (pL2a1 and pL2a2) diverge from the baseline model at a
very early stage and consistently show lower values of $L_\nu E_\nu^2$
prior to the onset of the explosion. This implies that for the
strongly perturbed post-shock flow in these models, the globally
asymmetric post-shock accretion flow onto the proto-neutron star leads
to a net decrease of the neutrino emission. While this slight
reduction of the total neutrino luminosity is a potentially important
phenomenon, we refrain from analyzing it in detail in this paper since
be believe that a more rigorous transport scheme is required to verify
the existence of this effect.

It is obvious that progenitor asphericities \emph{somehow} lead to
more violent turbulent motions in the gain region that help to push
the shock out, increase the mass in the gain region, and thus boost
the efficiency of neutrino heating. However, our model allows
us to describe this mechanism more precisely. 

It appears that the \emph{direct} injection of kinetic energy into the
gain region by the \emph{advection} of lateral velocity perturbations
to and through the shock plays at most a subdominant role. This is
suggested by a comparison of the density-weighted lateral velocity
dispersion $\sqrt{\langle v_\theta^2 \rangle}$,
\begin{equation}
\label{eq:velocity_dispersion}
\sqrt{\langle v_\theta^2 \rangle}
=\sqrt{\frac{\int \rho v_\theta^2 \, \ud \Omega}{\int \rho \, \ud \Omega}},
\end{equation}
in the pre- and post-shock region. If we assumed that the lateral
velocity perturbations were simply advected through a stationary
spherical shock, then $\sqrt{\langle v_\theta^2 \rangle}$ should be
continuous across the shock front and should scale roughly as
$\sqrt{\langle v_\theta^2 \rangle} \sim r^{-1}$ according to linear
perturbation theory \citep{lai_00,buras_06_b,takahashi_14}, which essentially reflects the
conservation of local angular momentum. However,
Fig.~\ref{fig:v_theta_rms} shows that the post-shock values of
$\sqrt{\langle v_\theta^2 \rangle}$ are significantly higher than the
pre-shock values, even for strongly perturbed models like pL2a2. This
implies that the mere quasi-spherical advection of kinetic energy in
lateral motions cannot directly account for the increased activity of
convection and/or the SASI, especially in the exploding models with
lower perturbation amplitudes.

The pre-collapse asphericities must therefore affect the growth and
saturation of non-radial hydrodynamic instabilities in a more subtle
way. A closer look at the flow geometry of a strongly perturbed model
provides further clues: Fig.~\ref{fig:sasi_p0_vs_p5a1} compares
several snapshots of the entropy and lateral velocity for the baseline
model p0 and model pL2a1 shortly before the onset of the explosion in
the latter. Model p0 exhibits the typical flow features of an $\ell=1$
SASI sloshing mode in the non-linear regime: Expanding bubbles form
and collapse alternately in both hemispheres, coherent
vorticity waves (clearly seen in the lateral velocity) propagate
inward to maintain the SASI cycle, and downflows develop from fast
lateral the flows immediately behind the shock, separating from the
shock at the triple points. The flow is distinctly quasi-periodic.

\begin{figure}
\includegraphics[width=\linewidth]{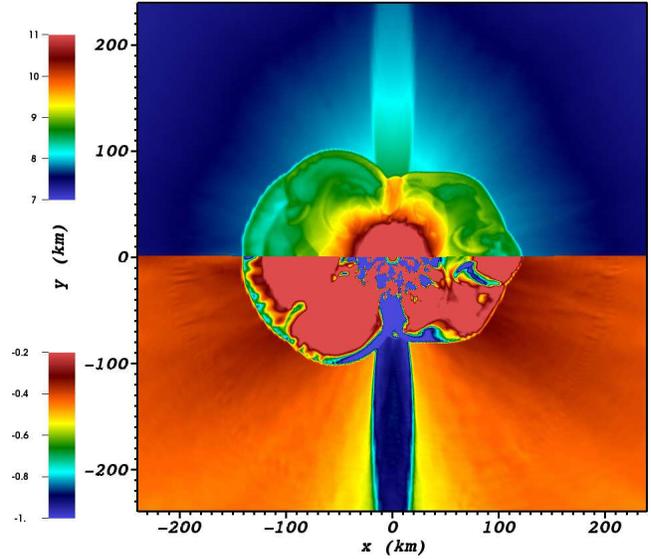}
\caption{Logarithm $\log_\mathrm{10} \rho$ of the density (top) and
  equivalent isotropic mass accretion rate $4\pi \rho v_r r^2$ in units of
  $M_\odot \ \mathrm{s}^{-1}$ (bottom) for model pL2a1 $403 \ \mathrm{ms}$
   after bounce.
\label{fig:accretion_asymmetry_pL2a1}
}
\end{figure}

The perturbed model pL2a1 differs from this pattern in some important
respects: The post-shock velocity field settles into a relatively
stable configuration with two lateral flows colliding near the equator
immediately behind the shock, where they form a single persistent
downflow. Sloshing motions on top of the stable prolate deformation of
the shock are not immediately evident, although a detailed look at the
Legendre coefficients of the shock surface still shows some
quasi-periodic $\ell=1$ oscillations. The structure of the post-shock
flow clearly mirrors the velocity perturbations in the pre-shock
region, where we likewise find lateral flows colliding near the equator
(and actually forming a double shock). However, the lateral velocities
in the pre-shock region are considerably amplified as they fall
through the shock at an oblique angle. The permanent prolate
deformation of the shock is maintained because the pre-shock flow is
strongly anisotropic.  Aside from the high lateral velocities in the
pre-shock region of the order of $5\times 10^8 \ \mathrm{cm}
\ \mathrm{s}^{-1}$, we find strong deviations from spherical symmetry
in the density and the equivalent isotropic mass accretion rate $4 \pi \rho
v_r r^2$ as shown in Fig.~\ref{fig:accretion_asymmetry_pL2a1}.  With
these pronounced asphericities in the pre-shock region, the
Rankine-Hugoniot jump conditions inevitably result in strong and
permanent shock deformation.

\begin{figure*}
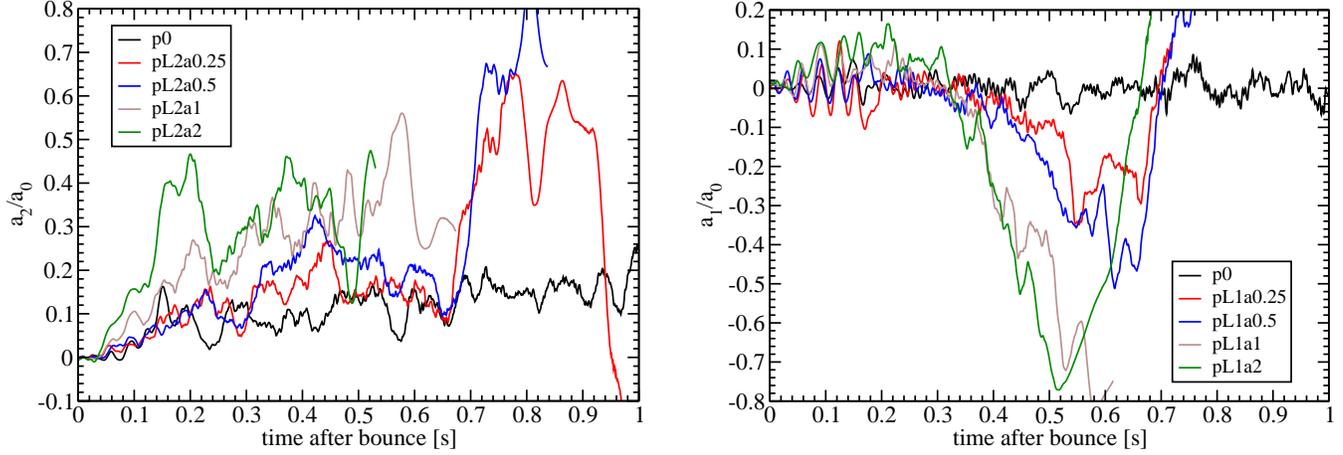

\includegraphics[width=0.48 \linewidth]{f17a.eps}
\hspace{0.03 \linewidth}
\includegraphics[width=0.48 \linewidth]{f17b.eps}
\caption{Normalized Legendre coefficients for the quadrupolar
  deformation of the shock in model p0 and pL2a0.25 to pL2a2 and for
  the dipolar deformation in model p0 and pL1a0.25 to pL1a2.
  Running averages over $20 \ \mathrm{ms}$ are applied to reduce
  high-frequency oscillations and show the secular evolution of
  the shock deformation more clearly.
\label{fig:shock_deformation_perturbed}}
\end{figure*}

\begin{figure*}
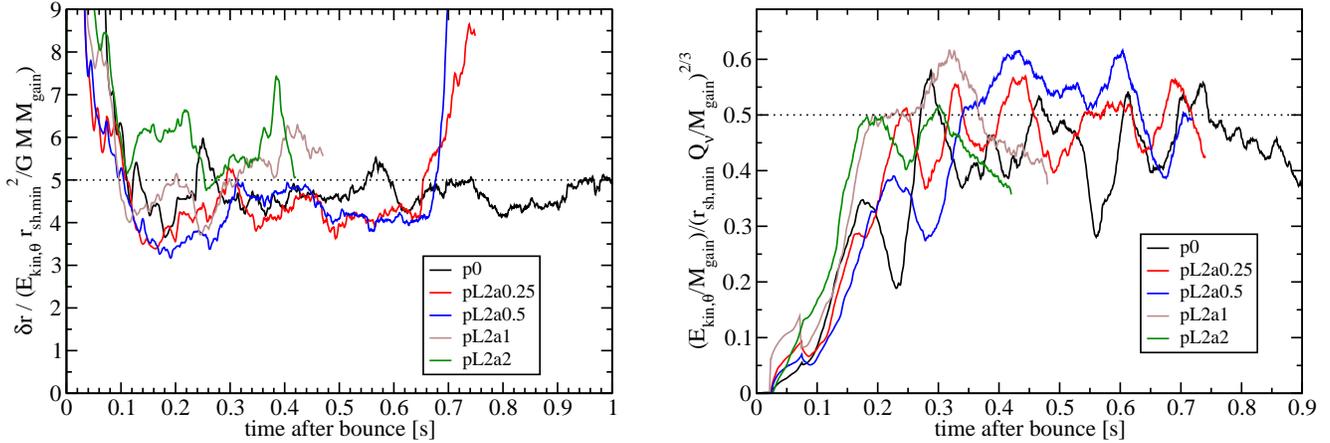

\includegraphics[width=0.48 \linewidth]{f18a.eps}
\hspace{0.03 \linewidth}
\includegraphics[width=0.48 \linewidth]{f18b.eps}
\caption{Saturation properties of non-radial instabilities for the
  perturbed models pL2a0.25 to pL2a2 compared to the baseline model
  p0. The left panel shows the ratio of the RMS shock deformation
  $\delta r$ and the quantity $\left(E_{\mathrm{kin},\theta}
  r_\mathrm{sh,min}^2\right)/(G M M_\mathrm{gain})$, and illustrates
  the approximate validity of equation~(\ref{eq:dr_sasi}) even in the
  presence of strong seed perturbations prior to the onset of the
  explosion.  Similarly, the right panel shows the ratio
  $(E_{\mathrm{kin},\theta}/M_\mathrm{gain})/(r_\mathrm{sh,min}\dot{Q}_\nu/M_\mathrm{gain})^{2/3}$
  to demonstrate that equation~(\ref{eq:kinetic_energy_vs_heating})
  for the relation between neutrino heating and the kinetic energy
  contained in lateral motions likewise remains valid in the
  pre-explosion phase. For the sake of clarity, all curves are
  terminated at the onset of the explosion, and we use running averages
  over $50 \ \mathrm{ms}$ for all quantities.
\label{fig:qval_pL2}
}
\end{figure*}

\begin{figure}
\includegraphics[width=\linewidth]{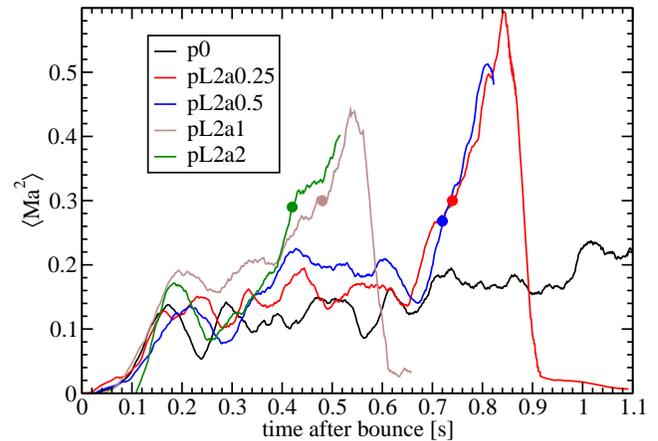}
\caption{Average squared Mach number of lateral motions in the
gain region for the baseline model p0 and models pL2a0.25 to
pL2a2. The onset of the explosion, defined as the time
when the criticality parameter $\tau_\mathrm{adv}/\tau_\mathrm{heat}$
reaches unity, is marked by a filled circle on each curve.
\label{fig:mach_p5}
}
\end{figure}

\begin{figure}
\includegraphics[width=\linewidth]{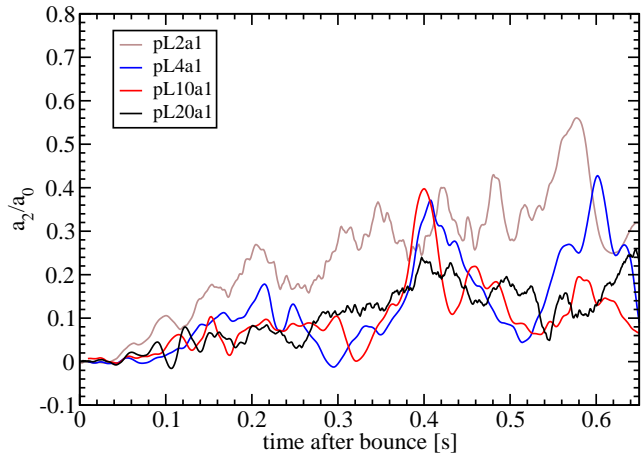}
\caption{Normalized Legendre coefficients for the quadrupolar
  deformation of the shock in models pL2a1, pL4a1, pL10a1, and pL20a2
  during the pre-explosion phase.  Running averages over $20
  \ \mathrm{ms}$ are applied to reduce high-frequency oscillations and
  show the secular evolution of the shock deformation more clearly.
  The perturbation pattern pL2a1 clearly leads to a stronger
  quadrupolar deformation of the shock than patterns with $\ell>2$.
\label{fig:quadrupole_different_l}}
\end{figure}

\begin{figure}
\includegraphics[width=\linewidth]{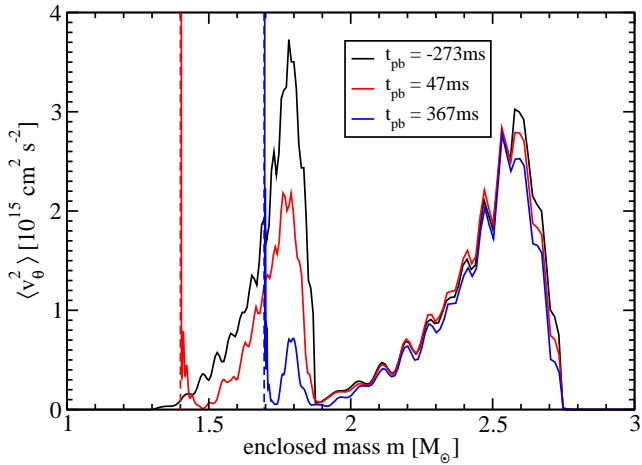}
\caption{Lateral velocity dispersion $\langle v_\theta^2 \rangle$ as a
  function of enclosed mass at different times for model pL20a1 with
  small-scale $\ell=20$ perturbations.  The velocity perturbations in
  the Si/O shell (between $m\approx 1.4 M_\odot$ and $m \approx 1.9
  M_\odot$) are strongly damped, whereas the outer convective zone is
  as yet little affected. Note that the curves show only the pre-shock
  region; the shock position is indicated by a dashed line. Also note
  that $\langle v_\theta^2 \rangle$ has been smoothed roughly over the
  radial extent of the individual convective cells.
\label{fig:damping_of_perturbations}
}
\end{figure}

The strong density variations arise despite the solenoidal nature
of the initial perturbations for two reasons: First, shocks may form
in the course of the non-linear evolution of the perturbations as
convective cells collide. More importantly, the spherical isodensity
surfaces in the progenitor are distorted during collapse as the infall
is accelerated or delayed depending on the initial radial
velocity. In order for matter in a convective updraft/downdraft to
reach the shock at the same time as matter at rest, it must originate
from a different radius with an initial displacement $\delta  r$
depending on the radial velocity perturbation $\delta v_r$ and the
infall time $t$:
\begin{equation}
\delta r \approx \delta v_r t.
\end{equation}
If the compression factor (i.e.\ the ratio between the initial density
and the pre-shock density) for fluid elements reaching the shock at
the same time is identical, this implies Eulerian density
perturbations\footnote{It is important to stress that Eulerian density
  perturbations at constant $r$ rather than Lagrangian density
  perturbations are relevant when we consider the deformation of the
  shock.} at the shock that depend on the initial density gradient in
the progenitor:
\begin{equation}
\label{eq:delta_rho}
\frac{\delta \rho}{\rho}
\sim \frac{\delta r}{r}\frac{\pd \ln \rho}{\pd \ln r}
\sim \frac{\delta v_r t}{r}\frac{\pd \ln \rho}{\pd \ln r}
\sim \frac{\delta v_r}{c_s}\frac{\pd \ln \rho}{\pd \ln r}.
\end{equation}
The differential infall thus effectively translates radial velocity
perturbations into density perturbation of order
$\mathcal{O}(\delta v_r/c_s)$ (since the infall time is of the order of the
  sound-crossing time) instead of $\mathcal{O}(\delta v_r/c_s)^2$ during
steady-state convection. Conceptually, this
amplification mechanism is slightly different from the generation
of density perturbations from radial velocity perturbations
as investigated by \citet{lai_00} and \citet{takahashi_14} for the
case of supersonic infall. In both cases, density perturbations
are essentially generated by the deformation of isodensity contours
during the infall, but the available time-scale is
$t \approx r/c_s$  in our case as opposed to $t \approx r/v_r$ 
($v_r$ being the initial radial velocity
in the unperturbed accretion flow) in the
setup of \citet{lai_00} and \citet{takahashi_14}.

We surmise that these two factors -- the ``forced oblate asphericity''
of the shock due to the anisotropic mass flux through the shock and
the generation of high lateral velocities by an oblique shock -- are
primarily responsible for increasing the kinetic energy contained in
non-spherical instabilities in the presence of strong perturbations.
All the exploding models in the pL2aX series are characterized by a
considerably stronger quadrupolar deformation than in the baseline
model prior to the onset of the explosion as can be seen from
Fig.~\ref{fig:shock_deformation_perturbed}, which shows the
normalized Legendre coefficient $a_2/a_0$ for several of these
models. This finding is not confined to the perturbation pattern pL2:
\emph{The onset of the explosion in perturbed models is always
  associated with a stronger $\ell=2$ (and/or $\ell=1$) deformation of the
  shock than in the baseline model.}

\begin{figure}
\includegraphics[width=\linewidth]{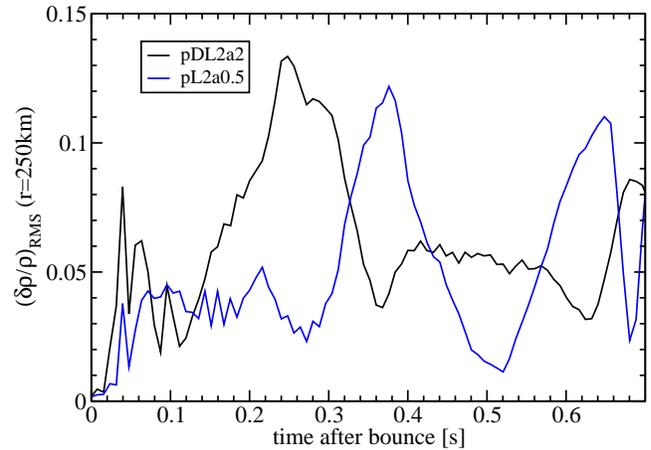}
\caption{RMS density fluctuations (equation~\ref{eq:def_delta_rho}
) ahead of the shock at $r=250
  \ \mathrm{km}$ for model pDL2a2 (black) and model pL2a0.5 (blue)
  with quadrupolar density and velocity perturbations, respectively.
  Overall, the density fluctuations in these two models are of similar magnitude,
  although there are phase differences). Consequently, these two models
  explode at a similar time shortly after $700 \ \mathrm{ms}$ after bounce.
\label{fig:delta_rho_quadrpuole}
}
\end{figure}

\subsection{Saturation of Instabilities in the Presence
of Strong Seed Asphericities} 
\label{sec:saturation_with_perturbations}
It is noteworthy that the saturation of the non-radial instabilities
in the perturbed models is still regulated by the amount of heating as
in the baseline model.  Fig.~\ref{fig:qval_pL2} shows that
equations~(\ref{eq:kinetic_energy_vs_heating}) and (\ref{eq:dr_sasi})
describe the relation between the volume-integrated neutrino heating
rate $\dot{Q}_\nu$, the lateral kinetic energy in the gain region
$E_{\mathrm{kin},\theta}$, and the average shock deformation $\delta
r$ reasonably well even in the presence of strong pre-shock
asphericities. This is not inconsistent with our assumption that the
forced asphericity of the shock is responsible for the increased
violence of SASI and/or convection, but rather suggests that the
saturation level is determined by a feedback process: The forced
deformation of the shock increases the kinetic energy in non-radial
motions in the gain region, this in turn leads to a larger shock
radius and a higher mass in the region, which in turn boosts the
activity of hydrodynamics instabilities (as reflected by the factor
$(r_\mathrm{sh,min}-r_\mathrm{gain})$ in
equation~(\ref{eq:kinetic_energy_vs_heating})).  
In the pre-explosion
phase, the positive feedback ceases
at some point when the further excursions of the shock
no longer help to boost the violence of convective and SASI
motions effectively enough to permanently sustain a large shock radius.
The saturation level will depend both on the overall parameters of
the accretion flow (accretion rate, proto-neutron star radius,
neutrino luminosity) and on the additional forcing due to seed
perturbations in the progenitor. The amplification of the shock
deformation by this feedback effect may become less important for very
strong forcing, however. Model pL2a2 may be a possible example for the
transition to this regime of very strong forcing; here the shock
deformation $\delta r$ is somewhat higher than suggested by
equation~(\ref{eq:dr_sasi}).

The fact that neutrino heating and the saturation level of aspherical
instabilities are still related by
equations~(\ref{eq:kinetic_energy_vs_heating}) and (\ref{eq:dr_sasi})
implies that it is impossible to distinguish whether the improvement
of the time-scale ratio $\tau_\mathrm{adv}/\tau_\mathrm{heat}$ due to
the expansion of the shock or the easier formation of large buoyant
bubbles is more crucial for facilitating the onset of the explosion in
the presence of strong progenitor asphericities.
Equations~(\ref{eq:kinetic_energy_vs_heating}) and (\ref{eq:dr_sasi})
show that the heating conditions, the shock deformation (and hence the
typical size of high-entropy bubbles) are inextricably linked to each
other and to the typical density perturbation $\delta \rho/\rho$ in
the post-shock flow through the turbulent Mach number
  through equation~(\ref{eq:dr_sasi_mach}) (since $ \delta \rho/\rho
\sim \langle \mathrm{Ma}^2\rangle$). In particular, there is a
remarkable correlation between the criticality parameter
$\tau_\mathrm{adv}/\tau_\mathrm{heat}$ and $\langle
\mathrm{Ma}^2\rangle$ at the onset of the explosion: At the time of
the explosion ($\tau_\mathrm{adv}/\tau_\mathrm{heat}=1$), all models
also appear to reach a ``critical Mach number'' $\langle
\mathrm{Ma}^2\rangle \approx 0.3$ in the gain
  layer (Fig.~\ref{fig:mach_p5} and Appendix~\ref{sec:toy_model}).

Incidentally, Fig.~\ref{fig:qval_pL2} also shows that the
  initial perturbations do not have a major effect on the approach to
  non-linear saturation. The seed for the SASI (which is clearly the
  instability that grows during the first $\mathord{\gtrsim} 100
  \ \mathrm{ms}$) is already provided by the asphericities left from
  prompt convection. Progenitor asphericities from the silicon and
  oxygen shell can only start to make themselves felt a few tens of
  milliseconds after bounce when they reach the shock, and at this
  stage, the SASI has already reached sizeable amplitudes.  Only
  afterwards do we see a weak trend (superimposed over considerably
  stochastic variations) towards slightly faster saturation in models
  with strong initial perturbations. Since the initial perturbations
  affect the heating conditions and the saturation of post-shock
  instabilities primarily through the conversion of initial
  \emph{radial} velocity perturbations into density perturbations,
  their effect becomes strongest only when the middle of the initial
  ``convection zone'' (where $|v_r|$ is highest) reaches the shock.  

\subsection{Sensitivity to Perturbation Patterns}

\subsubsection{Spatial Scale of the Seed Asphericities}
The mechanism outlined for the models of the pL2aX series works far
less efficiently for perturbations dominated by higher angular
wavenumbers $\ell >2$, and also somewhat less efficiently for $\ell=1$
than for $\ell=2$ (see Table~\ref{tab:perturbations}).

The behavior at $\ell>2$ may be related to the stability properties of
the standing accretion shock. Naturally, pre-shock asphericities will
more efficiently excite modes that are already unstable; they may
excite other modes as well, but if the damping time-scale (due
to linear damping or non-linear damping by parasitic instabilities) for such
stable modes is short, the resulting amplitude will be negligibly
small. This reasoning already suggests that only low-$\ell$ modes can
be efficiently excited in the SASI-dominated regime. In the
convectively dominated regime, modes with higher angular wavenumber
may be excited as well, but the most unstable wavenumber also shifts
towards lower $\ell$ as the ratio of the shock radius and gain radius
$r_\mathrm{sh}/r_\mathrm{gain}$ increases. We therefore expect that
the pre-shock asphericities need to be able to excite $\ell=1$ or
$\ell=2$ perturbations in that case as well in order to increase the
violence of convective motions up to the point of shock
revival. Perturbations with high $\ell$ may initially excite
small-scale convection in the gain region effectively, but the small
``overlap'' with large-scale modes will render further excitation
inefficient once increased convective overturn pushes the shock out.

Since all our velocity perturbation patterns with $\ell>2$ eventually
develop some $\ell=2$ component during the
infall,\footnote{This is due to non-linear effects, and
  also due to the fact that the perturbations are not constructed as a
  linear combination of eigenfunctions of the linearized
  perturbation equations with a single wavenumber $\ell$. Constructing
  the perturbations from a generalized stream function with a specific
  $\ell$ is not sufficient to guarantee this and can hence lead to the development
of an $\ell=2$ component even during the linear phase.} we still find explosions
in many of these cases. As for the pL2aX series, the explosion is
preceded by a strong quadrupolar deformation of the shock, but for
comparable kinetic energies contained in the velocity field of the
initial perturbations, the normalized quadrupole amplitude is
considerably lower (Fig.~\ref{fig:quadrupole_different_l}).  The
degree of ``overlap'' with the $\ell=2$ mode of the shock is probably
crucial for triggering the explosion, and since this overlap depends
on the detailed evolution of the asphericities during the infall, one
cannot expect a simple monotonic dependence on $\ell$. High-$\ell$
perturbations will generally develop less overlap with the $\ell=2$
mode, but on the other hand, an $\ell=2$ component may emerge faster
in these cases because the typical evolution time-scale of the
convective cells is smaller (see below). Moreover, the forcing changes
rapidly in time because of the small radial extent of the individual
convective eddies in the models with higher $\ell$.

The non-linear damping of the pre-collapse perturbations 
in the subsonic infall region is another
factor contributing to the less efficient excitation of instabilities
in the gain region in the case of high $\ell$. Non-linear
effects will start to come into play on the crossing time-scale
for convective cells, which is approximately
\begin{equation}
t_\mathrm{cell}\approx \frac{r \pi}{\ell \ |v_{\theta,\mathrm{max}} |}.
\end{equation}
For a model like pL20a1, we have $t_\mathrm{cell} \approx 0.5
\ \mathrm{s}$, and consequently, the pre-shock perturbations are
damped considerably over the course of the simulation. Since the
collapse time until core bounce is already $273 \ \mathrm{ms}$ for the $15 M_\odot$
progenitor, the initial perturbations in the Si/O shell are damped
considerably.  This is illustrated by
Fig.~\ref{fig:damping_of_perturbations}, which shows the evolution
of the lateral velocity dispersion $\langle v_\theta^2 \rangle$ in the
pre-shock region. 

Overall, our findings are compatible with 
the weak trend towards a more efficient excitation of
aspherical motions in the post-shock layer seen by \citet{couch_14}
 for initial perturbations with smaller $\ell$.
However, they suggest that i) the limited investigation of
perturbation of different scales by \citet{couch_14} may have
missed the most interesting case of $\ell=1$ and $\ell=2$, and that ii)
the dependence on $\ell$ for $\ell>2$ may not be monotonic and
arise from a combination of various factors.

While the small effect of perturbations with high $\ell$ is thus
relatively easy to understand, the case of perturbations with $\ell=1$
presents something of a conundrum. While the $\ell=1$ mode of the SASI
is clearly unstable, we do not obtain explosions for dipolar
perturbations below $\mathrm{Ma}_{\theta,\mathrm{expl}}=0.19$, whereas
  models with quadrupolar perturbations explode even for
  $\mathrm{Ma}_{\theta,\mathrm{expl}}=0.05$. Several credible explanations
for this behavior can be adduced:
\begin{enumerate}
\item Due to the different aspect ratio of the convective cells and
  the solenoidal constraint, the radial velocity perturbations are
  smaller by a factor of $\mathord{\sim} 2$ in the pL1aX series than
  in the pL2aX series for the same lateral velocity dispersion
  $\langle v_\theta^2 \rangle$ (see
    Table~\ref{tab:solenoidal_properties}). Consequently, smaller
  density perturbations are generated during the infall,
  cf.\ equation~(\ref{eq:delta_rho}).
\item In the pL2aX series, we encounter converging flows near the
  equator, which can steepen into a double shock
  (Figs.~\ref{fig:sasi_p0_vs_p5a1} and
  \ref{fig:accretion_asymmetry_pL2a1}). The formation of such a double
  shock results in a significantly increased mass flux onto the shock
  in a narrow wedge around the equator. Presumably, this helps to
  stabilize the quadrupolar shock deformation and the post-shock flow
  geometry with the very persistent, almost radial downflow near the
  equator.
\item Models with strong dipolar shock deformation develop a very
  pronounced neutrino emission asymmetry. The neutrino luminosities
  are considerably lower below the high-entropy bubble, and
  are enhanced in the hemisphere where the shock radius is smaller.
  Presumably, the reduction of the neutrino heating in the
  high-entropy bubble delays the explosive runaway compared to
  the case of a quadrupolar perturbation pattern. For quadrupolar
  perturbations, the matter funneled into the cooling region through
  the equatorial downflow can still be redistributed relatively efficiently
  to higher latitudes to radiate neutrinos into the polar high-entropy
  bubbles.
\end{enumerate}
While the hemispheric emission anisotropy may be a \emph{generic}
factor hampering shock revival in models with dipolar perturbations,
the more efficient generation of density perturbations in models with
quadrupolar perturbations depends very much on the precise convective
flow geometry in the progenitor. We therefore believe that it would be
premature to state that dipolar asphericities are generically less
effective as a means of facilitating shock revival. Only better models
for the multi-D flow structure of convection in the progenitor will
allow definite conclusions.

\subsubsection{Location and Extent of Convective Regions}
Our results clearly demonstrate that the location of the convective
regions is a crucial factor in determining whether pre-collapse
asphericities can aid shock revival or not: Models for which the
initial perturbations are restricted to the unstable regions according
to mixing-length theory (pPAaX, pPSaX, pPDaX) explode at late times at
best. 

While only moderate convective Mach numbers are required to bring
about an explosion in some of these models
(e.g.\ $\mathrm{Ma}_{\theta,\mathrm{expl}}=0.07$ in model pPSa2,
see Table~\ref{tab:perturbations}), we
find two problems that thwart early shock revival: As the inner
convective zones are extremely narrow, we expect that convection is
dominated by high-$\ell$ modes ($\ell \sim 10$) in these zones, which
makes it difficult to excite large-scale shock deformations. On the
other hand, the convective shell driven by neon burning is very wide
and is probably dominated by low-$\ell$ modes for that reason, but it
only reaches the shock at very late times. Between $\mathord{\sim} 200
\ \mathrm{ms}$ and $\mathord{\sim} 700 \ \mathrm{ms}$ after bounce, the
infalling mass shells are therefore essentially spherical (except for
acoustic waves and gravity waves generated at the convective
boundaries). Regardless of their amplitude, pre-collapse asphericities
therefore cannot trigger explosions much earlier than $\mathord{\sim} 700
\ \mathrm{ms}$ after bounce if convective activity is restricted
to the regions where mixing-length theory predicts instability.

The location and width of the convective regions may, however, be
highly sensitive to the zero-age main sequence mass and other stellar
parameters as well as to the treatment of convection, semiconvection,
convective overshoot, and angular momentum transport in stellar
evolution calculations. As for the flow geometry and the typical
scale of the convective eddies, we have to defer final answers until
multi-D models of supernova progenitors at the onset of collapse
become available.

\subsubsection{Non-solenoidal vs.\ Solenoidal Perturbations}
The purely transverse velocity perturbation pattern (pCOaX) inspired by
the setup of \citet{couch_13} proves about as efficient at triggering
shock revival as the solenoidal perturbation pattern with $\ell=2$,
which implies that the effective reduction of the critical luminosity
is also of the order of several tens of percent. Superficially, this
strong effect appears to be somewhat at odds with the results of
\citet{couch_13}, which suggest that the critical luminosity is
smaller by only $\mathord{\sim} 2\%$ in their perturbed models.

However, the huge impact of transverse velocity perturbations in our
simulations can easily be accounted for. In all pCOaX models, large
density anisotropies develop during the infall simply because the
initial perturbations are non-solenoidal as shown by a simple
rule-of-thumb estimate: The time derivative of the
density is given by the divergence of the momentum density field,
\begin{equation}
\frac{\pd \rho}{\pd t} = - \nabla \cdot (\rho \mathbf{v}+ \rho \mathbf{\delta v}),
\end{equation}
from which we can split off a component $(\pd \rho/\pd t)_\mathrm{infall}$ due
to compression, by the spherical background flow,
\begin{equation}
\frac{\pd \rho}{\pd t} =\left(\frac{\pd \rho}{\pd t}\right)_\mathrm{infall}- \rho \nabla \cdot \mathbf{\delta v}-
\mathbf{\delta v} \cdot \nabla \rho.
\end{equation}
Since $\mathbf{\delta v} \cdot \nabla \rho=0$ (i.e.\ the velocity perturbations
are orthogonal to the density gradient), density perturbations will develop during the infall
following
\begin{equation}
\frac{1}{\rho}\frac{\ud \delta \rho}{\ud t} = - \nabla \cdot \mathbf{\delta v}.
\end{equation}
With the typical size
$L$ of the convective cells, the density contrast grows like
\begin{equation}
\frac{\delta \rho}{\rho}
\sim
\frac{t \delta v}{L},
\end{equation}
or
\begin{equation}
\frac{\delta \rho}{\rho}
\sim 
\frac{t \ell \delta v}{\pi r}.
\end{equation}
if $\ell$ is the typical angular wavenumber. Since the sound crossing
time-scale is of the order of the infall time
for the small $\ell$ considered here, non-linear damping and
pressure equilibration can be neglected to zeroth order.

Although density perturbations at the shock arise for completely
different reasons, the expected level of density fluctuations is
similar to the one given by equation~(\ref{eq:delta_rho})
in the case of solenoidal perturbations for similar perturbation
amplitudes. Moreover, the density perturbations arising during the
infall have a strong $\ell=2$ component that can couple to the
$\ell=2$ mode of the shock.\footnote{This can be seen by taking the divergence of
the velocity perturbation ($\mathbf{\delta{v}} = f(r)\sin
4\theta \, \mathbf{e}_\theta$), which is given by
\begin{displaymath}
\nabla \cdot \mathbf{\delta{v}}=
f(r) \nabla \cdot (\sin 4\theta\, \mathbf{e}_\theta)=
\frac{f(r)}{r} \left(-\frac{8}{7} P_2(\cos\theta) +\frac{64}{7} P_4(\cos\theta)\right),
\end{displaymath}
where $P_2(x)=3x^2/2-1/2$ and $P_4(x)=35x^4/8-15 x^2/4+3/8$.}

Although these factors imply that purely lateral velocity
perturbations are very efficient at facilitating shock revival,
\citet{couch_13} nonetheless obtained only a minute effect in their 3D
simulations for several reasons. Whether or not the forced deformation
of the shock due to pre-shock asphericities is genuinely weaker in the
3D case is a moot point, but the perturbation pattern of
\citet{couch_13} differs from our pCOaX models in that there is an
additional modulation in the $\varphi$-direction, and the overlap with
the $\ell=2$ mode of the shock is presumably much smaller for this
reason.  Furthermore, \citet{couch_13} treat neutrino heating and
cooling by means of a simple leakage scheme, which results in a strong
and unfavorable temporal variation of the heating conditions during
the early accretion phase: In their unperturbed baseline model, the
time-scale ratio $\tau_\mathrm{adv}/\tau_\mathrm{heat}$ transiently
rises to $0.6$ around $130 \ \mathrm{ms}$ after bounce and then
plummets rather abruptly to only $0.1$ within a hundred milliseconds.
This leaves only a short time window for pre-shock perturbations to
achieve sufficient shock expansion to push the model above the critical
threshold. Obviously, it is difficult to quantify the impact of
progenitor asphericities for such highly non-stationary heating conditions,
and the reduction of the critical heating parameter $f$ of
\citet{couch_13} by $2\%$ may not fully reflect the potential of
pre-shock perturbations to aid shock revival.

\subsubsection{Density Perturbations vs. \ Velocity Perturbations}
The foregoing discussion has already made it clear that the efficient
conversion of velocity perturbations into density perturbations and
into an anisotropic mass flux onto the shock is the key to efficient
shock revival in the perturbed models. Since pure density
perturbations in the initial model only grow moderately during
collapse ($\delta\rho/\rho \propto r^{-1/2}$ in the linear regime
according to \citealt{lai_00,takahashi_14}), it is evident that
relatively strong initial density perturbations are required to
produce an appreciable effect.  According to
equation~(\ref{eq:delta_rho}), velocity perturbations translate into
density perturbations at the shock of the order of the initial
convective Mach number, i.e.\ $\delta \rho/\rho \propto \mathrm{Ma}_\mathrm{prog}$,
whereas the initial density perturbations should only be of the order
of $(\delta \rho/\rho)_\mathrm{ini} \sim \mathrm{Ma}_\mathrm{prog}^2$ (at least in
the interior of the convective zones), and even moderate amplification
during the infall will not create as large anisotropies in the mass
flux onto the shock as in the case of initial velocity perturbations of
the same convective Mach number.

However, if the initial density perturbations are large enough to
produce pre-shock density fluctuations of the same magnitude as in
models with velocity perturbations, the effect on shock revival is
similar. This can be illustrated by comparing the density
fluctuations ahead of the shock for the two models pL2a0.5 and pDL2a2,
which explode at a similar time. Fig.~\ref{fig:delta_rho_quadrpuole}
shows the RMS fluctuations of the density around the
spherical average $\bar{\rho}$,
\begin{equation}
\label{eq:def_delta_rho}
\left(\frac{\delta \rho}{\rho} \right)_\mathrm{rms}
=\frac{1}{\bar{\rho}}\sqrt{\frac{1}{4\pi}\int (\rho-\bar{\rho})^2\, \ud \Omega},
\end{equation}
at a radius of $250 \ \mathrm{km}$. Both models show a similar level
of pre-shock density fluctuations, especially at the time of shock
revival.  This can be interpreted as further evidence that the
anisotropic pre-shock density field is the primary factor responsible
for enhancing the heating conditions in strongly perturbed models.

Since initial density perturbations of the order of $ \delta \rho/\rho
\gg \mathrm{Ma}_\mathrm{prog}^2$ appear to be required in order to achieve any
appreciable effect, it seems likely that convective density
fluctuations in the progenitor play a subdominant role for supernova
dynamics compared to convective velocity perturbations. In order to obtain
sufficiently large density perturbations, physical mechanisms
other than convection probably need to be invoked, such as
rotation or unstable g-modes \citep{murphy_04}.

A detailed analysis of the infall dynamics reveals a
  further complication in the setup of initial perturbations.  In the
  models with density perturbations, we observe relatively strong
  acoustic waves propagating inside of and beyond the perturbed
  region. These acoustic waves can lead to a considerable modification
  of the initial angle-dependence of the density perturbations during
  the infall. Sometimes the sign of the density perturbations is even
  reversed, e.g.\ the initial model may show a density enhancement at
  the poles, while the perturbations arriving at the shock show higher
  densities in the equatorial region.  This is simply a consequence of
  the fact that we perturb the density only and keep the temperature
  fixed, which results in pressure fluctuations $\delta P/P \sim 0.5
  \delta \rho/\rho$ for the moderate entropies in the perturbed
  regions. However, this does not strongly change the dependence of
  the typical amplitude of the pre-shock density perturbations on the
  initial perturbations $\delta \rho/\rho \sim \mathrm{Ma}_\mathrm{prog}^2$. The
  acoustic component of the perturbations will only reach the shock
  faster and with a different phase than the advected density
  perturbations. 

The actual level of acoustic wave activity present in convective zones
in the progenitor cannot be easily predicted. Within mixing-length
theory, one assumes complete pressure equilibration of convective bubbles
with their surroundings, i.e.\ $\delta P=0$. By contrast, turbulence theory  would suggest that
pressure and density fluctuations are similar, i.e.\
$\delta P/P \propto \delta \rho/\rho \propto \mathrm{Ma}_\mathrm{prog}^2$ (cp.\ equation~(31.4) 
and \textsection~10 in \citealt{landau_fluid} in \citealt{landau_fluid}). However, this is
a minor concern here, since we expect the role of density perturbations
in the progenitor for shock revival to be subdominant anyway.

Interestingly, it appears to be important whether densities are
enhanced in the equatorial region or in the polar region. In model
pDL2a2m, the sign of $\delta \rho/\rho$ is reversed in the initial
perturbation pattern, resulting in a higher mass inflow rate at the
poles compared to the equator (which is reversed compared
  to the initial angular dependence of the perturbations, see our
  earlier remarks on the role of acoustic waves in models with density
  perturbations). Different from model pDL2a2, this model fails to
explode, indicating that the excitation of an oblate deformation of
the shock is less conducive to runaway bubble expansion. Incidentally,
the fact that an oblate shock deformation does not boost the heating
conditions as effectively as a prolate shock deformation also explains
the counterintuitive cases where a \emph{larger} perturbation
amplitude prevents shock revival (as for model pPSa4 compared to
pPSa2, see~Table~\ref{tab:perturbations}): In model pPSa2, a
quadrupolar density perturbation with lower densities near the axis of
the spherical polar grid is present around the onset of the explosion
due to the interaction of waves generated at the inner boundary of the
outermost convection zone, whereas these secondary waves do not give
rise to density perturbations that would lead to an oblate shock. In
the outermost convective zone itself, the perturbation patterns leads
to overdensities near the axis, i.e.\ to a hurtful, oblate shock
deformation. Because the different fate of these two models is
entirely due to the different \emph{geometry} of \emph{secondary}
perturbations arising from non-linear interactions at the convective
boundary (e.g.\ acoustic waves excited as the artificially
  imposed eddies distort the boundary), a higher initial perturbation
amplitude can indeed prove harmful in such special cases.

\section{Summary and Conclusions}
\label{sec:conclusions}
Using relativistic 2D simulations with multi-group neutrino transport,
we have performed an extensive parameter study to investigate 
whether progenitor asphericities arising during convective
burning could play a crucical role in the supernova explosion mechanism.  Our investigation is
based on a detailed quantitative analysis of the interplay of neutrino
heating and aspherical instabilities, which provides a framework for
understanding the role of seed perturbations, but is also an important
step towards understanding shock revival in multi-D in its own right.
Different from the recent works of \citet{couch_13,couch_14}, we
systematically vary the amplitude and geometry of the initial
perturbations and investigate both velocity and density perturbations.
We also discuss some of the physical principles governing the typical
velocities, the density contrast, and the flow geometry in the inner
shells of supernova progenitors, and attempt to incorporate some of
these principles into our models. Our simulations indicate that even
moderate velocity perturbations in the progenitor can aid shock
revival rather effectively if low-$\ell$ modes dominate the convective
flow in the shells outside the iron core.  Furthermore, the analysis
of our simulations has unearthed some interesting semi-empirical
scaling laws that govern the relation between neutrino heating and the
activity of aspherical instabilities in our 2D models.

The main results of our study may be summarized as follows:
\begin{enumerate}
\item In our quantitative analysis of the interplay of neutrino
  heating and non-radial instabilities, we find evidence for
  quasi-stationary saturation of the SASI and/or convection in the
  gain region in all our models. The neutrino heating, the shock
  deformation, the typical turbulent velocities and the turbulent Mach
  number, $\sqrt{\langle \mathrm{Ma}^2\rangle}$, in the gain region
  are all related to each other by simple scaling laws: At saturation,
  the dispersion of the lateral velocity in the gain region, $\langle
  v_\theta^2 \rangle$, is related to the neutrino heating per unit
  mass, $\dot{q}_\nu$, and the width of the gain region, $\delta
  r_\mathrm{gain}$, as $\langle v_\theta^2 \rangle \propto (\delta
  r_\mathrm{gain} \dot{q}_\nu)^{2/3}$.  The deformation of the shock
  is related to $\langle v_\theta^2 \rangle$ and the gravitational
  acceleration at the shock $g$ through another scaling relation
  $\delta r \propto \langle v_\theta^2 \rangle/g_\mathrm{shock}
  \propto \langle \mathrm{Ma}^2\rangle$.
\item Based both on the analysis of our numerical simulations and on
  analytic estimates, we argue that the key towards easier shock
  revival in multi-D lies in achieving high turbulent Mach numbers in
  the post-shock flow. We also argue that it is very difficult to
  ascribe the reduced luminosity threshold for an explosive runaway to
  a single cause such as more efficient neutrino heating, turbulent
  stresses, or the formation of large high-entropy bubbles, because
  in the saturation limit
  these phenomena are all inextricably related to each other and
  regulated by the violence of aspherical motions in the post-shock
  region (for which the turbulent Mach number is a useful measure).
\item Despite these ambiguities, we show that one can construct a
  simple analytic model that takes into account how turbulent pressure
  in the post-shock region pushes the shock further out and thereby
  enhances the heating conditions. Our analytic estimate suggests that
  the critical luminosity is lowered by roughly $25\%$ in multi-D
  compared to 1D, and that an explosive runaway occurs when the
  squared turbulent Mach number $\langle \mathrm{Ma}^2 \rangle$ in
  gain region reaches a value of roughly $0.46$ (compared to
  $\mathord{\sim}0.3$ in the numerical models), both of which is
  in reasonable agreement with simulations \citep{murphy_08b,hanke_12,couch_12b}.
\item {Concerning initial perturbations due to convection in the
  progenitor,} we argue that velocity perturbations ought to reflect
  the subsonic nature of the flow in the convective zones outside the
  iron core, and should approximately obey the anelastic condition
  $\nabla \cdot (\rho \, \mathbf{\delta v})=0$, at least away from the
  convective boundaries. The perturbation pattern of \citet{couch_13}
  strongly violates this condition. Density perturbations $\delta
  \rho/\rho$ should be of the order of the square of the convective
  Mach number $\mathrm{Ma}_\mathrm{prog}$ in the interior of convective regions.
\item Asphericities in the progenitor enhance the heating conditions
  primarily because they result in a permanent, ``forced''
  deformation of the accretion shock
due to directional variations of the mass
infall rate. The shock deformation results in
  a larger average shock radius and helps to channel the kinetic
  energy of the infalling material into more violent aspherical
  motions in the post-shock region as the matter hits the shock at an oblique
  angle. The very efficient conversion of velocity perturbations
  $\delta v$ into large density perturbations $\delta \rho/\rho \sim
  \mathrm{Ma}_\mathrm{prog}$ ahead of the shock due to differential infall appears to be a
  key element of this mechanism because it causes
the anisotropy in the mass infall rate.
\item For a given typical amplitude of velocity perturbations,
  quadrupolar and, to a lesser extent, dipolar perturbations are most
  efficient at triggering an explosion. For quadrupolar velocity
  perturbations, convective Mach numbers in the progenitor as low as $0.05$ yield an
  appreciable effect and reduce the critical luminosity required for
  shock revival by $\gtrsim 10\%$ according to our estimate,
  with stronger perturbations having a proportionately larger
  effect. Even in this conservative case with low convective Mach
  numbers, shock revival already occurs when the critical time-scale
  ratio $\tau_\mathrm{adv}/\tau_\mathrm{heat}$ approaches a value of
  $\approx 0.5$ in the corresponding \emph{unperturbed} model. 
\item On the other hand, one would have to invoke convective Mach
  numbers that are probably unrealistically high to achieve similar
  effects with perturbations dominated by higher angular wavenumbers
  $\ell$ for several reasons: These modes are inefficient at exciting
  a permanent $\ell=1$ or $\ell=2$ deformation of the shock, and non-linear
  damping during the infall comes into play much earlier for small-scale
  perturbations.
\item Only large density perturbations $\delta \rho/\rho \gtrsim 0.1$
  have a significant impact on the heating conditions. Such strong
  perturbations would probably also require inordinately high
  convective Mach numbers.
\end{enumerate}

These results have implications on several levels. They suggest that
convective perturbations in the progenitor can aid shock revival down
to much lower Mach numbers than recently demonstrated by
\citet{couch_13}. Relatively small convective velocities below $ 10^{8}
\ \mathrm{cm} \ \mathrm{s}^{-1}$, which are more in line with 3D
simulations of oxygen shell burning (but still higher than predicted
by \citealt{kuhlen_03}) may already have a significant impact on the
heating conditions. With our present limited knowledge of the
multi-dimensional structure of supernova progenitors just prior
to collapse, it is therefore still conceivable that asphericities
arising from convective burning may be one of the key elements
for obtaining robust supernova explosions.

There are, however, some caveats. Our present study is limited to 2D,
and it remains to be seen whether the forced deformation of the
accretion shock can play the same beneficial role for shock revival in
3D as in 2D. In particular, the relatively efficient excitation of
quadrupole modes could be ascribed to the presence of an (artificial)
symmetry axis. Unfortunately, a direct comparison with the 3D
simulations of \citet{couch_13,couch_14} is hampered by the fact that
we cannot mimic their inherently non-axisymmetric setup in any of our
2D simulations. It is reassuring that the trend towards a
  more efficient excitation of turbulent motions in the post-shock
  region for small $\ell$ found by \citet{couch_14} is compatible with
  our findings; however, aside from their use of 
initial velocity perturbations
  that lead to an unphysically strong contamination by acoustic waves, their study
  missed the spot in parameter space ($\ell=1,2$) that emerged as most
  interesting in our simulations.  Clearly, a more systematic study
of 3D perturbation geometries with a neutrino treatment on par with or
better than our FMT scheme would be highly desirable.

Moreover, our results already place relatively tight constraints on
the required properties of the convective flow in the shells outside
the iron core. If progenitor asphericities are to have an impact on
shock revival, convective Mach numbers need to be of the order of at
least $\gtrsim 0.05$, and large-scale dipolar or quadrupolar modes
should dominate the flow. Furthermore, extended regions in the
progenitor need to be convective, which is by no means a trivial
requirement given that 1D stellar evolution calculations predict
rather narrow convective shells at least in some cases.  Clearly, only
3D simulations covering full $4 \pi$ in solid angle and multiple
burning shells will finally tell us whether this is indeed the case.
They may also reveal whether sufficiently large velocity and density
perturbations may arise for other reasons, e.g.\ because of
rotation effects or g-mode activity \citep{murphy_04}.  The case of
rotating progenitors may be particularly interesting because rotation
could at least help to organize the flow into large-scale modes
(depending on the convective Rossby number).  At any rate,
self-consistent non-stationary multi-D models of supernova progenitors
are urgently needed to replace the ordered laminar flow patterns
used in this study, which can never fully capture reality.

Further work is also needed on the conceptual level. While we have
been able to provide a qualitative description of the interaction of
perturbations in the pre-shock region with the shock and with the
hydrodynamic instabilities active in the post-shock region, our
results prompt a number of questions: Is the susceptibility of the
shock to a forced deformation by $\ell=1$ and $\ell=2$ perturbations
more intimately linked to the SASI, which is also a low-$\ell$
instability?  How does the saturation of convection
and/or the SASI depend on the presence of strong progenitor asphericities? As
we consider the role of convective perturbations from the pre-collapse
phase in the neutrino-driven mechanism, we are even brought back to
some of the more basic questions in supernova modeling: 
What is the precise mechanism by which aspherical instabilities in
the supernova core aid shock revival? Is it by generating turbulent
stresses that push the shock out
\citep{murphy_12,mueller_12b,couch_14} or by facilitating the
formation of large high-entropy bubbles
\citep{thompson_00,fernandez_09,dolence_13}, or by more efficient
neutrino heating due to longer dwell times in the gain region
\citep{buras_06_b,murphy_08b,marek_09}? In this paper, we have touched
many and partially answered some of these questions,
collecting and sharpening ideas about the interplay of neutrino
heating and hydrodynamic instabilities from the literature and
combining them with our analysis of a large suite of 2D simulations
with and without initial perturbations. At this junction, it is of
course impossible to present a complete picture of the complicated
feedback mechanisms linking neutrino heating, 
  non-spherical instabilities, and initial perturbations.
Nonetheless, we hope that the ideas presented here may prove fruitful
for the analysis of supernova simulations in the future and spark
further urgently needed work on the hydrodynamics of the supernova
engine.

We acknowledge fruitful exchange with M.~Viallet, J.~Guilet,
Th.~Foglizzo, A.~Heger, W.~Hillebrandt, O.~Pejcha,
and T.~Thompson.  This work was supported by
the Deutsche Forschungsgemeinschaft through the Transregional
Collaborative Research Center SFB/TR 7 ``Gravitational Wave
Astronomy'' and the Cluster of Excellence EXC 153 ``Origin and
Structure of the Universe''
(http://www.universe-cluster.de). B.M. acknowledges support by the
Alexander von Humboldt Foundation through a Feodor Lynen
fellowship and by the Australian Research Council through
a Discovery Early Career Researcher Award (grant DE150101145).
The computations were performed on the IBM iDataPlex
system \emph{hydra} at the Rechenzentrum of the Max-Planck Society
(RZG), with support by the European PRACE
  infrastructure on the \emph{Curie} supercomputer of the Grand
  \'Equipement National de Calcul Intensif (GENCI) and on
  \emph{SuperMUC} at the Leibniz-Rechenzentrum (LRZ) (with additional
  computer time from the Gauss Centre for Supercomputing e.V. on
  that platform), as well as on \emph{Raijin} at the NCI National
  Facility (project fh6) and \emph{swinStar} using our ASTAC
  allocation for Q2/2014, and on the Monash eGrid Cluster.

\appendix

\section{Fast Multigroup Transport Scheme}
\label{sec:numerics}
Currently, the state of the art in multi-D simulations of
core-collapse supernovae is defined by multi-group neutrino
hydrodynamics simulations relying on various approximations to reduce
the complexity of the general relativistic Boltzmann equation
\citep{livne_04,swesty_09,mueller_10,bruenn_13,zhang_13}. While there
has been a lot of debate about the merits and demerits of the
approximations involved (ray-by-ray-approximation vs.\ multi-angle
transport, general relativity vs.\ the Newtonian approximation,
inclusion or non-inclusion of energy-exchanging scattering reactions, etc.),
all these schemes pose similar challenges from the computational point
of view. Typically, core-collapse supernova simulations covering
several hundreds of milliseconds of the post-bounce phase require
$\mathord{\sim} 10^7$ core-h in 3D and $\mathord{\sim} 10^5$ core-h in 2D on modern
supercomputers. Even in 2D, exhaustive parameter studies with
several dozen models are hardly feasible within a reasonable
time-frame with these state-of-the-art methods in the light of such
extraordinary
computational demands.

For our present study with $\mathord{\sim} 40$ axisymmetric models, we therefore
introduce a new multi-group neutrino transport scheme that captures
many of the essential features seen in simulations with more
sophisticated methods at a fraction of the computational cost.  It is
designed as a compromise between more elaborate multi-group schemes
and more severe approximations like gray transport
\citep{fryer_06,scheck_06} or the light bulb and leakage schemes
\citep{ruffert_96,rosswog_03,murphy_08b,oconnor_11} used in many
recent studies of multi-dimensional instabilities in core-collapse
supernovae. While this new scheme is similar to the IDSA
approximation of \citet{liebendoerfer_09} in this respect, its
derivation from the Boltzmann equations is more in line with
traditional approximation schemes for the radiative transfer
equations relying on a closure of the moment equations. Different
from current implementations of the IDSA approximation, there
is also no need to fall back onto a leakage scheme for
the heavy flavor neutrinos.

\subsection{Solution of the Monochromatic Neutrino Energy Equation}
Our fast multi-group (FMT) scheme solves the \emph{stationary}
neutrino transport problem in the so-called ray-by-ray approximation
\citep{buras_06_a} with the help of a closure relation for the flux
factor in the monochromatic neutrino energy equation. As further
approximations, we neglect velocity-dependent terms in the transport
equation and confine ourselves to isoenergetic scattering.

For a more transparent explanation of the FMT scheme, we also work in
the Newtonian approximation in this section and disregard neutrino
pair reactions for the moment. The generalization of the FMT scheme to
the relativistic case is given in Section~\ref{sec:fmt_gr}, and the
neutrino physics input (including a derivation of an effective
one-particle rate for nucleon bremsstrahlung) is described in
Section~\ref{sec:fmt_rates}.

Using all these approximations, the monochromatic neutrino
energy equation reduces to the simple form
\begin{equation}
\frac{1}{r^2}\frac{\pd H r^2}{\pd r}
=\kappa_a \left(J_\mathrm{eq}-J\right),
\end{equation}
or
\begin{equation}
\label{eq:momeq0}
\frac{1}{r^2}\frac{\pd H r^2}{\pd r}
=\kappa_a \left(J_\mathrm{eq}-\frac{H}{h}\right),
\end{equation}
where $J$ and $H$ are the zeroth and first moment of
the neutrino intensity, respectively, $J_\mathrm{eq}$
is the zeroth moment of the equilibrium distribution function, and $h=H/J$ is the flux factor. $\kappa_a$
is the absorption opacity including phase-space
blocking effects.

\subsubsection{Flux Factor -- Interior Solution}
\label{sec:two_stream}
At high and intermediate optical depths, we provide the required closure for
equation~(\ref{eq:momeq0}) by solving the Boltzmann equation using a
two-stream approximation with a radially outgoing and and radially
ingoing ray. For the sake of simplicity, we assume purely isotropic
scattering. With $f_o$ (outgoing) and $f_i$ (ingoing) denoting the
value of the distribution function in the direction of these two rays,
we end up with the following two equations:
\begin{eqnarray}
\label{eq:boltzmann_out}
\frac{\pd f_o}{\pd r}
&=&
\kappa_a \left(f_\mathrm{eq}-f_o\right)
+\kappa_s \left(f_i-f_o\right),
\\
\label{eq:boltzmann_in}
-\frac{\pd f_i}{\pd r}
&=&
\kappa_a \left(f_\mathrm{eq}-f_i\right)
+\kappa_s \left(f_o-f_i\right).
\end{eqnarray}
Here, $f_\mathrm{eq}$ denotes the equilibrium value of the distribution
function, and $\kappa_s$ is the scattering opacity.
Thanks to the choice of the ray directions, there are no
angular advection terms in the equations, which therefore remain
``quasi-planar'' despite the spherical geometry of the transport problem.

Equations (\ref{eq:boltzmann_out},\ref{eq:boltzmann_in}) can be solved 
by means of a Riccati transformation. After transforming
to new variables $f_+=(f_o+f_i)/2$ and $f_-=(f_o-f_i)/2$,
we obtain
\begin{eqnarray}
\label{eq:two_stream_plus}
\frac{\pd f_+}{\pd r}
&=&
-\left(\kappa_a+\kappa_s\right) f_-,
\\
\label{eq:two_stream_minus}
\frac{\pd f_-}{\pd r}
&=&
\kappa_a \left(f_\mathrm{eq}-f_+\right).
\end{eqnarray}
Using the ansatz $f_- = \psi + \chi f_+$, we obtain 
a valid solution for $f_o$ and $f_i$ from
three equations for $\psi$, $\chi$ and $f_+$
\begin{eqnarray}
\label{eq:chi}
\frac{\pd \chi}{\pd r}
&=&
-\kappa_a + \left(\kappa_a + \kappa_s \right) \chi^2,
\\
\label{eq:psi}
\frac{\pd \psi}{\pd r}
&=&
\kappa_a f_\mathrm{eq} + \left(\kappa_a + \kappa_s \right) \chi \psi,
\\
\label{eq:fplus}
\frac{\pd f_+}{\pd r} 
&=&
-\left(\kappa_a+\kappa_s\right) \left(\psi + \chi f_+ \right).
\end{eqnarray}
In order to satisfy the correct boundary conditions
for $f_o$ and $f_i$, namely $f_i(R_\mathrm{max})=0$
and $f_i(0)=f_o(0)$, we impose the following
boundary conditions on $f_+$, $\chi$, and $\psi$:
\begin{equation}
\chi(R_\mathrm{max})=1,
\quad
\psi(R_\mathrm{max})=0,
\quad
f_+(0)=-\frac{\psi(0)}{\chi(0)}.
\end{equation}
The actual flux factor $h$ that we feed into equation~(\ref{eq:momeq0})
is then computed from $f_o$ and $f_i$. However, we do not simply
compute $h$ using the two-stream approximation as $h=(f_o-f_i)/(f_o+f_i)$.
Instead we assume a continuous distribution function of the form
\begin{equation}
f(\mu)=e^{a (\mu - \eta)},
\end{equation}
which we fit to the values on the outgoing and ingoing ray ($\mu=\pm 1$)
in order to obtain better agreement in the diffusive regime,
where $\mathcal{I}(\mu)=J+3 \mu H$. The resulting flux factor is
\begin{equation}
\label{eq:flux_factor_two_stream}
h = 1 + 2 \frac{2 f_i/f_o}{1-f_i/f_o} + \frac{2}{\ln f_i/f_o}.
\end{equation}
In the vicinity of the removable singularity at $f_i/f_o=1$, we 
use the series expansion
\begin{equation}
h = \frac{1}{6} \left(1-\frac{f_i}{f_o}\right) + 
\frac{1}{12} \left(1-\frac{f_i}{f_o}\right)^2 
\ldots .
\end{equation}

\begin{figure}
\includegraphics[width=\linewidth]{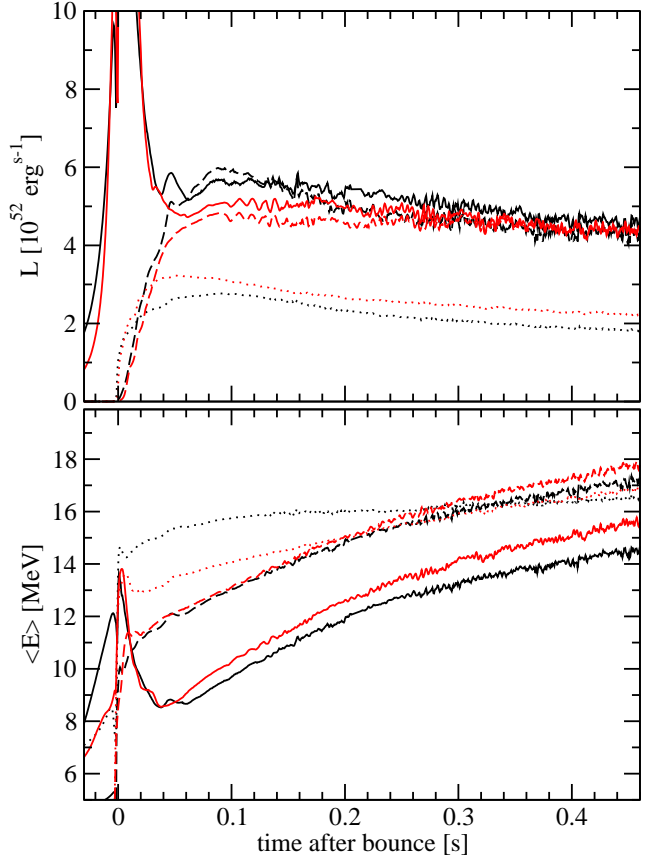}
\caption{
\label{fig:neutrino_vertex}
Total neutrino luminosities (top panel) and angle-averaged
mean energies (bottom panel) for the baseline model p0 as simulated
with the FMT scheme (black) and with \textsc{Vertex-CoCoNuT} (red).
Solid, dashed, and dotted curves are used for $\nu_e$, $\bar{\nu}_e$,
and $\nu_{\mu/\tau}$, respectively. All quantities are measured at a
radius of $400 \ \mathrm{km}$. }
\end{figure}

\begin{figure}
\includegraphics[width=\linewidth]{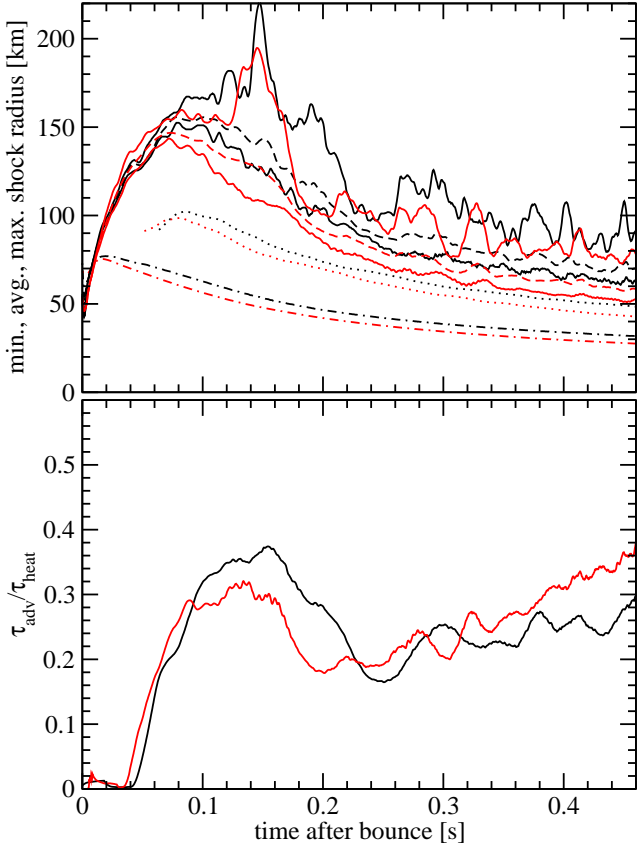}
\caption{
\label{fig:shock_vertex}
\emph{Top:} Time evolution of the maximum, minimum (solid curves), and average
(dashed) shock radius, the gain radius (dotted), and the proto-neutron
star radius (dash-dotted,
defined by a fiducial density
of $10^{11} \ \mathrm{g} \ \mathrm{cm}^{-3}$)  for the baseline model p0 as
simulated with the FMT scheme (black) and with \textsc{Vertex-CoCoNuT} (red).
\emph{Bottom:} The time-scale ratio $\tau_\mathrm{adv}/\tau_\mathrm{heat}$
in the simulations with the FMT scheme (black) and with \textsc{Vertex-CoCoNuT} (red).  }
\end{figure}

\subsubsection{Flux Factor -- Transition to Free Streaming}
It can easily be seen that the two-stream approximation fails to
reproduce the gradual transition of the neutrino radiation field from
an isotropic distribution to a forward-peaked distribution towards low
optical depths: Instead, equation~(\ref{eq:flux_factor_two_stream})
gives a flux factor of unity as soon as the intensity on the ingoing
ray vanishes ($f_i=0$), i.e.\ as soon as emission and scattering
reactions cease. In the worst case (e.g.\ when accretion has died down
after the explosion), this would imply that the flux factor abruptly jumps to
$h=1$  at the neutrinosphere. On the other hand, the
opposite problem can be encountered at moderate optical depths inside
the neutrinosphere. Here, the outward propagation of neutrinos in a
spherically stratified medium may lead to a faster transition to
forward peaking than the two-stream approximation would suggest.
This is a consequence of the last term on the left-hand side
of the transfer equation in spherical symmetry
\begin{equation}
\frac{\pd \mathcal{I}}{\pd t}+\mu \frac{\pd \mathcal{I}}{\pd
  r}+\frac{1-\mu^2}{r}\frac{\pd \mathcal{I}}{\pd \mu}=C,
\end{equation}
where $\mu$ is the angle cosine with respect to the radial
direction and $C$ is the collision integral. The last term
can essentially be understood as an advection term shifting
radiation intensity towards $\mu=1$.

In order to avoid or at least mitigate such unphysical effects
inherent in the quasi-planar two-stream approximation, we make
the following two modifications: First, we replace the total
opacity in equation~(\ref{eq:two_stream_plus})
with a reduced opacity corrected for
the advection towards $\mu=1$ in phase space,
\begin{equation}
\kappa_a+\kappa_s \rightarrow \mathrm{max}\left(\kappa_a+\kappa_s-\frac{2}{5r},0\right).
\end{equation}
Here, the term $2/5r$ is conceived of as an appropriate average over
the term $(1-\mu^2) r^{-1}\pd \mathcal{I} /\pd \mu$ for a nearly isotropic
radiation field.

The second modification is more important and consists in matching
the ``interior solution'' for the flux factor to an ``exterior solution''
once the flux factor from the two-stream approximation exceeds an
(adjustable) threshold value $h_\mathrm{match}$, which we set to
$h_\mathrm{match}=0.51$ in this present study.  The exterior solution
is computed using a closure relation for the Eddington factor $k=K/J$
(where K is the second angular moment of the neutrino intensity).
This is achieved by converting the first two moment equations
\begin{equation}
\frac{1}{r^2}\frac{\pd r^2 H}{\pd r}
=
\kappa_a \left(J_\mathrm{eq}-\frac{H}{h}\right),
\end{equation}
\begin{equation}
\frac{\pd k J}{\pd r}
+\frac{\left(3 k-1\right) J}{r}
=
-\left(\kappa_a+\kappa_s\right) H,
\end{equation}
into an ODE for the flux factor,
\begin{equation}
\label{eq:two_moment}
\frac{\pd h}{\pd r}
=\frac{{r^{-1}}h \left[k(h)-1\right]+\kappa_a \left[h^2-k(h)+k(h) J_\mathrm{eq}/J\right]
+\kappa_s h^2}
{k(h)-h k'(h)},
\end{equation}
where the derivative $k'$ of the Eddington factor
with respect to the flux factor enters in the denominator.
Equation~(\ref{eq:two_moment}) is integrated outward from the
point where the interior solution reaches $h=h_\mathrm{match}$. Obviously,
the existence of a singular point of the ODE~(\ref{eq:two_moment}) presents
a potential complication \citep{koerner_92}. We overcome this problem
by choosing $h_\mathrm{match}$ and the closure $k(h)$ such that no
singular point is encountered in the exterior domain. Specifically,
we use
\begin{equation}
k(h)=\frac{1-2h+4h^2}{3},
\end{equation}
which puts the singular point at $h=1/2$. Our choice of
$h_\mathrm{match}=0.51$ then ensures that we can integrate
equation~(\ref{eq:two_moment}) without encountering any
singularity. The closure relation employed here gives a somewhat lower
Eddington factor $k(h)$ than other closures proposed in the literature
\citep{minerbo_78,pomraning_81,levermore_84,janka_phd,janka_92,janka_92b,cernohorksy_94}
over a wide range of flux factors $h$. This is an unavoidable
compromise: Due to the breakdown of the two-stream approximation at
low optical depths, it is advisable to move the matching point between
the interior and exterior solution to a low flux factor while still
avoiding the critical point -- even at the expense of a slightly
suboptimal choice for the Eddington factor $k(h)$.

\subsection{Numerical Solution of the Equations}
The numerical solution of
equations~(\ref{eq:momeq0},\ref{eq:chi},\ref{eq:psi},\ref{eq:fplus},
\ref{eq:two_moment}) presents few difficulties. We use the implicit
Euler method for the stiff ODEs
(\ref{eq:momeq0},\ref{eq:chi},\ref{eq:psi},\ref{eq:fplus}), whereas
equation~(\ref{eq:two_moment}) can be integrated using an explicit
scheme. Care must be taken to ensure the correct direction of
integration: The integration must proceed inward for $\chi$ and
$\psi$, and outward for $f_+$ in order to respect the boundary
conditions.  For equation~(\ref{eq:momeq0}), the direction of
integration depends on the sign of the flux factor $h$, and implicit
finite-differencing automatically ensures that the solutions in
patches with $h>0$ and $h<0$ join smoothly at the singular points
($h=0$).

In this study, we solve the FMT equations for 21 exponentially spaced
energy group, with the energy at bin centers ranging from $3
\ \mathrm{MeV}$ to $197 \ \mathrm{MeV}$.

\subsection{General Relativistic Case}
\label{sec:fmt_gr}
In the general relativistic case, we assume a stationary metric $g_{\mu\nu}$
with vanishing shift and adopt the isotropic gauge,
\begin{equation}
g_{\mu\nu}=\mathrm{diag}(-\alpha^2,\phi^4,r^2 \phi^4, r^2 \sin^2\theta \phi^4),
\end{equation}
where $\alpha$ and $\phi$ are the lapse function and conformal factor.
These assumptions are generally valid to good accuracy in supernova
cores, and allow us to decouple the solution of the transport equation
for different energy groups as in the Newtonian case because the
redshifted energy $\hat{\epsilon}=\alpha \epsilon$ ($\epsilon$ being
the energy measured by a local observer) is a constant of motion.  By
grouping neutrinos according to $\hat{\epsilon}$, we can replace
equation~(\ref{eq:momeq0}) with a balance equation of the type
$\nabla_\mu j^\mu=s$ for the neutrino number within the $k$-th energy
group,
\begin{equation}
\frac{1}{\sqrt{-g}}\frac{\pd}{\pd r}\left(\sqrt{-g} \phi^{-2} \epsilon_k^{-1} H_k \Delta \epsilon_k\right)
=
 \kappa_a \epsilon_k^{-1} \left( J_\mathrm{eq,k}-\frac{H_k}{h}\right).
\end{equation}
In this equation, $\epsilon_k=\alpha^{-1} \hat{\epsilon}_k $ and
$\Delta \epsilon_k=\alpha^{-1} \Delta \hat{\epsilon}_k$ are now functions of
radius. The factor $\alpha^{-1}$ accounts for the conversion from
monochromatic energy densities and fluxes into number densities and fluxes, and the
the factor $\phi^{-2}$ on the left-hand side converts the flux density from a local, orthonormal observer
frame\footnote{We deliberately avoid the term ``comoving frame'' in this section, 
because we neglect velocity-dependent terms in our approximation.} to the 
non-orthonormal coordinate frame (whose basis vectors
are given by the coordinate derivatives $\pd/\pd x^i$).

Equations~(\ref{eq:boltzmann_out},\ref{eq:boltzmann_in}) for the two-stream solution of the
Boltzmann equation need to be generalized as well. The relativistic Boltzmann equation
reads \citep{lindquist_66,ehlers_71,stewart_71},
\begin{equation}
\frac{\ud f}{\ud \lambda} = 
\frac{\pd f}{\pd x^\mu} \frac{\ud x^\mu}{\ud \lambda}+
\frac{\pd f}{\pd p_\mu} \frac{\ud p_\mu}{\ud \lambda}
=
u^\mu v_\mu C_\mathrm{local},
\end{equation}
where $u^\mu$ is the neutrino four-velocity and $v^\mu$ is the
four-velocity of the observer in whose frame the collision integral
$C_\mathrm{local}$ is computed. In our case we have
$v^\mu=(\alpha^{-1},0,0,0)$ (as we neglect velocity-dependent terms),
$u_o^\mu=(\alpha^{-1}, \phi^{-2},0,0)$ for the four-velocity
of the outgoing ray, and $u_i^\mu=(\alpha^{-1},-\phi^{-2},0,0)$
for the ingoing ray.  $\ud p_\mu/\ud \lambda$ vanishes for the outgoing and
ingoing ray if we assume that the metric is nearly stationary
and neglect non-radial derivatives of the metric function in
keeping with the ray-by-ray-approximation. Hence, the equations
for the two-stream approximation take on a very simple form,
\begin{eqnarray*}
\frac{1}{\phi^2} \frac{\pd f_o}{\pd r}
&=&
\left[
\kappa_a \left(f_\mathrm{eq}-f_o\right)
+\kappa_s \left(f_i-f_o\right)
\right],
\\
-\frac{1}{\phi^2} \frac{\pd f_i}{\pd r}
&=&
\left[
\kappa_a \left(f_\mathrm{eq}-f_i\right)
+\kappa_s \left(f_o-f_i\right)
\right].
\end{eqnarray*}
Essentially, the only difference to equations~(\ref{eq:boltzmann_out},\ref{eq:boltzmann_in})
is the factor in front of the radial derivative that accounts for the conversion
between coordinate distances and physical distances. The system can be solved
exactly along the same lines as in Section~\ref{sec:two_stream}.

In our current implementation, the exterior solution for the flux
factor is not modified in the relativistic case. At least during the
first second of the post-bounce phase, the matching radius is located
at a fairly large radius so that relativistic effects are already
small and can be neglected for the computation of the flux factor
in practice, in particular in an approximative scheme as ours.

\subsection{Computation of Source Terms}
Once the solution for the energy-dependent stationary transport
problem has been computed, the neutrino source terms can in principle be calculated
directly from the stationary radiation field and the absorption
opacity. However, we prefer an indirect method for computing the
neutrino source terms for the equations of hydrodynamics. In
the CFC metric used in \textsc{CoCoNuT}, the frequency-integrated
zeroth moment of the collision integral 
\begin{equation}
C_\mathrm{tot}^{(0)}=\sum_k C_{\mathrm{local},k} \Delta \epsilon_k,
\end{equation}
can be obtained from what is essentially the radial derivative of the
redshift luminosity $L_\infty=(4 \pi)^2 \alpha^2 \phi^4 r^2
H_\mathrm{tot}$,
\begin{equation}
\frac{\pd \phi \alpha^2 \phi^4 r^2 H_\mathrm{tot}}{\pd r}
=\alpha^2 \phi^6 r^2 C_\mathrm{tot}^{(0)},
\end{equation}
where $H_\mathrm{tot}$ is the frequency-integrated first moment of the
neutrino intensity for a stationary solution of the transport
equation.\footnote{This can easily be verified by setting the radial
  velocity $v_r$ and the radial component of the shift $\beta^r$ to
  zero in equation~(27) of \citet{mueller_10} and integrating over
neutrino energy.} The neutrino source term for the energy equation is
  then computed from $C_\mathrm{tot}^{(0)}$ as in
  \citet{mueller_10}. Although we neglect velocity-dependent effects
  in our solution for the radiation field, we also follow
  \citet{mueller_10} in including a Lorenz boost from the fluid frame
  to the Eulerian frame, which we found to increase the robustness
  of the scheme.

The source term in the equation for the electron fraction can be
obtained in a completely analogous manner. The required weighted sum
$\mathcal{C}_\mathrm{tot}^{(0)}$ over the collision integral,
\begin{equation}
\mathcal{C}_\mathrm{tot}^{(0)}=\sum_k \epsilon_k^{-1} C_k \Delta
\epsilon_k,
\end{equation}
is computed as
\begin{equation}
\frac{\pd \alpha \phi^4 r^2 \mathcal{H}_\mathrm{tot}}{\pd r}
=\alpha \phi^6 r^2 \mathcal{C}_\mathrm{tot}^{(0)},
\end{equation}
where $\mathcal{H}_\mathrm{tot}=\sum_k \epsilon_k^{-1} H_k \Delta \epsilon_k$.
The source term for $Y_e$ is then again obtained from $\mathcal{C}_\mathrm{tot}^{(0)}$
exactly as in \citet{mueller_10}.

In the present study, the momentum source term due to neutrino
interactions is neglected because its effect is small, although its
computation is straightforward in principle: The momentum source term
in the comoving frame is given by
\begin{equation}
Q_M=4 \pi \sum_k (\kappa_s+\kappa_a) H_k \Delta \epsilon_k,
\end{equation}
and must be boosted to the Eulerian frame to obtain the
source terms in the momentum and energy equation.

It is worth noting that unlike time-dependent neutrino transport
methods, a naive ray-by-ray implementation  of the FMT scheme that neglects the lateral advection of neutrinos
with the fluid does not
lead to spurious convective instability at high optical depths. In
time-dependent transport schemes, lateral fluid motions effectively
lead to an energy and lepton number exchange between different fluid
elements if the neutrinos are not advected with the fluid. In the FMT
scheme, on the other hand, neutrinos cannot ``lag behind'' laterally
moving fluid elements, and no instability can occur.

\subsection{Treatment of Neutrino Reactions}
\label{sec:fmt_rates}
The FMT scheme assumes as very simple form for the collision integral
that includes only the absorption, emission, and isoenergetic
scattering of \emph{single} neutrinos. This places certain
restrictions on the neutrino reactions that can be accommodated within the
scheme. In our present implementation, we include charged-current
reactions of electron neutrinos and antineutrinos neutrinos with
nucleons and nuclei, isoenergetic neutrino scattering off nucleons and
nuclei, and the production of $\nu_\mu$ and $\nu_\tau$ by
nucleon-nucleon bremsstrahlung. The energy transfer between neutrinos
and the medium in the scattering of heavy flavor neutrinos is taken
into account approximatively. In the following, we provide a brief
explanation of our implementation of these rates, addressing in
particular the case of the heavy flavor neutrinos, where we construct
effective single-particle absorption opacities for the relevant
reactions.

\subsubsection{Electron Neutrinos and Antineutrinos}
For electron neutrinos and antineutrinos, the dominant opacity sources
are absorption on nuclei (during collapse) and nucleons, and the
corresponding scattering reactions, which are almost isoenergetic, and
the restriction to the simple form of the collision integral is not a
severe limitation. In the FMT scheme, these
reactions are currently included as in \citet{rampp_02}
(cf.\ \citealp{bruenn_85,mezzacappa_93}), i.e.\ the recoil energy
transfer in scattering reactions is neglected, as are correlation
effects at high densities \citep{burrows_98,burrows_99}, weak
magnetism \citep{horowitz_97}, and the effects of nucleon interaction
potentials \citep{martinez_12,roberts_12c} (although this
particular effect can easily be included). Different from
\citet{rampp_02}, we assume all scattering reactions to be isotropic,
i.e.\ we truncate the Legendre expansion of the scattering kernels at
the zeroth moment.  Neutrino-electron scattering is neglected
completely, although it plays an important role in the deleptonization
of the core during the collapse phase \citep{mezzacappa_93b}.

\subsubsection{Heavy Flavor Neutrinos}
The situation is different for the heavy flavor neutrinos; here, the
principal production processes are pair processes, and
non-isoenergetic scattering on nucleons also plays an important role in
thermalizing $\mu/\tau$ neutrinos during the accretion phase. A
special treatment is therefore required to accommodate these processes
even with the simple form for the collision integral assumed in the
FMT scheme.

As a production process, we include nucleon-nucleon bremsstrahlung
in the limit of vanishing nucleon degeneracy. We compute an effective
single particle rate in order to reduce the collision term
to the simple form $\kappa_a (f_\mathrm{eq}-f)$. Our starting
point is the form of the collision integral
derived by \citet{thompson_00_c} under the assumption of an isotropic
radiation field for non-degenerate nucleons (cp.\ their
equation 3.47),
\begin{eqnarray}
\label{eq:brems2}
\left(\frac{\pd f_\nu}{\pd t}\right)_\mathrm{brems}
&=&
\mathcal{A} \int\limits_0^\infty  \ud \epsilon_{\bar{\nu}}
\Big\{
K_1 \left(\frac{\epsilon}{2k_b T}\right)
e^{-\epsilon/(2k_b T)}
\\
\nonumber
&& 
\left[
(1-f_\nu)(1-f_{\bar{\nu}})-f_\nu f_{\bar{\nu}} e^{\epsilon/(k_b T)}
\right]
\Big\}
.
\end{eqnarray}
Here $f_\nu$ and $f_{\bar{\nu}}$ designate the value of the
distribution function of neutrinos of energy $\epsilon_\nu$ and
antineutrinos of energy $\epsilon_{\bar{\nu}}$, respectively, and
$\epsilon=\epsilon_\nu+\epsilon_{\bar{\nu}}$.  $T$ is the matter
temperature, and $K_1$ is a modified Bessel function.  Various weak
interaction constants, non-dimensional factors, and the dependence on
the thermodynamic quantities are lumped into the pre-factor
$\mathcal{A}$, for which we refer the reader to the original paper of
\citet{thompson_00_c}. We reduce equation~(\ref{eq:brems2}) for $\pd
f_\nu /\pd t$ to the desired form by assuming a thermal Fermi-Dirac
distribution with zero chemical potential for the antineutrino
participating in the reaction (and vice versa for $\pd f_{\bar{\nu}}
/\pd t$). The term in square brackets then becomes
\begin{eqnarray}
&&
(1-f_\nu)(1-f_{\bar{\nu},\mathrm{eq}})-f_\nu f_{\bar{\nu},\mathrm{eq}} e^{\epsilon/(k_b T)}
\\
\nonumber
&&=
(1-f_\nu)\frac{e^{\epsilon_{\bar{\nu}}/(k_b T)}}{1+e^{\epsilon_{\bar{\nu}}/(k_b T)}}
-f_\nu \frac{1}{1+e^{\epsilon_{\bar{\nu}}/(k_b T)}} 
e^{\epsilon/(k_b T)}
\\
\nonumber
&&=
(1-f_\nu)\frac{e^{\epsilon_{\bar{\nu}}/(k_b T)}}{1+e^{\epsilon_{\bar{\nu}}/(k_b T)}}
-f_\nu e^{\epsilon_\nu/(k_b T)} \frac{e^{\epsilon_{\bar{\nu}}/(k_b T)}}{1+e^{\epsilon_{\bar{\nu}}/(k_b T)}}
\\
\nonumber
&&=
\left[1-\left(1+e^{\epsilon_\nu/(k_b T)}\right) f_\nu\right]
\frac{e^{\epsilon_{\bar{\nu}}/(k_b T)}}{1+e^{\epsilon_{\bar{\nu}}/(k_b T)}}
\\
\nonumber
&&=
\left(1+e^{\epsilon_\nu/(k_b T)}\right)
\left(\frac{1}{1+e^{\epsilon_\nu/(k_b T)}}- f_\nu\right)
\frac{e^{\epsilon_{\bar{\nu}}/(k_b T)}}{1+e^{\epsilon_{\bar{\nu}}/(k_b T)}}
\\
\nonumber
&&=
\left(1+e^{\epsilon_\nu/(k_b T)}\right)
\frac{e^{\epsilon_{\bar{\nu}}/(k_b T)}}{1+e^{\epsilon_{\bar{\nu}}/(k_b T)}}
\left(f_{\nu,\mathrm{eq}}- f_\nu\right).
\end{eqnarray}
By pulling the term $\left(f_{\nu,\mathrm{eq}}- f_\nu\right)$ out of the
integral in equation~(\ref{eq:brems2}), the collision term
assumes the desired form $\kappa_a (f_\mathrm{eq}-f)$ with an
effective single-particle opacity $\kappa_a$ given by
\begin{equation}
\label{eq:kappa_brems}
\kappa_a=
\mathcal{A} 
\int\limits_0^\infty \ud \epsilon_{\bar{\nu}}
K_1 \left(\frac{\epsilon}{2k_b T}\right)
e^{-\epsilon/(2k_b T)}
\left(1+e^{\epsilon_\nu/(k_b T)}\right)
\frac{e^{\epsilon_{\bar{\nu}}/(k_b T)}}{1+e^{\epsilon_{\bar{\nu}}/(k_b T)}}.
\end{equation}
For the numerical evaluation of equation~(\ref{eq:kappa_brems}), it is
sufficient to replace $K_1$ with a low-order asymptotic expansion, as the
singularity $\epsilon=0$ is never encountered when the integral is computed
on an energy grid with finite resolution.

In addition to bremsstrahlung as a production process, we include the
scattering reactions on nucleons and nuclei as for electron neutrinos
and antineutrinos.  However, for the heavy flavor neutrinos, we also
take the small recoil energy transfer in neutrino-nucleon scattering
into account by means of the following approximation: Since a neutrino
of energy $\epsilon$ scattering off a nucleon transfers a fraction of
its energy of about $(\epsilon-3 k_b T)/(m_n c^2)$ to the medium (or
gains energy for $\epsilon<3 k_bT$) \citep{tubbs_79,janka_phd}, we
include an effective absorption opacity $\kappa_{a,\mathrm{\nu N}}$
for this process, which is defined as a fraction of the
neutrino-nucleon scattering opacity $\kappa_{s,\mathrm{\nu N}}$.
\begin{equation}
\kappa_{a,\mathrm{\nu N}}
=
\frac{\left|\epsilon_\nu-3 k_b T\right|}{m_n c^2}
\kappa_{s,\mathrm{\nu N}}.
\end{equation}
In the outer layers of the proto-neutron star, where high-energy
neutrinos originating from deeper inside the core partly thermalize
by scattering on nucleons in a relatively cool medium (i.e.\ $\langle
\epsilon\rangle > 3 k_b T$), this definition ensures an energy
transfer to the matter of the right order, and hence an attenuation
of the $\nu_\mu$ and $\nu_\tau$ neutrino luminosity of $\mathord{\sim} 5 \ldots 10 \%$
outside the primary production region as observed in simulations
with a rigorous treatment of neutrino-nucleon scattering
\citep{mueller_12a}.

\subsection{Comparison with the \textsc{Vertex-CoCoNuT} Code}
To illustrate that the FMT scheme provides a good
  approximation to more sophisticated transport schemes (such as
  Boltzmann transport or two-moment transport with a variable
  Eddington factor closure), we briefly compare the neutrino emission
  as well as the dynamical evolution of model p0 to a 2D model
  computed with the \textsc{Vertex-CoCoNuT} code \citep{mueller_10}
  using the ``full set'' of neutrino opacities from
  \citet{mueller_12a}: Figure~\ref{fig:neutrino_vertex} shows the
  neutrino luminosities and mean energies for all flavours both using
  both the FMT scheme and the relativistic \textsc{Vertex} transport
  module; different from Figure~\ref{fig:neutrino_p0}, the plot covers
  only the phase up to $460 \ \mathrm{ms}$ after bounce for which
  \textsc{Vertex-CoCoNuT} simulation data is available. The evolution
  of the maximum, minimum, and average shock radius, the gain radius,
  the proto-neutron star radius, and the time-scale ratio
  $\tau_\mathrm{adv}/\tau_\mathrm{heat}$ is depicted in
  Figure~\ref{fig:shock_vertex}.

Overall, the neutrino luminosities and mean energies obtained with the
FMT scheme follow those obtained with \textsc{Vertex} reasonably
well. There most conspicuous differences concern the emission of heavy
flavour neutrinos, for which we obtain smaller luminosities and higher
mean energies -- a result which is not unexpected because our
treatment of the heavy flavour neutrino interaction rates is
simplified considerably.  Interestingly (but again unsurprisingly), the
FMT scheme yields a somewhat larger spread between electron neutrino
and electron antineutrino mean energies.

It is presumably the weaker cooling due to the reduced emission of
heavy flavour neutrinos that mostly drives the dynamical differences
between the FMT model and the \textsc{Vertex} model.  The proto-neutron
star contracts more slowly with the FMT scheme, which in turn results
in larger gain radius and shock-radius. Furthermore, the slower
contraction of the proto-neutron star leads to a somewhat weaker
increase of the electron neutrino and antineutrino mean energies with
time. Despite the different contraction of the proto-neutron star, the
differences in the heating conditions remain relatively similar,
however.  The time-scale ratio $\tau_\mathrm{adv}/\tau_\mathrm{heat}$
is slightly better in the FMT run at early times, but eventually the
heating conditions become more more optimistic in the \textsc{Vertex}
model. 

While a detailed verification of the FMT scheme by cross-comparisons
with other neutrino transport codes in the vein of
\citet{liebendoerfer_05,liebendoerfer_09,mueller_10}, and
\citet{oconnor_14} is beyond the scope of this paper, there is
evidently quite good agreement with \textsc{Vertex-CoCoNuT}, especially
considering that we compare multi-D runs where differences
in the neutrino treatment can lead to important feedback
effects. We conclude that the FMT scheme provides a reliable alternative
to more sophisticated transport schemes \emph{at least for the purpose of our current paper}.

\section{A Toy Model for the Reduction of the Critical Luminosity in Multi-D}
\label{sec:toy_model}
In this paper, we make frequent use of simple power-law expressions
for the time-scales $\tau_\mathrm{adv}$ and $\tau_\mathrm{heat}$ and
for other quantities relevant to shock revival in supernovae
(in particular in Section~\ref{sec:critical_luminosity}).  Most of
these relations were derived in \citet{janka_01,janka_12}, but many of them are
also found in other papers scattered across the
literature. \citet{janka_12} also used the power law expressions for
$\tau_\mathrm{adv}$ and $\tau_\mathrm{heat}$ to reformulate the
time-scale criterion $\tau_\mathrm{adv}/\tau_\mathrm{heat}$ as a
critical condition for the neutrino luminosity.

In the interest of clarity, we provide a concise summary of the
assumptions and simplifications entering the derivation of these power
laws and the critical condition in this Appendix.  Furthermore, we show
how multi-D effects can be incorporated into the resulting model for
the heating conditions by treating them as an isotropic turbulent
pressure that aids shock expansion, following the ideas put forth by
\citet{murphy_12} and \citet{mueller_12b} in their analyses of multi-D
simulations.

\subsection{Spherically Symmetric Toy Model of the Gain Region}
Our basic ingredient consists of a simple stationary 1D model for the
gain region \citep{janka_01}, which is assumed to be radiation-dominated
with $P \propto T^4$ and adiabatically stratified
with power law profile of the density $\rho$ and the
pressure $P$,
\begin{equation}
P \propto r^{-4}, \quad \rho \propto r^{-3}.
\end{equation}
Hydrodynamic boundary conditions both for $P$ and $\rho$ are required at the shock.
The post-shock quantities are given in terms of the pre-shock
density $\rho_\mathrm{pre}$ and velocity $v_\mathrm{pre}^2$
and the compression factor $\beta$ at the shock as,
\begin{equation}
\rho_\mathrm{sh}=\beta \rho_\mathrm{pre},
\end{equation}
\begin{equation}
P_\mathrm{sh}=\frac{\beta-1}{\beta} \rho_\mathrm{pre} v_\mathrm{pre}^2.
\end{equation}
For $v_\mathrm{pre}$, we use a large fraction of the free-fall velocity
compatible that is roughly compatible with numerical simulations,
\begin{equation}
v_\mathrm{pre}
\sim
\sqrt{\frac{2 G M}{r_\mathrm{sh}}},
\end{equation}
and $\rho_\mathrm{pre}$ is then obtained from the accretion rate $\dot{M}$ as
$\rho_\mathrm{pre}= \dot{M}/(4\pi r^2 v_\mathrm{pre})$.

In order to fix the shock radius, \emph{one} additional boundary
condition is required. We fix the shock radius by requiring
equilibrium between neutrino energy heating and cooling at
the gain radius $r_\mathrm{gain}$.
Since the cooling and heating rates per baryon roughly scale
with $T^6\propto P^{3/2}$ and $L_\nu E_\nu^2/r_\mathrm{gain}^2$,
respectively, the required boundary condition is
\begin{equation}
P_\mathrm{gain}^{3/2} \propto \frac{L_\nu E_\nu^2}{r_\mathrm{gain}^2},
\end{equation}
where $L_\nu$ and $E_\nu$ are the neutrino luminosity and mean energy
(cf.\ equation~\ref{eq:avg_energy}).
The flux factor at the gain radius is implicitly assumed to be
fixed.

With these boundary conditions and the assumption of power-law
profiles for the density and pressure, one obtains
equation~(\ref{eq:shock_radius}) for the scaling of the shock radius:
\begin{equation}
r_\mathrm{sh} \propto 
\frac{(L_\nu  E_\nu^2)^{4/9} r_\mathrm{gain}^{16/9}}
{\dot{M}^{2/3} M^{1/3}}.
\end{equation}
Once the shock radius is determined, the mass in the gain region
$M_\mathrm{gain}$ and the advection time-scale
$\tau_\mathrm{adv}=M_\mathrm{gain}/ \dot{M}$ can be calculated by
analytic integration, which gives
\begin{equation}
M_\mathrm{gain} \propto \dot{M} r_\mathrm{sh}^{3/2} \ln (r_\mathrm{sh}/r_\mathrm{gain}),
\end{equation}
\begin{equation}
\label{eq:tadv_log}
\tau_\mathrm{adv} \propto  r_\mathrm{sh}^{3/2} \ln (r_\mathrm{sh}/r_\mathrm{gain}).
\end{equation}
In this paper, we drop the logarithmic correction and work with an empirical
power-law instead (cp.\ equation~\ref{eq:tadv}),
\begin{equation}
\label{eq:tadv_long}
\tau_\mathrm{adv} \propto \frac{r_\mathrm{sh}^{3/2}}{\sqrt{M}} \approx
5 \ \mathrm{ms} \times \left(\frac{r}{100 \ \mathrm{km}}\right)^{3/2}
  \left(\frac{M}{M_\odot}\right)^{-1/2}.
\end{equation} 
Here the normalization is chosen such that we get a reasonable
fit with simulation data even though we neglect the logarithmic term
in equation~(\ref{eq:tadv_log}).
By contrast, the estimate for the heating time-scale (equation~\ref{eq:theat}),
\begin{equation}
\tau_\mathrm{heat}
\propto
\frac{|e_\mathrm{gain}| r_\mathrm{gain}^2}{L_\nu E_\nu^2},
\end{equation}
is based on a zeroth-order approximation for the volume-integrated
neutrino heating. Essentially, we assume that the entire mass of the
gain region is located close to the gain radius and neglect neutrino cooling.

\subsection{Incorporation of Turbulent Stresses and Their Effect on Shock Revival}
As a zeroth-order approximation, we assume that convection and/or the
SASI alter the shock position and hence the runaway condition by
providing isotropic turbulent stresses $P_\mathrm{turb}$ everywhere.
These turbulent stresses are ultimately provided by
$P\, dV$ work exerted by neutrino-heated matter as it expands
(and is then accelerated by buoyancy due to the density contrast
with the ambient medium), i.e.\ in principle the reservoir
of thermal energy in the gain region is permanently tapped to
maintain the turbulent motions. The turbulent energy is permanently
dissipated back into thermal energy. Although 
$P\, dV$ work (or buoyant driving) and dissipation merely balance each
other in the steady state, the \emph{total reservoir of
(thermal+kinetic) energy} stored in the gain region, will be higher
than without turbulent motions. Moreover, due to the short thermal equilibration time-scale
at the gain radius, the thermodynamic boundary condition
at the gain radius remains unchanged; and as turbulent motions are
expected to flatten the entropy gradient (albeit not completely),
the \emph{temperature, and thermal pressure
profiles} remain essentially unchanged (except for the change in
shock radius) compared to the case without turbulent stresses
on the level of precision that our simple analysis can aim for.

Under these assumptions, we can therefore just add
the turbulent pressure to the (unchanged) thermal pressure
to obtain the total effective pressure $P_\mathrm{tot}$
that will then be used to determine the shock position.
Using a constant turbulent Mach number for
  the entire gain region and an adiabatic index $\Gamma=4/3$, the effective total pressure becomes
\begin{equation}
P_\mathrm{tot}=P+P_\mathrm{turb}=P \left(1+\frac{4
  \langle \mathrm{Ma} ^2\rangle }{3}\right).
\end{equation}
With the turbulent pressure included the outer boundary condition
for the \emph{thermal} pressure at the shock becomes
\begin{equation}
P_\mathrm{sh}\left(1+\frac{4
  \langle \mathrm{Ma}^2 \rangle}{3}\right)=\frac{\beta-1}{\beta} \rho_\mathrm{pre}
v_\mathrm{pre}^2.
\end{equation}
On the other hand, the inner boundary condition remains unchanged,
because it is a \emph{thermodynamic} boundary condition reflecting
the balance between heating and cooling at $r_\mathrm{gain}$.

With this new boundary condition, one can derive
equation~(\ref{eq:shock_radius_stress}) for the shock radius,
\begin{equation}
r_\mathrm{sh} \propto 
\frac{(L_\nu  E_\nu^2)^{4/9} r_\mathrm{gain}^{16/9}
\left(1+\frac{4\langle \mathrm{Ma}^2 \rangle}{3}\right)^{2/3}}
{\dot{M}^{2/3} M^{1/3}},
\end{equation}
as well as equation~(\ref{eq:critical_luminosity_stresses}) for
the critical luminosity in the presence of an isotropic turbulent pressure,
\begin{equation}
\label{eq:critical_luminosity_stresses2}
L_\nu E_\nu^2 \propto (\dot{M} M)^{3/5}
r_\mathrm{gain}^{-2/5} \left(1+\frac{4\langle \mathrm{Ma}^2 \rangle}{3}\right)^{-3/5}.
\end{equation}

In order to estimate the reduction of the critical luminosity due to
turbulent stresses, we must relate the correction term
$\left(1+\frac{4\langle \mathrm{Ma}^2 \rangle}{3}\right)^{-3/5}$ to the time-scale
criterion $\tau_\mathrm{adv}/\tau_\mathrm{heat}$, and plug in its
value at the runaway threshold (which will still be given by
$\tau_\mathrm{adv}/\tau_\mathrm{heat}=1$). To this end, we compute the
ratio of the turbulent kinetic energy per unit mass in the gain region
and the post-shock sound speed using
equation~(\ref{eq:kinetic_energy_vs_heating}) and
$c_\mathrm{s,post}^2 \approx GM/(3 r_\mathrm{sh})$ from
equation~(\ref{eq:c_s_shock})
as approximation for the sound speed,
\begin{eqnarray}
\langle \mathrm{Ma}^2 \rangle
&\approx &
\frac{2E_{\mathrm{kin},\theta}}{M_\mathrm{gain}}
\frac{3r_\mathrm{sh}}{G M}
\approx
\left[\frac{(r_\mathrm{sh}-r_\mathrm{gain}) \dot{Q}_\nu}{M_\mathrm{gain}}\right]^{2/3}
\frac{3r_\mathrm{sh}}{G M}
\\
\nonumber
&\approx &
\left(\frac{r_\mathrm{sh} |e_\mathrm{gain}|}{3\tau_\mathrm{heat}}\right)^{2/3}
\frac{3r_\mathrm{sh}}{G M}.
\end{eqnarray}
 Note that we have used $r_\mathrm{sh}-r_\mathrm{gain} \approx
r_\mathrm{sh}/3$ to simplify further calculations. The neutrino
heating rate has been eliminated in favor the heating time-scale
$\tau_\mathrm{heat}$ and the average total energy per unit mass in the
gain region, $|e_\mathrm{gain}|$ (equation~(\ref{eq:theat})), in order
to express $\langle \mathrm{Ma}^2\rangle$ in terms of the critical
time-scale ratio.  To arrive at the desired expression, we approximate
\begin{equation}
|e_\mathrm{gain}| \approx  \frac{GM}{3r_\mathrm{sh}},
\end{equation}
although this is slightly inconsistent with our former assumption $|e_\mathrm{gain}|=\mathrm{const.}$,
and use equation~(\ref{eq:tadv_long}) to obtain our final result,
\begin{eqnarray}
\nonumber
\langle \mathrm{Ma}^2 \rangle 
&\approx&
\left(\frac{G M}{9\tau_\mathrm{heat}}\right)^{2/3}
\times
100 \ \mathrm{km} \times \left(\frac{\tau_\mathrm{adv}}{5 \ \mathrm{ms}}\right)^{2/3}
\left(\frac{M}{M_\odot}\right)^{1/3}
\frac{3}{G M}
\\
&\approx&
0.4649 \times \left(\frac{\tau_\mathrm{adv}}{\tau_\mathrm{heat}}\right)^{2/3}.
\end{eqnarray}
The squared Mach number at the onset of the explosion is $0.46$
according to this prediction, which is somewhat higher than the value
of $\langle \mathrm{Ma}^2 \rangle\approx 0.3$ found in
Section~\ref{sec:saturation_with_perturbations}, but it is still in
the right ballpark. Plugging this result into
equation~(\ref{eq:critical_luminosity_stresses}) or
(\ref{eq:critical_luminosity_stresses2}) and comparing to the 1D
result without the correction term $\left(1+\frac{4\langle
  \mathrm{Ma}^2 \rangle}{3}\right)^{-3/5}$ immediately gives the
reduction of the critical luminosity. Our simple toy model predicts
that it should be around $75\%$ of the critical luminosity in 1D,
which is roughly consistent with numerical simulations
\citep{murphy_08b,nordhaus_10,hanke_12,couch_12a,couch_12b}.

We note, however, that there is an important loophole in our
derivation of the critical luminosity in 2D: As soon as the turbulent
stresses can no longer be modeled as isotropic and as soon as the
boundary conditions at the shock change due to an anisotropic mass
flux onto the shock, the reduction compared to the 1D case is no
longer given by the correction factor $\left(1+4\langle \mathrm{Ma}^2
\rangle/3\right)^{-3/5}$ alone. This precludes any application of this
simple model to the case of asphericities in the progenitor. A
simplified analytic description of shock expansion due to global
anisotropies in the pre-shock region along similar lines would be a
highly desirable goal for the future.

\bibliography{paper}

\begin{thebibliography}{116}
\expandafter\ifx\csname natexlab\endcsname\relax\def\natexlab#1{#1}\fi

\bibitem[{{Arnett}(1994)}]{arnett_94}
{Arnett} D., 1994, \apj, 427, 932

\bibitem[{{Arnett}, {Meakin} \& {Young}(2009){Arnett}, {Meakin}, \&
  {Young}}]{arnett_09}
{Arnett} D., {Meakin} C., {Young} P.~A., 2009, \apj, 690, 1715

\bibitem[{{Arnett} \& {Meakin}(2011)}]{arnett_11}
{Arnett} W.~D., {Meakin} C., 2011, \apj, 733, 78

\bibitem[{{Asida} \& {Arnett}(2000)}]{asida_00}
{Asida} S.~M., {Arnett} D., 2000, \apj, 545, 435

\bibitem[{{Bazan} \& {Arnett}(1994)}]{bazan_94}
{Bazan} G., {Arnett} D., 1994, \apjl, 433, L41

\bibitem[{{Bazan} \& {Arnett}(1998)}]{bazan_98}
{Bazan} G., {Arnett} D., 1998, \apj, 496, 316

\bibitem[{{Blondin}, {Mezzacappa} \& {DeMarino}(2003){Blondin}, {Mezzacappa},
  \& {DeMarino}}]{blondin_03}
{Blondin} J.~M., {Mezzacappa} A., {DeMarino} C., 2003, \apj, 584, 971

\bibitem[{{Bruenn}(1985)}]{bruenn_85}
{Bruenn} S.~W., 1985, \apjs, 58, 771

\bibitem[{{Bruenn}(1986)}]{bruenn_86}
{Bruenn} S.~W., 1986, \apjl, 311, L69

\bibitem[{{Bruenn} {et~al}\mbox{.}(2013){Bruenn}, {Mezzacappa}, {Hix}, {Lentz},
  {Bronson Messer}, {Lingerfelt}, {Blondin}, {Endeve}, {Marronetti}, \&
  {Yakunin}}]{bruenn_13}
{Bruenn} S.~W. {et~al.}, 2013, \apjl, 767, L6

\bibitem[{{Buras} {et~al}\mbox{.}(2006{\natexlab{a}}){Buras}, {Janka}, {Rampp},
  \& {Kifonidis}}]{buras_06_b}
{Buras} R., {Janka} H.-T., {Rampp} M., {Kifonidis} K., 2006{\natexlab{a}},
  \aap, 457, 281

\bibitem[{{Buras} {et~al}\mbox{.}(2006{\natexlab{b}}){Buras}, {Rampp}, {Janka},
  \& {Kifonidis}}]{buras_06_a}
{Buras} R., {Rampp} M., {Janka} H.-T., {Kifonidis} K., 2006{\natexlab{b}},
  \aap, 447, 1049

\bibitem[{{Burrows}(2013)}]{burrows_13}
{Burrows} A., 2013, Reviews of Modern Physics, 85, 245

\bibitem[{{Burrows}, {Dolence} \& {Murphy}(2012){Burrows}, {Dolence}, \&
  {Murphy}}]{burrows_12}
{Burrows} A., {Dolence} J.~C., {Murphy} J.~W., 2012, \apj, 759, 5

\bibitem[{{Burrows} \& {Fryxell}(1992)}]{burrows_92}
{Burrows} A., {Fryxell} B.~A., 1992, Science, 258, 430

\bibitem[{{Burrows} \& {Goshy}(1993)}]{burrows_93}
{Burrows} A., {Goshy} J., 1993, \apjl, 416, L75+

\bibitem[{{Burrows} \& {Hayes}(1996)}]{burrows_96}
{Burrows} A., {Hayes} J., 1996, Physical Review Letters, 76, 352

\bibitem[{{Burrows}, {Hayes} \& {Fryxell}(1995){Burrows}, {Hayes}, \&
  {Fryxell}}]{burrows_95}
{Burrows} A., {Hayes} J., {Fryxell} B.~A., 1995, \apj, 450, 830

\bibitem[{{Burrows} \& {Sawyer}(1998)}]{burrows_98}
{Burrows} A., {Sawyer} R.~F., 1998, \prc, 58, 554

\bibitem[{{Burrows} \& {Sawyer}(1999)}]{burrows_99}
{Burrows} A., {Sawyer} R.~F., 1999, \prc, 59, 510

\bibitem[{{Carter} \& {Prakash}(2002)}]{carter_02}
{Carter} G.~W., {Prakash} M., 2002, Physics Letters B, 525, 249

\bibitem[{{Cernohorsky} \& {Bludman}(1994)}]{cernohorksy_94}
{Cernohorsky} J., {Bludman} S.~A., 1994, \apj, 433, 250

\bibitem[{{Chandrasekhar}(1961)}]{chandrasekhar_61}
{Chandrasekhar} S., 1961, {Hydrodynamic and Hydromagnetic Stability}.
  Clarendon, Oxford

\bibitem[{{Chatzopoulos}, {Graziani} \& {Couch}(2014){Chatzopoulos},
  {Graziani}, \& {Couch}}]{chatzopoulos_14}
{Chatzopoulos} E., {Graziani} C., {Couch} S.~M., 2014, ArXiv e-prints 1405.4873

\bibitem[{{Chen}, {Heger} \& {Almgren}(2013){Chen}, {Heger}, \&
  {Almgren}}]{chen_13}
{Chen} K.-J., {Heger} A., {Almgren} A.~S., 2013, Astronomy and Computing, 3, 70

\bibitem[{{Couch}(2013{\natexlab{a}})}]{couch_12b}
{Couch} S.~M., 2013{\natexlab{a}}, \apj, 775, 35

\bibitem[{{Couch}(2013{\natexlab{b}})}]{couch_12a}
{Couch} S.~M., 2013{\natexlab{b}}, \apj, 765, 29

\bibitem[{{Couch} \& {Ott}(2013)}]{couch_13}
{Couch} S.~M., {Ott} C.~D., 2013, \apjl, 778, L7

\bibitem[{{Couch} \& {Ott}(2014)}]{couch_14}
{Couch} S.~M., {Ott} C.~D., 2014, ArXiv e-prints 1408.1399

\bibitem[{{Dimmelmeier}, {Font} \& {M{\"u}ller}(2002){Dimmelmeier}, {Font}, \&
  {M{\"u}ller}}]{dimmelmeier_02_a}
{Dimmelmeier} H., {Font} J.~A., {M{\"u}ller} E., 2002, \aap, 388, 917

\bibitem[{{Dolence} {et~al}\mbox{.}(2013){Dolence}, {Burrows}, {Murphy}, \&
  {Nordhaus}}]{dolence_13}
{Dolence} J.~C., {Burrows} A., {Murphy} J.~W., {Nordhaus} J., 2013, \apj, 765,
  110

\bibitem[{{Ehlers}(1971)}]{ehlers_71}
{Ehlers} J., 1971, in General Relativity and Cosmology, {Sachs} R.~K., ed., pp.
  1--70

\bibitem[{{Fern{\'a}ndez}(2012)}]{fernandez_12}
{Fern{\'a}ndez} R., 2012, \apj, 749, 142

\bibitem[{{Fern{\'a}ndez} {et~al}\mbox{.}(2014){Fern{\'a}ndez}, {M{\"u}ller},
  {Foglizzo}, \& {Janka}}]{fernandez_14}
{Fern{\'a}ndez} R., {M{\"u}ller} B., {Foglizzo} T., {Janka} H.-T., 2014,
  \mnras, 440, 2763

\bibitem[{{Fern{\'a}ndez} \& {Thompson}(2009)}]{fernandez_09}
{Fern{\'a}ndez} R., {Thompson} C., 2009, \apj, 703, 1464

\bibitem[{{Foglizzo}, {Scheck} \& {Janka}(2006){Foglizzo}, {Scheck}, \&
  {Janka}}]{foglizzo_06}
{Foglizzo} T., {Scheck} L., {Janka} H.-T., 2006, \apj, 652, 1436

\bibitem[{{Fryer}, {Holz} \& {Hughes}(2004){Fryer}, {Holz}, \&
  {Hughes}}]{fryer_04}
{Fryer} C.~L., {Holz} D.~E., {Hughes} S.~A., 2004, \apj, 609, 288

\bibitem[{{Fryer}, {Rockefeller} \& {Warren}(2006){Fryer}, {Rockefeller}, \&
  {Warren}}]{fryer_06}
{Fryer} C.~L., {Rockefeller} G., {Warren} M.~S., 2006, \apj, 643, 292

\bibitem[{{Goldreich}, {Lai} \& {Sahrling}(1997){Goldreich}, {Lai}, \&
  {Sahrling}}]{goldreich_97}
{Goldreich} P., {Lai} D., {Sahrling} M., 1997, in Unsolved Problems in
  Astrophysics, {Bahcall} J.~N., {Ostriker} J.~P., eds., pp. 269--280

\bibitem[{{Guilet}, {Sato} \& {Foglizzo}(2010){Guilet}, {Sato}, \&
  {Foglizzo}}]{guilet_10}
{Guilet} J., {Sato} J., {Foglizzo} T., 2010, \apj, 713, 1350

\bibitem[{{Hanke} {et~al}\mbox{.}(2012){Hanke}, {Marek}, {M{\"u}ller}, \&
  {Janka}}]{hanke_12}
{Hanke} F., {Marek} A., {M{\"u}ller} B., {Janka} H.-T., 2012, \apj, 755, 138

\bibitem[{{Hanke} {et~al}\mbox{.}(2013){Hanke}, {M{\"u}ller}, {Wongwathanarat},
  {Marek}, \& {Janka}}]{hanke_13}
{Hanke} F., {M{\"u}ller} B., {Wongwathanarat} A., {Marek} A., {Janka} H.-T.,
  2013, \apj, 770, 66

\bibitem[{{Heger}, {Langer} \& {Woosley}(2000){Heger}, {Langer}, \&
  {Woosley}}]{heger_00}
{Heger} A., {Langer} N., {Woosley} S.~E., 2000, \apj, 528, 368

\bibitem[{{Herant}, {Benz} \& {Colgate}(1992){Herant}, {Benz}, \&
  {Colgate}}]{herant_92}
{Herant} M., {Benz} W., {Colgate} S., 1992, \apj, 395, 642

\bibitem[{{Herant} {et~al}\mbox{.}(1994){Herant}, {Benz}, {Hix}, {Fryer}, \&
  {Colgate}}]{herant_94}
{Herant} M., {Benz} W., {Hix} W.~R., {Fryer} C.~L., {Colgate} S.~A., 1994,
  \apj, 435, 339

\bibitem[{{Horowitz}(1997)}]{horowitz_97}
{Horowitz} C.~J., 1997, \prd, 55, 4577

\bibitem[{{H\"udepohl} {et~al}\mbox{.}(2009){H\"udepohl}, {M\"uller}, {Janka},
  {Marek}, \& {Raffelt}}]{huedepohl_10}
{H\"udepohl} L., {M\"uller} B., {Janka} H., {Marek} A., {Raffelt} G.~G., 2009,
  \prl

\bibitem[{{Janka}(1991)}]{janka_phd}
{Janka} H.-T., 1991, PhD thesis, Technische Universit{\"a}t M{\"u}nchen, (1991)

\bibitem[{{Janka}(1992)}]{janka_92}
{Janka} H.-T., 1992, \aap, 256, 452

\bibitem[{{Janka}(2001)}]{janka_01}
{Janka} H.-T., 2001, \aap, 368, 527

\bibitem[{{Janka}(2012)}]{janka_12}
{Janka} H.-T., 2012, Annual Review of Nuclear and Particle Science, 62, 407

\bibitem[{{Janka}, {Dgani} \& {van den Horn}(1992){Janka}, {Dgani}, \& {van den
  Horn}}]{janka_92b}
{Janka} H.-T., {Dgani} R., {van den Horn} L.~J., 1992, \aap, 265, 345

\bibitem[{{Janka} {et~al}\mbox{.}(2012){Janka}, {Hanke}, {H{\"u}depohl},
  {Marek}, {M{\"u}ller}, \& {Obergaulinger}}]{janka_12b}
{Janka} H.-T., {Hanke} F., {H{\"u}depohl} L., {Marek} A., {M{\"u}ller} B.,
  {Obergaulinger} M., 2012, Progress of Theoretical and Experimental Physics,
  2012, 010000

\bibitem[{{Janka} \& {M\"uller}(1996)}]{janka_96}
{Janka} H.-T., {M\"uller} E., 1996, \aap, 306, 167

\bibitem[{{Kippenhahn} \& {Weigert}(1990)}]{kippenhahn}
{Kippenhahn} R., {Weigert} A., 1990, {Stellar Structure and Evolution}.
  Springer, Berlin

\bibitem[{{Kiziltan} {et~al}\mbox{.}(2013){Kiziltan}, {Kottas}, {De Yoreo}, \&
  {Thorsett}}]{kiziltan_13}
{Kiziltan} B., {Kottas} A., {De Yoreo} M., {Thorsett} S.~E., 2013, \apj, 778,
  66

\bibitem[{{K{\"o}rner} \& {Janka}(1992)}]{koerner_92}
{K{\"o}rner} A., {Janka} H.-T., 1992, \aap, 266, 613

\bibitem[{{Kuhlen}, {Woosley} \& {Glatzmaier}(2003){Kuhlen}, {Woosley}, \&
  {Glatzmaier}}]{kuhlen_03}
{Kuhlen} M., {Woosley} W.~E., {Glatzmaier} G.~A., 2003, in Astronomical Society
  of the Pacific Conference Series, Vol. 293, 3D Stellar Evolution, {Turcotte}
  S., {Keller} S.~C., {Cavallo} R.~M., eds., p. 147

\bibitem[{{Lai} \& {Goldreich}(2000)}]{lai_00}
{Lai} D., {Goldreich} P., 2000, \apj, 535, 402

\bibitem[{{Landau} \& {Lifshitz}(1959)}]{landau_fluid}
{Landau} L.~D., {Lifshitz} E.~M., 1959, {Course of theoretical physics}, Vol.
  VI, Fluid mechanics. Pergamon Press, Oxford

\bibitem[{{Lattimer} \& {Swesty}(1991)}]{lattimer_91}
{Lattimer} J.~M., {Swesty} F.~D., 1991, {\it Nucl.~Phys.~A}, 535, 331

\bibitem[{{Levermore}(1984)}]{levermore_84}
{Levermore} C.~D., 1984, \jqsrt, 31, 149

\bibitem[{{Liebend{\"o}rfer}(2005)}]{liebendoerfer_05_b}
{Liebend{\"o}rfer} M., 2005, \apj, 633, 1042

\bibitem[{{Liebend{\"o}rfer} {et~al}\mbox{.}(2005){Liebend{\"o}rfer}, {Rampp},
  {Janka}, \& {Mezzacappa}}]{liebendoerfer_05}
{Liebend{\"o}rfer} M., {Rampp} M., {Janka} H.-T., {Mezzacappa} A., 2005, \apj,
  620, 840

\bibitem[{{Liebend{\"o}rfer}, {Whitehouse} \&
  {Fischer}(2009){Liebend{\"o}rfer}, {Whitehouse}, \&
  {Fischer}}]{liebendoerfer_09}
{Liebend{\"o}rfer} M., {Whitehouse} S.~C., {Fischer} T., 2009, \apj, 698, 1174

\bibitem[{{Lindquist}(1966)}]{lindquist_66}
{Lindquist} R.~W., 1966, {\it Ann.~Phys.}, 37, 487

\bibitem[{{Livne} {et~al}\mbox{.}(2004){Livne}, {Burrows}, {Walder},
  {Lichtenstadt}, \& {Thompson}}]{livne_04}
{Livne} E., {Burrows} A., {Walder} R., {Lichtenstadt} I., {Thompson} T.~A.,
  2004, \apj, 609, 277

\bibitem[{{Marek} \& {Janka}(2009)}]{marek_09}
{Marek} A., {Janka} H., 2009, \apj, 694, 664

\bibitem[{{Marek}, {Janka} \& {M{\"u}ller}(2009){Marek}, {Janka}, \&
  {M{\"u}ller}}]{marek_08}
{Marek} A., {Janka} H., {M{\"u}ller} E., 2009, \aap, 496, 475

\bibitem[{{Mart{\'{\i}}nez-Pinedo}
  {et~al}\mbox{.}(2012){Mart{\'{\i}}nez-Pinedo}, {Fischer}, {Lohs}, \&
  {Huther}}]{martinez_12}
{Mart{\'{\i}}nez-Pinedo} G., {Fischer} T., {Lohs} A., {Huther} L., 2012,
  Physical Review Letters, 109, 251104

\bibitem[{{Meakin} \& {Arnett}(2006)}]{meakin_06}
{Meakin} C.~A., {Arnett} D., 2006, \apjl, 637, L53

\bibitem[{{Meakin} \& {Arnett}(2007{\natexlab{a}})}]{meakin_07_b}
{Meakin} C.~A., {Arnett} D., 2007{\natexlab{a}}, \apj, 665, 690

\bibitem[{{Meakin} \& {Arnett}(2007{\natexlab{b}})}]{meakin_07}
{Meakin} C.~A., {Arnett} D., 2007{\natexlab{b}}, \apj, 667, 448

\bibitem[{{Mezzacappa} \& {Bruenn}(1993{\natexlab{a}})}]{mezzacappa_93}
{Mezzacappa} A., {Bruenn} S.~W., 1993{\natexlab{a}}, \apj, 405, 669

\bibitem[{{Mezzacappa} \& {Bruenn}(1993{\natexlab{b}})}]{mezzacappa_93b}
{Mezzacappa} A., {Bruenn} S.~W., 1993{\natexlab{b}}, \apj, 410, 740

\bibitem[{{Minerbo}(1978)}]{minerbo_78}
{Minerbo} G.~N., 1978, \jqsrt, 20, 541

\bibitem[{{M{\"u}ller}, {Janka} \& {Dimmelmeier}(2010){M{\"u}ller}, {Janka}, \&
  {Dimmelmeier}}]{mueller_10}
{M{\"u}ller} B., {Janka} H., {Dimmelmeier} H., 2010, \apjs, 189, 104

\bibitem[{{M{\"u}ller} \& {Janka}(2014)}]{mueller_14}
{M{\"u}ller} B., {Janka} H.-T., 2014, \apj, 788, 82

\bibitem[{{M{\"u}ller}, {Janka} \& {Heger}(2012){M{\"u}ller}, {Janka}, \&
  {Heger}}]{mueller_12b}
{M{\"u}ller} B., {Janka} H.-T., {Heger} A., 2012, \apj, 761, 72

\bibitem[{{M{\"u}ller}, {Janka} \& {Marek}(2012){M{\"u}ller}, {Janka}, \&
  {Marek}}]{mueller_12a}
{M{\"u}ller} B., {Janka} H.-T., {Marek} A., 2012, \apj, 756, 84

\bibitem[{{M\"uller} \& {Janka}(1997)}]{mueller_97}
{M\"uller} E., {Janka} H.-T., 1997, \aap, 317, 140

\bibitem[{{Murphy} \& {Burrows}(2008)}]{murphy_08b}
{Murphy} J.~W., {Burrows} A., 2008, \apj, 688, 1159

\bibitem[{{Murphy}, {Burrows} \& {Heger}(2004){Murphy}, {Burrows}, \&
  {Heger}}]{murphy_04}
{Murphy} J.~W., {Burrows} A., {Heger} A., 2004, \apj, 615, 460

\bibitem[{{Murphy}, {Dolence} \& {Burrows}(2013){Murphy}, {Dolence}, \&
  {Burrows}}]{murphy_12}
{Murphy} J.~W., {Dolence} J.~C., {Burrows} A., 2013, \apj, 771, 52

\bibitem[{{Murphy} \& {Meakin}(2011)}]{murphy_11}
{Murphy} J.~W., {Meakin} C., 2011, \apj, 742, 74

\bibitem[{{Nordhaus} {et~al}\mbox{.}(2010){Nordhaus}, {Burrows}, {Almgren}, \&
  {Bell}}]{nordhaus_10}
{Nordhaus} J., {Burrows} A., {Almgren} A., {Bell} J., 2010, \apj, 720, 694

\bibitem[{{O'Connor}(2014)}]{oconnor_14}
{O'Connor} E., 2014, ArXiv e-prints

\bibitem[{{O'Connor} \& {Ott}(2011)}]{oconnor_11}
{O'Connor} E., {Ott} C.~D., 2011, \apj, 730, 70

\bibitem[{{{\"O}zel} {et~al}\mbox{.}(2012){{\"O}zel}, {Psaltis}, {Narayan}, \&
  {Santos Villarreal}}]{oezel_12}
{{\"O}zel} F., {Psaltis} D., {Narayan} R., {Santos Villarreal} A., 2012, \apj,
  757, 55

\bibitem[{{Pejcha} \& {Thompson}(2012)}]{pejcha_12}
{Pejcha} O., {Thompson} T.~A., 2012, \apj, 746, 106

\bibitem[{{Pomraning}(1981)}]{pomraning_81}
{Pomraning} G.~C., 1981, \jqsrt, 26, 385

\bibitem[{{Rampp} \& {Janka}(2002)}]{rampp_02}
{Rampp} M., {Janka} H.-T., 2002, \aap, 396, 361

\bibitem[{{Reddy} {et~al}\mbox{.}(1999){Reddy}, {Prakash}, {Lattimer}, \&
  {Pons}}]{reddy_99}
{Reddy} S., {Prakash} M., {Lattimer} J.~M., {Pons} J.~A., 1999, \prc, 59, 2888

\bibitem[{{Roberts}, {Reddy} \& {Shen}(2012){Roberts}, {Reddy}, \&
  {Shen}}]{roberts_12c}
{Roberts} L.~F., {Reddy} S., {Shen} G., 2012, \prc, 86, 065803

\bibitem[{{Rosswog} \& {Liebend{\"o}rfer}(2003)}]{rosswog_03}
{Rosswog} S., {Liebend{\"o}rfer} M., 2003, \mnras, 342, 673

\bibitem[{{Ruffert}, {Janka} \& {Schaefer}(1996){Ruffert}, {Janka}, \&
  {Schaefer}}]{ruffert_96}
{Ruffert} M., {Janka} H.-T., {Schaefer} G., 1996, \aap, 311, 532

\bibitem[{{Scheck} {et~al}\mbox{.}(2006){Scheck}, {Kifonidis}, {Janka}, \&
  {M{\"u}ller}}]{scheck_06}
{Scheck} L., {Kifonidis} K., {Janka} H.-T., {M{\"u}ller} E., 2006, \aap, 457,
  963

\bibitem[{{Schwab}, {Podsiadlowski} \& {Rappaport}(2010){Schwab},
  {Podsiadlowski}, \& {Rappaport}}]{schwab_10}
{Schwab} J., {Podsiadlowski} P., {Rappaport} S., 2010, \apj, 719, 722

\bibitem[{{Smartt}(2009)}]{smartt_09b}
{Smartt} S.~J., 2009, \araa, 47, 63

\bibitem[{{Stewart}(1971)}]{stewart_71}
{Stewart} J.~M., ed., 1971, Lecture Notes in Physics, Berlin Springer Verlag,
  Vol.~10, {Non-equilibrium relativistic kinetic theory}

\bibitem[{{Suwa} {et~al}\mbox{.}(2010){Suwa}, {Kotake}, {Takiwaki},
  {Whitehouse}, {Liebend{\"o}rfer}, \& {Sato}}]{suwa_10}
{Suwa} Y., {Kotake} K., {Takiwaki} T., {Whitehouse} S.~C., {Liebend{\"o}rfer}
  M., {Sato} K., 2010, \pasj, 62, L49+

\bibitem[{{Suwa} {et~al}\mbox{.}(2013){Suwa}, {Takiwaki}, {Kotake}, {Fischer},
  {Liebend{\"o}rfer}, \& {Sato}}]{suwa_13}
{Suwa} Y., {Takiwaki} T., {Kotake} K., {Fischer} T., {Liebend{\"o}rfer} M.,
  {Sato} K., 2013, \apj, 764, 99

\bibitem[{{Swesty} \& {Myra}(2009)}]{swesty_09}
{Swesty} F.~D., {Myra} E.~S., 2009, \apjs, 181, 1

\bibitem[{{Takahashi} \& {Yamada}(2014)}]{takahashi_14}
{Takahashi} K., {Yamada} S., 2014, ArXiv e-prints 1408.3503

\bibitem[{{Takiwaki}, {Kotake} \& {Suwa}(2014){Takiwaki}, {Kotake}, \&
  {Suwa}}]{takiwaki_14}
{Takiwaki} T., {Kotake} K., {Suwa} Y., 2014, \apj, 786, 83

\bibitem[{{Tamborra} {et~al}\mbox{.}(2014){Tamborra}, {Hanke}, {Janka},
  {M{\"u}ller}, {Raffelt}, \& {Marek}}]{tamborra_14}
{Tamborra} I., {Hanke} F., {Janka} H.-T., {M{\"u}ller} B., {Raffelt} G.~G.,
  {Marek} A., 2014, \apj, 792, 96

\bibitem[{{Tanaka} {et~al}\mbox{.}(2009){Tanaka}, {Tominaga}, {Nomoto},
  {Valenti}, {Sahu}, {Minezaki}, {Yoshii}, {Yoshida}, {Anupama}, {Benetti},
  {Chincarini}, {Della Valle}, {Mazzali}, \& {Pian}}]{tanaka_09}
{Tanaka} M. {et~al.}, 2009, \apj, 692, 1131

\bibitem[{{Thompson}(2000)}]{thompson_00}
{Thompson} C., 2000, \apj, 534, 915

\bibitem[{{Thompson}, {Burrows} \& {Horvath}(2000){Thompson}, {Burrows}, \&
  {Horvath}}]{thompson_00_c}
{Thompson} T.~A., {Burrows} A., {Horvath} J.~E., 2000, \prc, 62, 035802

\bibitem[{{Tubbs}(1979)}]{tubbs_79}
{Tubbs} D.~L., 1979, \apj, 231, 846

\bibitem[{{Utrobin} \& {Chugai}(2011)}]{utrobin_11}
{Utrobin} V.~P., {Chugai} N.~N., 2011, \aap, 532, A100

\bibitem[{{Valentim}, {Rangel} \& {Horvath}(2011){Valentim}, {Rangel}, \&
  {Horvath}}]{valentim_11}
{Valentim} R., {Rangel} E., {Horvath} J.~E., 2011, \mnras, 414, 1427

\bibitem[{{Weiss} {et~al}\mbox{.}(2004){Weiss}, {Hillebrandt}, {Thomas}, \&
  {Ritter}}]{cox}
{Weiss} A., {Hillebrandt} W., {Thomas} H.-C., {Ritter} H., 2004, Cox and
  Giuli's Principles of Stellar Structure, by A.~Weiss, W.~Hillebrandt,
  H-C.~Thomas, H.~Ritter.~Cambridge, UK: Princeton Publishing Associates Ltd,
  2004.

\bibitem[{{Woosley} \& {Heger}(2007)}]{woosley_07}
{Woosley} S.~E., {Heger} A., 2007, \physrep, 442, 269

\bibitem[{{Yakunin} {et~al}\mbox{.}(2010){Yakunin}, {Marronetti}, {Mezzacappa},
  {Bruenn}, {Lee}, {Chertkow}, {Hix}, {Blondin}, {Lentz}, {Bronson Messer}, \&
  {Yoshida}}]{yakunin_10}
{Yakunin} K.~N. {et~al.}, 2010, Classical and Quantum Gravity, 27, 194005

\bibitem[{{Zhang} {et~al}\mbox{.}(2013){Zhang}, {Howell}, {Almgren}, {Burrows},
  {Dolence}, \& {Bell}}]{zhang_13}
{Zhang} W., {Howell} L., {Almgren} A., {Burrows} A., {Dolence} J., {Bell} J.,
  2013, \apjs, 204, 7

\end{thebibliography}

\end{document}